\documentclass[aps, 12pt, preprint, preprintnumbers,  
aps, amsmath,amssymb,nofootinbib,superscriptaddress,hyperref
]{revtex4-1} 
\usepackage{amssymb}
\usepackage{slashed}
\usepackage{amsmath}
\usepackage{braket}
\usepackage{float}
\usepackage[utf8]{inputenc}
\usepackage{graphicx}
\usepackage{subfigure}
\usepackage{latexsym}
\usepackage{epic}
\usepackage{eepic}
\usepackage{epsfig}
\usepackage{color}
\usepackage{natbib}
\usepackage[colorlinks=true,citecolor=darkred,urlcolor=darkred, pdfborder={0 0 0}]{hyperref}
\usepackage[normalem]{ulem}

\definecolor{darkred}{rgb}{0.6,0,0}

\definecolor{linkcolor}{rgb}{0,0,0.5}

\begin{document}
 \title{Sensitivity of  Lepton Number Violating Meson Decays in Different Experiments}
\author{Eung Jin Chun}\email{ejchun@kias.re.kr}
\affiliation{ Korea Institute for Advanced Study, Seoul 130-722, Korea}
\author{Arindam Das}\email{arindam.das@het.phys.sci.osaka-u.ac.jp}
\affiliation{Department of Physics, Osaka University, Toyonaka, Osaka 560-0043, Japan}
\author{Sanjoy Mandal}\email{smandal@iopb.res.in}
\affiliation{Institute of Physics, Sachivalaya Marg, Bhubaneswar 751005, India}
\affiliation{Homi Bhabha National Institute, BARC Training School Complex,
Anushakti Nagar, Mumbai 400094, India}
\author{Manimala Mitra}\email{manimala@iopb.res.in}
\affiliation{Institute of Physics, Sachivalaya Marg, Bhubaneswar 751005, India}
\affiliation{Homi Bhabha National Institute, BARC Training School Complex,
Anushakti Nagar, Mumbai 400094, India}
\author{Nita Sinha}
\email{nita@imsc.res.in}
\affiliation{The Institute of Mathematical Sciences,
C.I.T Campus, Taramani, Chennai 600 113, India}
\affiliation{Homi Bhabha National Institute, BARC Training School Complex, Anushakti Nagar, Mumbai 400094, India}
\preprint{\textbf{OU-HEP-1016}}
\preprint{\textbf{IP/BBSR/2019-4}}
\bibliographystyle{unsrt} 
\begin{abstract}
We study the discovery prospect of different three body lepton number violating~(LNV) meson decays $M_{1}^{-}\to\ell_{1}^{-}\ell_{2}^{-}M_{2}^{+}$ in the framework of right handed~(RH) neutrino extended Standard Model~(SM). We consider  a number of ongoing experiments, such as, 
NA62 and LHCb at CERN, Belle II at SuperKEK, as well as at the proposed future experiments, SHiP, MATHUSLA and FCC-ee.
The RH Majorana neutrino $N$ mediating these meson decays provides a resonant enhancement of the rates,
if the mass of $N$ lies in the range $(100\, \text{MeV}-6\, \text{GeV})$. We  consider  the effect of parent  mesons velocity, as well as, the effect of finite detector size. Using the expected upper limits on the number of events for the LNV decay modes, $M_{1}^{-} \to\ell_1^{-}\ell_2^{-}\pi^{+}$~($M_{1}=B, B_c,D, D_{s}\,\text{and}\,K$), we analyze the sensitivity reach of
the mixing angles $|V_{e N}|^{2}$, $|V_{\mu N}|^{2}$, $|V_{\tau N}|^{2}$, $|V_{e N}V_{\mu N}|$, $|V_{e N}V_{\tau N}|$ and $|V_{\mu N}V_{\tau N}|$ as a function of heavy  neutrino mass $M_{N}$. We show that, inclusion of parent meson velocity can account  to a  large difference for active-sterile mixing, specially for $D$, $D_s$ meson decay at SHiP and $K$ meson decay at NA62. Taking into account the velocity of the $D_s$ meson, the  future beam dump experiment SHiP can probe  $|V_{eN}|^2 \sim 10^{-9}$. For RH neutrino mass in between 2 - 5 GeV,   MATHUSLA can provide best sensitivity reach of 
active-sterile mixings.
\end{abstract}
\vspace{-3cm}
\maketitle
\section{INTRODUCTION}
The discovery of neutrino oscillations~\cite{GonzalezGarcia:2007ib} in a series of oscillation experiments have confirmed that neutrinos have non-zero masses and non-zero mixings. The solar and atmospheric mass splittings 
are $\mathcal{O}(10^{-5})$ and $\mathcal{O}(10^{-3})$ $\rm{eV}^2$, while the three mixings are $\theta_{12} \sim 33^{\circ}$, $\theta_{23} \sim 45^{\circ}$ and $\theta_{13} \sim 9^{\circ}$.  These observations indicate, at least two of the three SM neutrinos have non-zero masses. The absolute scale of the neutrino masses are yet unknown. The sum of  masses of three active neutrinos are bounded from cosmological observation  as  $\sum_i m_{\nu_i}<0.23$~eV~\cite{Lattanzi:2017ubx}. One of the most natural ansatz to explain  small neutrino masses is the seesaw mechanism, where the dimension-5 operator \cite{Weinberg:1979sa} with lepton and Higgs doublets generates Majorana mass of light neutrinos through electroweak symmetry breaking. This operator breaks global lepton number symmetry of the SM by two units. The other possibility is to generate  Dirac mass terms for  the SM neutrinos by including gauge singlet RH neutrinos in the theory. However, to explain eV mass of the neutrinos,  this requires unnatural fine-tuning of Yukawa coupling to a very small value  $Y_{\nu} \sim \mathcal{O}(10^{-11})$.  The seesaw mechanism on the other hand is most appealing, as the tiny  Majorana mass of the light neutrinos are  inversely proportional to the cut-off scale of the  theory. This large cut-off scale naturally explains eV mass of neutrinos. Seesaw  can be realised in  different beyond standard model~(BSM) extensions, such as Type-I~\cite{Minkowski:1977sc,type1seesaw2,Ramond:1979py,Mohapatra:1979ia}/Inverse seesaw~\cite{Mohapatra:1986aw,Mohapatra:1986bd} with gauge singlet Majorana neutrinos, Type-II~\cite{Magg:1980ut,Cheng:1980qt,Lazarides:1980nt,Mohapatra:1980yp} seesaw with Higgs triplets, and Type-III seesaw~\cite{Foot:1988aq,Ma:1998dn,Bajc:2006ia,Perez:2007rm} with fermionic triplet. For Type-I/Inverse seesaw, RH neutrinos can have Majorana/Quasi-Dirac masses, that can vary  in wide ranges, starting from GUT scale  down to  GeV. The low scale seesaw models, that inherits lighter  RH neutrino states   have higher discovery prospect, as they  can be tested   in a wide range of experiments. 
 
  Heavy SM gauge singlet RH neutrinos of mass GeV to TeV can be searched at LHC, via di-lepton \cite{Keung:1983uu, Sirunyan:2018xiv, ATLAS-CONF-2016-051, delAguila:2008cj,  Das:2017nvm,Das:2017gke}, as well as, tri-lepton final states \cite{delAguila:2008hw, Pascoli:2018heg, Das:2015toa, Das:2016hof,Sirunyan:2018mtv, Das:2017kkm}. The 13 TeV LHC tri-lepton search has constrained the mixing of the RH neutrinos with active neutrinos down to $|V_{\ell N}|^2 < 10^{-5}$ \cite{Sirunyan:2018mtv}, for the mass of RH  neutrino $M_N $ in between 10 GeV to 50 GeV.  For other future colliders, such as, $e^+ e^-$, FCC-hh,  and discussions on a future $e^- p$ collider, such as, LHeC, see \cite{Banerjee:2015gca, Mondal:2016kof, Antusch:2016ejd, Chakraborty:2018khw, Mandal:2018qpg, Das:2018usr, Ruiz:2017yyf, Cai:2017mow}.  For  heavy  RH neutrino of mass $M_N \sim $ TeV, the final decay products will be collimated and will produce fat-jets \cite{Bhardwaj:2018lma, Chakraborty:2018khw, Das:2018usr,Das:2017gke}. For the discussion on sub weak scale  RH neutrino state, that can produce lepton-jet, see 
  \cite{Izaguirre:2015pga}. The other mass ranges of RH  neutrinos,  such as, MeV-GeV  are tightly constrained from 
 the lepton number violating neutrinoless double beta decay ($0\nu\beta \beta$-decay) \cite{Avignone:2007fu,KlapdorKleingrothaus:2000sn, Benes:2005hn, Mitra:2011qr, Pascoli:2013fiz}, while in  GeV - TeV range, lepton flavor violating process $\mu \to e \gamma$~\cite{Ma:1980gm,Duerr:2011ks,Heusch:1993qu} can give significant constraint. RH  neutrinos have also been searched for in the laboratory experiment through peak searches in leptonic decays of pions and kaons~\cite{Shrock:1980vy,Atre:2009rg}.   
 
 Another interesting probe for Majorana neutrinos of  hundreds of  MeV- few GeV masses and their mixings are the lepton number violating (LNV) rare meson decay processes, such as $M_{1}^{-}\to\ell_1^{-}\ell_2^{-}\pi^{+}$~\cite{Atre:2009rg,Helo:2010cw,Cvetic:2010rw,Cvetic:2019shl,Milanes:2018aku,Li:2018pag,Abada:2017jjx,Yuan:2017uyq,Mejia-Guisao:2017gqp,Cvetic:2017vwl,Mandal:2017tab,Mandal:2016hpr,Cvetic:2016fbv,Milanes:2016rzr, Antusch:2017hhu}. For  RH  neutrino search, this process has an
  advantage as compared to the LNV $0\nu \beta \beta$-decay due to the lesser   uncertainties in the meson decay constant. In the later process, the nuclear matrix elements~(NME) uncertainty can make difference in the prediction of active-sterile mixing. We consider a number of  three body $\Delta L=2$ meson decays $M_1^{-}\to\ell_1^{-}\ell_2^{-}\pi^{+}$($M_1=B, B_c, D, D_s\,\text{and}\,K$) and derive sensitivity reach of the active-sterile mixing parameter in a number of ongoing and future experiments, such as,   NA62, LHCb at CERN, Belle II, SHiP, MATHUSLA and FCC-ee.  Note that, the light neutrino contribution to these  meson decays are extremely suppressed. However,   for  RH  neutrinos in the mass range of $0.140\,\text{GeV}<M_N<6\,\text{GeV}$, the intermediate Majorana  neutrinos can be produced on-shell. This results  in  resonant enhancement of these decay rates. In addition, large number of decaying mesons in these experiments, in-particular at SHiP will facilitate to improve the sensitivity reach significantly. In deriving these results,  we consider   parent meson velocity, that affects the probability of RH neutrino decay  inside the detector. We show, that inclusion of parent meson velocity can give one or two orders of magnitude shift $\mathcal{O}(10^1-10^2)$ in the results obtained. \\

The paper is organized as follows. In sec.~\ref{model}, we  very briefly review the basic features of the three RH neutrino framework, following which in sec.~\ref{sec:amplitude}, we
discuss in detail the contributions of the RH neutrino in meson decays. In sec.~\ref{Decay width}, we then compute the total
decay width of RH neutrino $N$ in the mass range~$0.140\,\text{GeV}\leq M_{N} \leq 6\,\text{GeV}$. In sec.~\ref{velocity and finite detector size effect}, we study the effects of parent meson
velocity in the RH neutrino decay probability. In sec.~\ref{signal events} and \ref{input for different experiments}, we give the formalism to calculate the signal events and give different inputs
for various experiments which we consider.
In sec.~\ref{results}, we derive the limits on the mixing angle $|V_{\ell N}|^{2}$, $\ell=e,\mu,\tau$ and $|V_{\ell_1 N}V_{\ell_2 N}|$, $\ell_1,\ell_2=e,\mu,\tau,\ell_1\neq\ell_2$ that are expected from the upper limits on the number of signal events of various LNV meson
decays,  that may be achievable in some of the ongoing and future experiments. In secs.~\ref{combined limit}, we present our combined limit from all the considered meson decays and give the comparison with other existing constraints on the mixing angles. Finally in sec.~\ref{conclusion}, we provide our conclusions. In the Appendix, we provide  
details of the RH neutrino decay width calculations.
\section{The Model}
\label{model}
We extend the SM to include additional RH neutrinos $N$. The heavy neutrinos can generate light neutrino masses through seesaw. For simplicity, we consider only one RH neutrino and carry out our analysis. The mixing of $N$ with the active neutrinos are given by the following expression,
\begin{align}
 \nu_{\ell}=\sum_{m=1}^{3}U_{\ell m}\nu_{m}+V_{\ell N}N_{m'}^{c},
\end{align}
where $\nu_{m}$ and $N_{m'}$ are the mass eigenstates.
We denote the mixing between the standard flavour neutrino
$\nu_{\ell}$~($\ell=e,\mu,\tau$) and the heavy mass eigenstate $N$ by $V_{\ell N}$. Due to the mixing, the charged and neutral currents in the lepton sector gets modified and can be written as
\begin{align}\label{interaction lagrangian}
&\mathcal{L}_{\ell}^{CC}=-\frac{g}{\sqrt{2}}W_{\mu}^{+}\left(\sum_{\ell=e}^{\tau}\sum_{m=1}^{3}U_{\ell  m}^{*}\bar{\nu}_{m}\gamma^{\mu}P_{L}\ell+
\sum_{\ell=e}^{\tau}V_{\ell N}^{*}\overline{N_{m'}^{c}}\gamma^{\mu}P_{L}\ell\right)+h. c,\\
&\mathcal{L}_{\ell}^{NC}=-\frac{g}{2\cos\theta_{W}}Z_{\mu}\left(\sum_{\ell =e}^{\tau}\sum_{m=1}^{3}U_{\ell  m}^{*}\bar{\nu}_{m}\gamma^{\mu}P_{L}\nu_{\ell}+
\sum_{\ell=e}^{\tau}V_{\ell N}^{*}\overline{N_{m'}^{c}}\gamma^{\mu}P_{L}\nu_{\ell}\right)+h.c.
\end{align}
For the purpose of phenomenology, we consider the mass and mixings of $N$ as free parameters,
constrained only by experimental observations. Note that, adding only one RH neutrino is not enough to correctly reproduce the neutrino oscillations parameters, namely two mass square differences and the mixings. In our considered model, we can add two more RH neutrinos to generate the neutrino masses and consider two of them to be heavy enough such that only one RH neutrino lies in the mass range $0.140\,\text{GeV}\leq M_N \leq 6\,\text{GeV}$. 
\section{Process}
\label{sec:amplitude}
\begin{figure}
	\centering
	\includegraphics[width=0.45\textwidth]{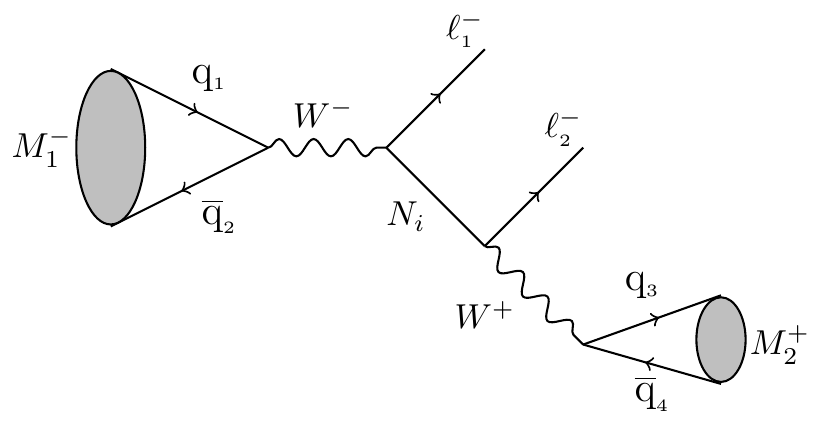}
        \includegraphics[width=0.45\textwidth]{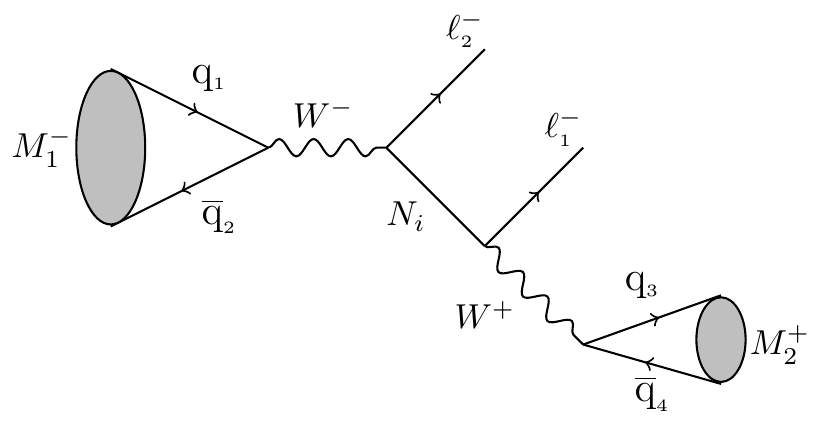}
	\caption{\small{The Feynman diagrams for the lepton number violating meson decays. These processes can produce resonance enhancement. See text for details}}
	\label{feynman diagram for LNV meson decay}
\end{figure}

\begin{figure}[h]
	\centering
        \includegraphics[width=0.45\textwidth]{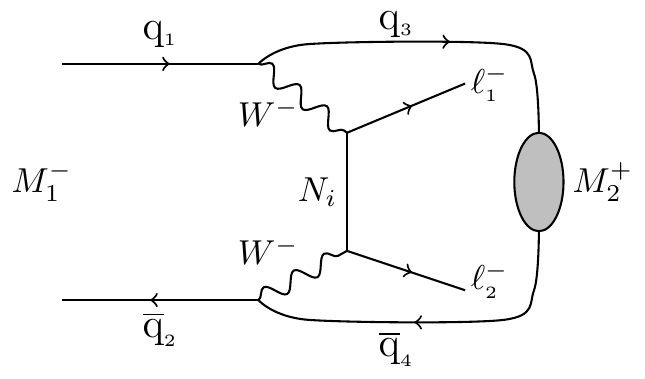} 
	\caption{\small{The t-channel diagram for the lepton number violating meson decay. See text for details.}}
	\label{t-channel diagram}
\end{figure}

The RH neutrino $N$, if a Majorana state can mediate the LNV process $M_{1}^{-}\to\ell_{1}^{-}\ell_{2}^{-}M_{2}^{+}$.
The Feynman diagrams for these decays are shown in Figs.~\ref{feynman diagram for LNV meson decay} and ~\ref{t-channel diagram}.
The diagram in Fig.~\ref{t-channel diagram} will give a very small contribution, as this is not a resonance production diagram. Note that the diagrams with light neutrino exchange are also present but the contributions will be negligibly small as they will not be
resonantly enhanced.
The decay amplitude for the processes $M_{1}^{-}(p)\to\ell_{1}^{-}(k_{1})\ell_{2}^{-}(k_{2})M_{2}^{+}(k_{3})$ depicted in Fig.\ref{feynman diagram for LNV meson decay} can be expressed as
\begin{align}
\mathcal{M}&=G_{F}^{2}V^{CKM}_{M_{1}}V^{CKM}_{M_{2}}f_{M_{1}}f_{M_{2}}(V_{\ell_{1}N}V_{\ell_{2}N})\frac{M_{N}}{(p-k_{1})^{2}-M_{N}^{2}+iM_{N}\Gamma_{N}}\nonumber \\
&\bar{u}(k_{1})\slashed{p}\slashed{k}_{3}(1+\gamma_{5})v(k_{2}),
\label{amplitude}
\end{align}
where $M_{1}$ and $M_{2}$ are both pseudo-scalar mesons. Though $M_2$ can also be a vector meson, we have considered only the case of pseudo-scalar meson.
In Eq.~\ref{amplitude}, $G_{F}$ is the Fermi coupling constant, {$V_{\ell_{j}N}$ are the mixing angles between the neutrino of flavor state $\nu_{\ell_{j}}$ and mass eigenstate $N$. 
$V_{M_{1}}^{CKM}~(V_{M_{2}}^{CKM})$ are the Cabbibo-Kobayashi-Maskawa (CKM) matrix elements at the annihilation
(creation) vertex of the meson $M_1$($M_2$).  $f_{M_{1}}$ and $f_{M_{2}}$ are the decay constants of $M_{1}$ and $M_{2}$, respectively. We use the values $f_{D}=0.204$~GeV, $f_{D_{s}}=0.258$~GeV,
$f_{K}=0.156$~GeV, $f_{B}=0.188$~GeV and $f_{B_c}=0.436$~GeV~\cite{Nakamura:2010zzi}. $M_{N}$, $\Gamma_{N}$ are
the mass and decay width of the heavy 
neutrino $N$. Finally, the total decay rate is given by
\begin{align}
\Gamma\left(M_{1}\to\ell_{1}\ell_{2}\pi\right)=\frac{1}{n!}\sum|\mathcal{M}|^{2}d_{3}(\text{PS}).
\label{decay rate}
\end{align}
In Eq.~\ref{decay rate}, $n=2$ for identical final leptons, otherwise $n=1$ and $d_{3}(\text{PS})$ is the three body phase space.
\section{TOTAL DECAY WIDTH OF $N$}
\label{Decay width}
Although the RH neutrino $N$ is a SM singlet, due to mixing with active neutrinos it can decay via charged and neutral current interactions. 
For the mass range~$0.140\,\text{GeV}\leq M_{N}\leq 6\,\text{GeV}$, RH neutrino can be produced as an intermediate on mass shell state in the LNV meson decays being considered here.
We consider only tree level diagrams in the calculation of RH neutrino total decay width. In the relevant mass range~$0.140\,\text{GeV}\leq M_{N}\leq 6\,\text{GeV}$, the following channels contribute to
the total decay width of heavy neutrino:
\begin{itemize}
 \item $N$ $\to\ell^{-}P^{+}$, where $\ell=e,\mu,\tau$, and $P^{+}$ $=\pi^{+},\,K^{+},\,D^{+},\,D_{s}^{+},\,B^{+}$~(for $\ell=e,\mu$). 
\item $N\rightarrow\nu_{\ell}P^{0}$, where $\nu_{\ell}$ are the flavor eigenstates $\nu_{e},\,\nu_{\mu},\,\nu_{\tau}$ and $P^{0}=\pi^{0},\,\eta,\,\eta',\,\eta_{c}$.
\item $N\rightarrow\ell^{-}V^{+}$, where $\ell~=~e,\,\mu,\,\tau$, and $V^{+}=\rho^{+},
K^{*+},\,D^{*+},\,D_{s}^{*+},\,B^{*+}$~(for $\ell=e,\mu$).
\item $N\rightarrow\nu_{\ell}V^{0}$, where $\nu_{\ell}=\nu_{e},\,\nu_{\mu},\,\nu_{\tau}$ and $V^{0}=\rho^{0},\,\omega,\,\phi,\,J/\psi$. 
\item $N\rightarrow\ell_{1}^{-}\ell_{2}^{+}\nu_{\ell_{2}}$, where $\ell_{1},\,\ell_{2}=e,\,\mu,\,\tau$, $\ell_{1}\neq
\ell_{2}.$
\item $N\rightarrow\nu_{\ell_{1}}\ell_{2}^{-}\ell_{2}^{+}$, where $\ell_{1},\,\ell_{2}=e,\,\mu,\,\tau$. 
\item $N\rightarrow\nu_{\ell_{1}}\nu\overline{\nu}$, where $\nu_{\ell_{1}}=\nu_{e},\,\nu_{\mu},\,\nu_{\tau}$.
\end{itemize}
Hence, the total decay width is given by
\begin{equation}
\begin{split}
&\Gamma_{N}=\sum_{\ell,P^{+}}2\Gamma^{\ell P^{+}}+\sum_{\ell,P^{0}}\Gamma^{\nu_{\ell}P^{0}}+\sum_{\ell,V^{+}}2\Gamma^{\ell V^{+}}+\sum_{\ell,V^{0}}\Gamma^{\nu_{\ell}V^{0}}\\
&+\sum_{\ell_{1},\ell_{2}(\ell_{1}\neq\ell_{2})}2\Gamma^{\ell_{1}\ell_{2}\nu_{\ell_{2}}}+\sum_{\ell_{1},\ell_{2}}\Gamma^{\nu_{\ell_{1}}\ell_{2}\ell_{2}}+\sum_{\nu_{\ell_{1}}}\Gamma^{\nu_{\ell_{1}}\nu\overline{\nu}}.
\end{split}
\label{Total decay width of N}
\end{equation}
As the RH neutrino is Majorana, the charge conjugate processes are also allowed and the decay rate is same, hence the 2 factor is included for some of the channels.
We can parameterize the above decay width as follows
\begin{align}
\Gamma_{N}=a_{e}(M_{N})|V_{eN}|^{2}+a_{\mu}(M_{N})|V_{\mu N}|^{2}+a_{\tau}(M_{N})|V_{\tau N}|^{2},
\end{align}
where, $a_e$, $a_\mu$ and $a_\tau$ are functions of the Majorana neutrino mass and hence will differ from mode  to  mode. 
We show the total decay width of heavy neutrino $N$ in the left panel of Fig.~\ref{N decay width and al coefficient}, for the choice of mixing $|V_{e N}|^{2}=|V_{\mu N}|^{2}=|V_{\tau N}|^{2}=1$.
Even with mixing angle equal to 1, the decay width lies in the range $10^{-16}\,\text{GeV}\leq\Gamma_{N}\leq 10^{-7}\,\text{GeV}$ for the $M_{N}$ mass range,
$0.140\,\text{GeV}\leq M_{N}\leq 6\,\text{GeV}$.
Hence, we can safely use the narrow-width approximation, 
$\frac{1}{(p_{N}^{2}-M_{N}^{2})^{2}+M_{N}^{2}\Gamma_{N}^{2}}\approx\frac{\pi}{M_{N}\Gamma_{N}}\delta(p_{N}^{2}-M_{N}^{2})$
and $\Gamma(M_{1}\to\ell_1\ell_2 M_{2})$ can be approximated as, $\Gamma(M_{1}\to\ell_1 N)\text{Br}(N\to\ell_2 M_{2})$.
 \begin{figure}[h]
 	\centering
 	\includegraphics[width=0.47\textwidth]{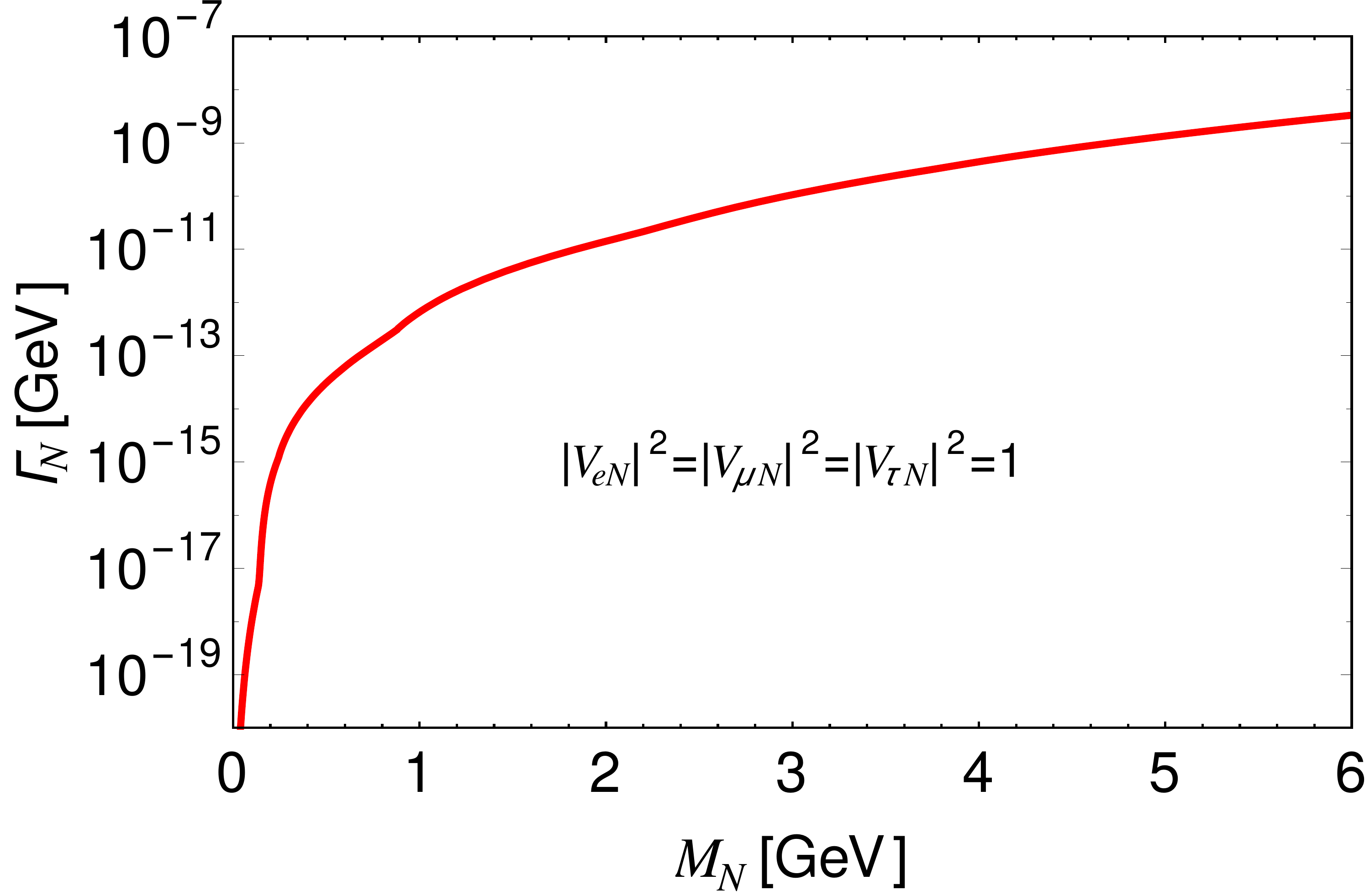}
 	\includegraphics[width=0.47\textwidth]{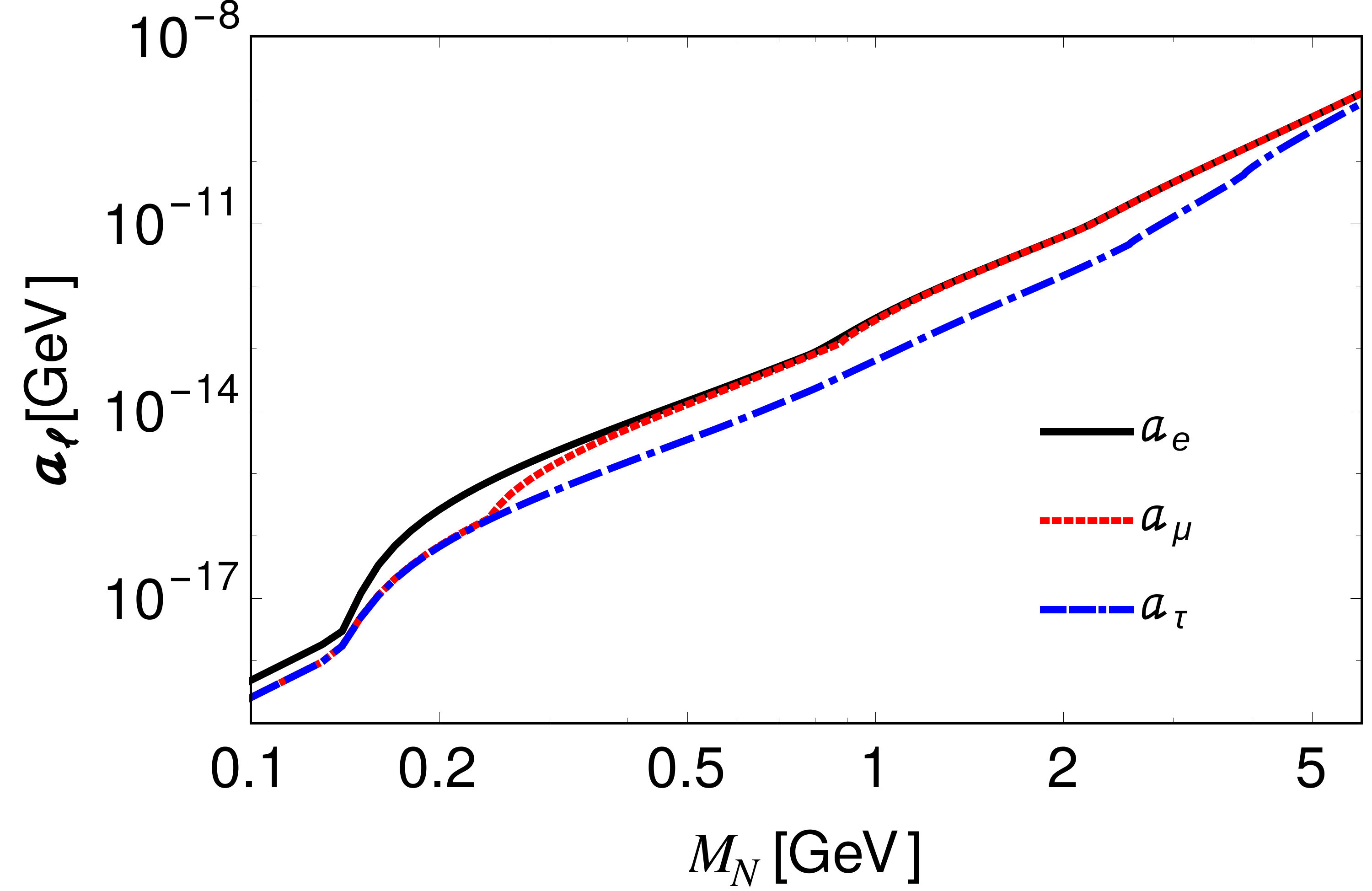}
 	\caption{\small{Left panel: The total decay width of heavy neutrino N with the assumption, $|V_{e N}|^{2}$=$|V_{\mu N}|^{2}$=$|V_{\tau N}|^{2}=1$. Right panel: $a_{e}$, $a_{\mu}$ and $a_{\tau}$ as function of mass $M_{N}$.}}
 	\label{N decay width and al coefficient}
 	\end{figure}
We also show in the right panel of Fig.~\ref{N decay width and al coefficient}, the different coefficients $a_e$, $a_\mu$ and $a_\tau$ as a function of RH neutrino mass $M_N$.
For the RH neutrino mass $M_N$ upto 0.25~GeV, $a_\mu\approx a_\tau$ and for mass $M_N>0.4$~GeV, $a_e\approx a_\mu$.
\section{Parent meson velocity and finite detector size effect}
\label{velocity and finite detector size effect}
For the mass range~$0.140\,\text{GeV}\leq M_{N}\leq 6\,\text{GeV}$, the RH neutrino produced in these LNV meson decays are on shell. The RH neutrino produced in meson decays
$M_{1}\to\ell_{1}N$, propagates and decays after traveling some distance from the production point. This is the decay length $L_{N}$ of the RH neutrino $N$ and it depends on the 
total decay width of $N$. If $L_{N}$ is greater than the actual size of the detector, then $N$ will decay outside the detector and the signature $M_{1}\to\ell_{1}\ell_{2}\pi$ cannot
be observed. For a particular experiment the detector size is finite. Hence, when calculating the signal events, we need to take into account this finite size detector effect by the probability factor $\mathcal{P}_{N}$, of $N$ to decay within the detector. In general, this probability factor can be written as
\begin{align}
 \mathcal{P}_{N}&=1-\text{exp}\left(-L_{D}\Gamma_{N}\frac{M_{N}}{p_{N}}\right)=1-\text{exp}\left(-\frac{L_{D}}{L_{N}}\right),
\end{align}
where $L_{N}=\frac{p_{N}}{M_N\Gamma_N}$, $L_{D}$ is the detector length, $p_{N}$ is the three momentum of $N$. Defining $x=\frac{L_{D}}{L_N}$, it is obvious that for a very large detector length $L_D$ and small decay length $L_N$, $\mathcal{P}_N=1-\text{exp}(-x)\approx 1$.
\begin{figure}
	\centering
	\includegraphics[width=0.45\textwidth]{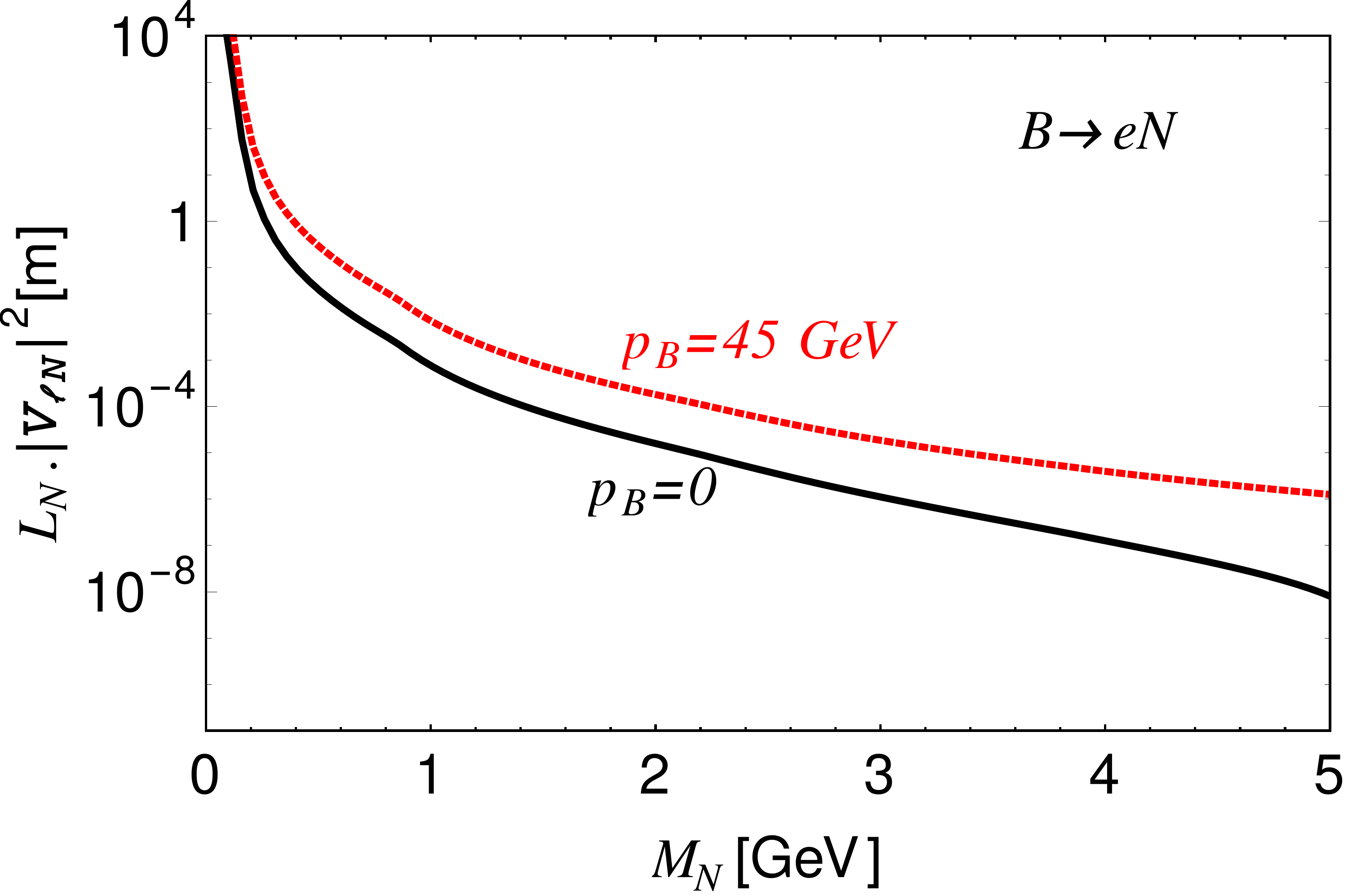}
	\includegraphics[width=0.45\textwidth]{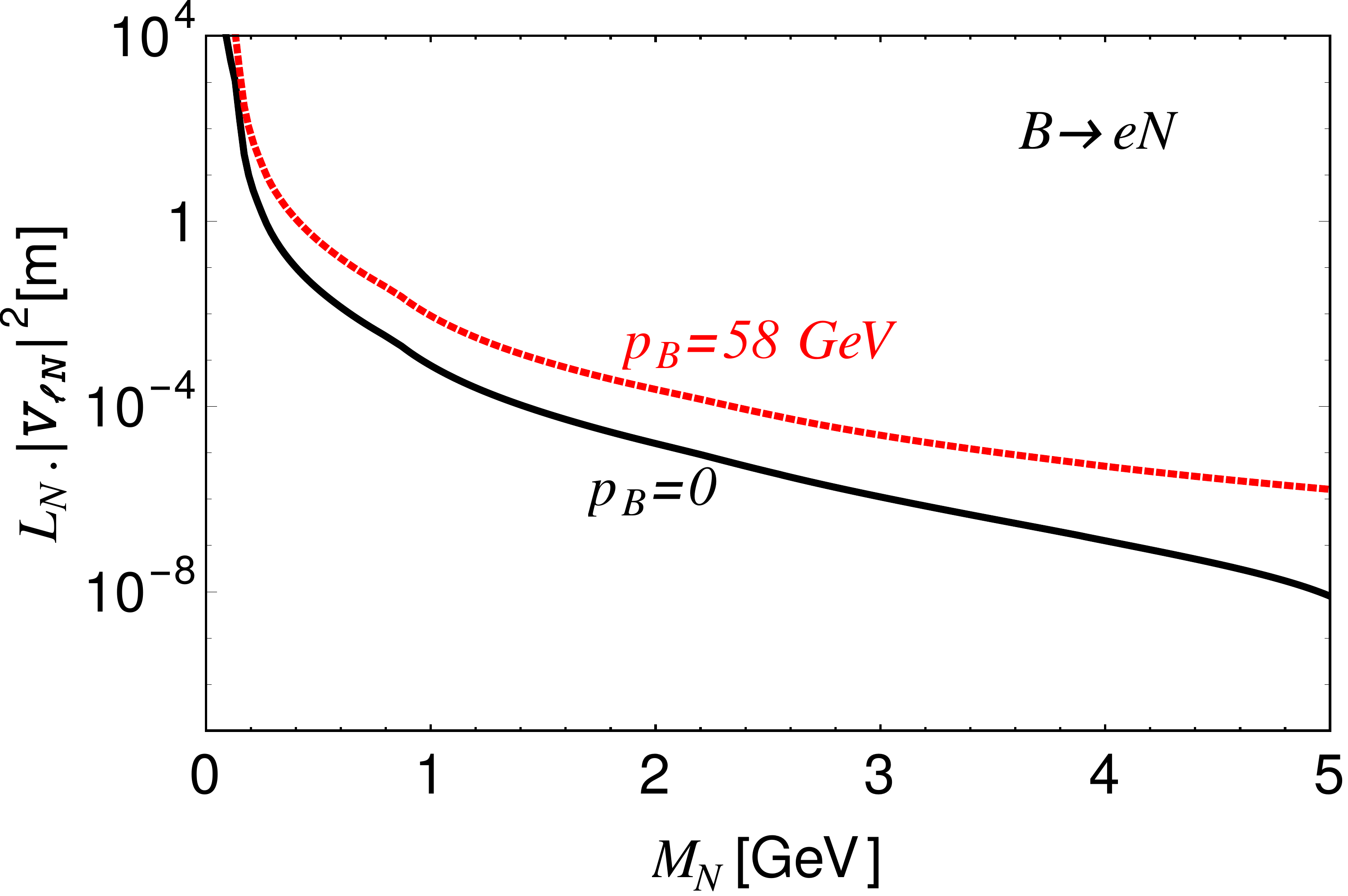}
	\includegraphics[width=0.45\textwidth]{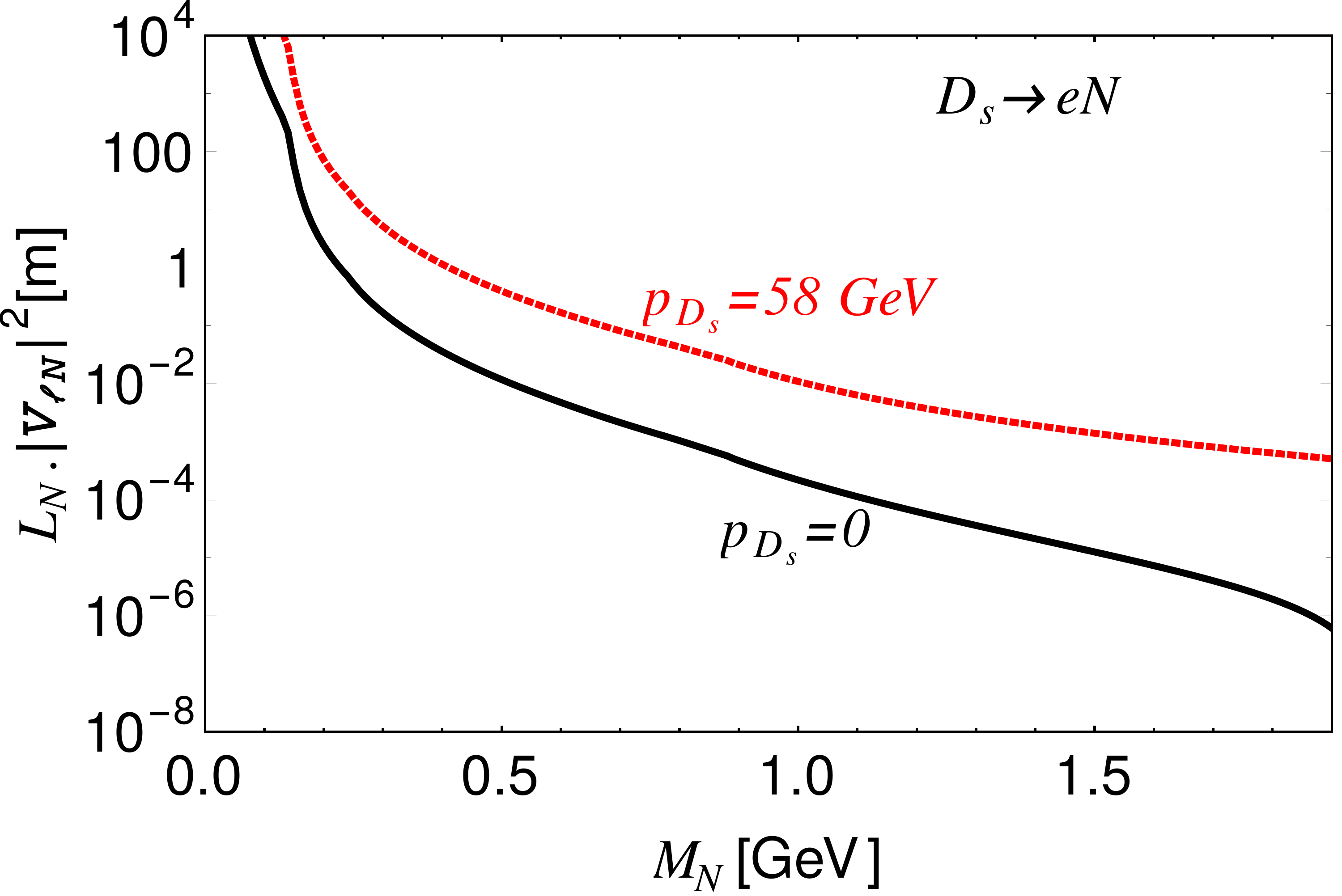}
		\includegraphics[width=0.45\textwidth]{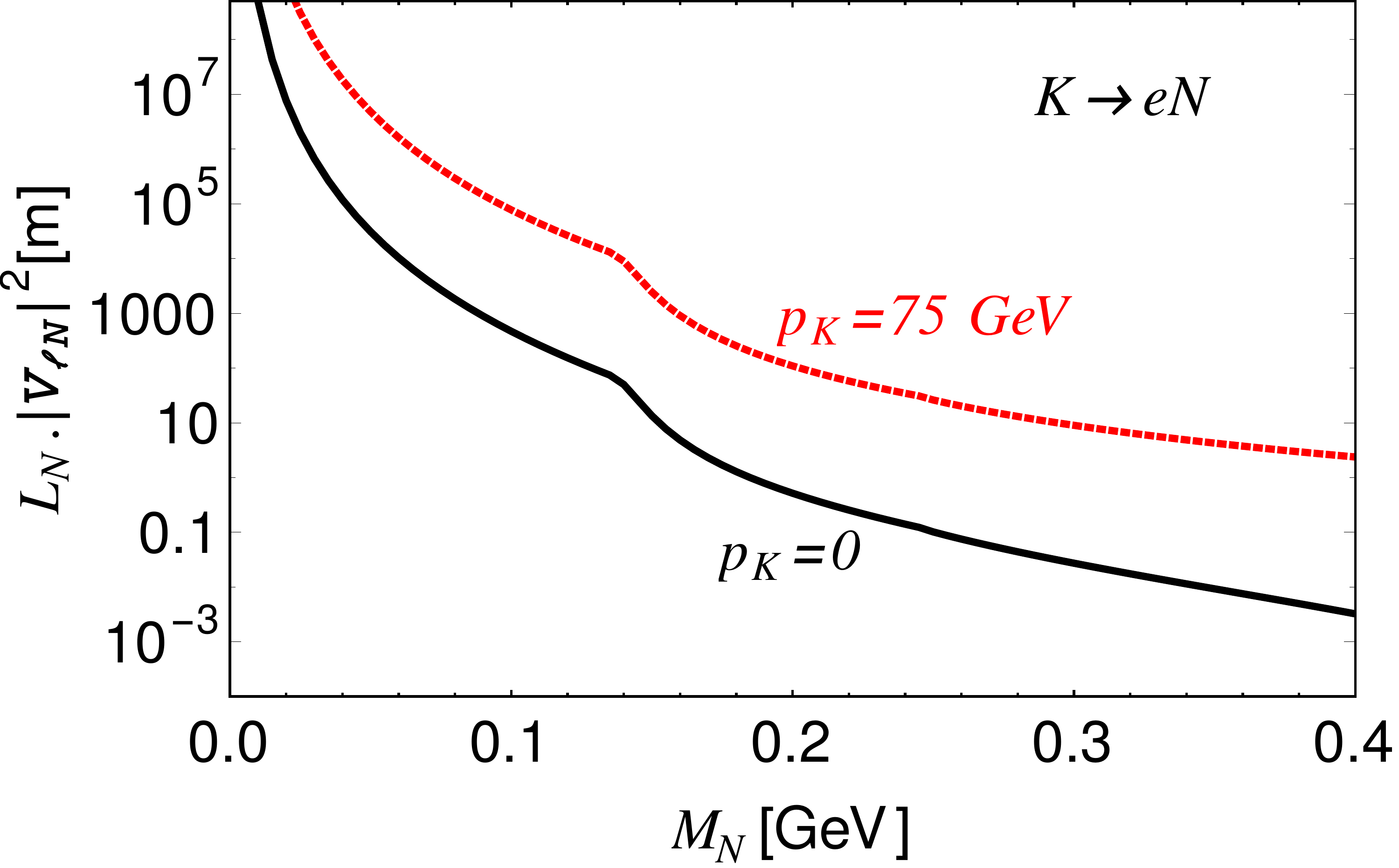}
	\caption{\small{Dependence of the decay length $L_{N}$ on parent meson velocity as a function of RH neutrino mass $M_N$. The upper panel is for $B$ meson decay at Belle-II~(left) and SHiP~(right). The left figure of lower panel is for $D_s$ meson decay at SHiP and the right figure of lower panel is for $K$ meson decay at NA62.}}
	\label{LN with velocity}
\end{figure}
Note that the probability factor depends on three momentum $p_{N}$, which in turn depends on the velocity of decaying meson $M_1$. Hence to incorporate the probability factor correctly, we need to use the correct velocity of the parent
meson $M_{1}$ in each of the experiments. If the parent meson $M_{1}$ decays at rest, three momentum $p_{N}$ is fixed and is given by $p_{N}^{*}=\frac{m_{M_{1}}}{2}\lambda^{\frac{1}{2}}\big(1,\frac{m_{\ell}^{2}}{m_{M_{1}}^{2}},
\frac{M_{N}^{2}}{m_{M_{1}}^{2}}\big)$. For the case of parent meson $M_{1}$ produced with fixed boost $\vec{\beta}$, the energy of $N$ is then given by,
\begin{align*}
E_{N}=E_{N}^{*}\left(\gamma+\frac{p_{N}^{*}}{E_{N}^{*}}\sqrt{\gamma^{2}-1}~\text{cos}~\theta_{N}^{*}\right),
\end{align*}
where $E_{N}^{*}$ is the energy of $N$ in rest frame of $M_{1}$ which is given as $E_{N}^{*}=\sqrt{p_{N}^{*2}+M_{N}^{2}}$. $\gamma=\frac{E_{M_{1}}}{m_{M_{1}}}$ and
$\theta_{N}^{*}$ is the emission
angle of particle $N$ in the rest frame of $M_{1}$, which is measured from the boost direction $\vec{\beta}$.
The energy  $E_{N}$ of the $N$ in the boosted $M_1$ frame lies within the range, 
\begin{align}
{E_{N}}\in [E_{N}^{-},\,E_{N}^{+}]=\left[\big(\gamma E^*_{N}-{p_{N}^{*}}\sqrt{\gamma^{2}-1}\big),\big(\gamma E^*_{N}+{p_{N}^{*}}\sqrt{\gamma^{2}-1}\big)\right]
\label{range of EN}
\end{align}
Similarly we can derive the range of $p_{N}\in [p_{N}^{-},\,p_{N}^{+}]$ from Eq.~\ref{range of EN} using the relation $p_{N}^{\pm}=\sqrt{E_{N}^{\pm 2}-M_{N}^{2}}$.
\begin{figure}
	\centering
		\includegraphics[width=0.45\textwidth]{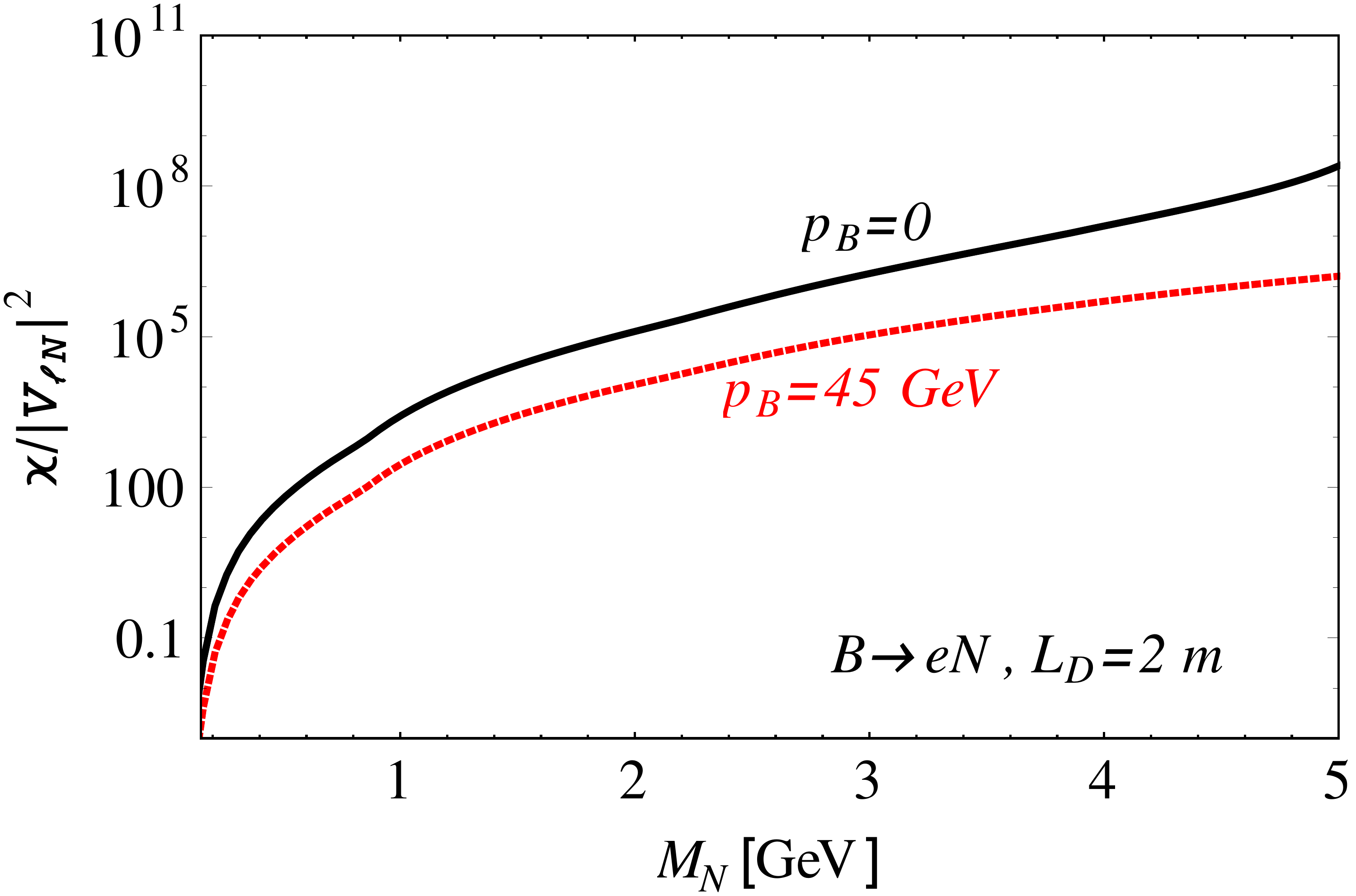}
	\includegraphics[width=0.45\textwidth]{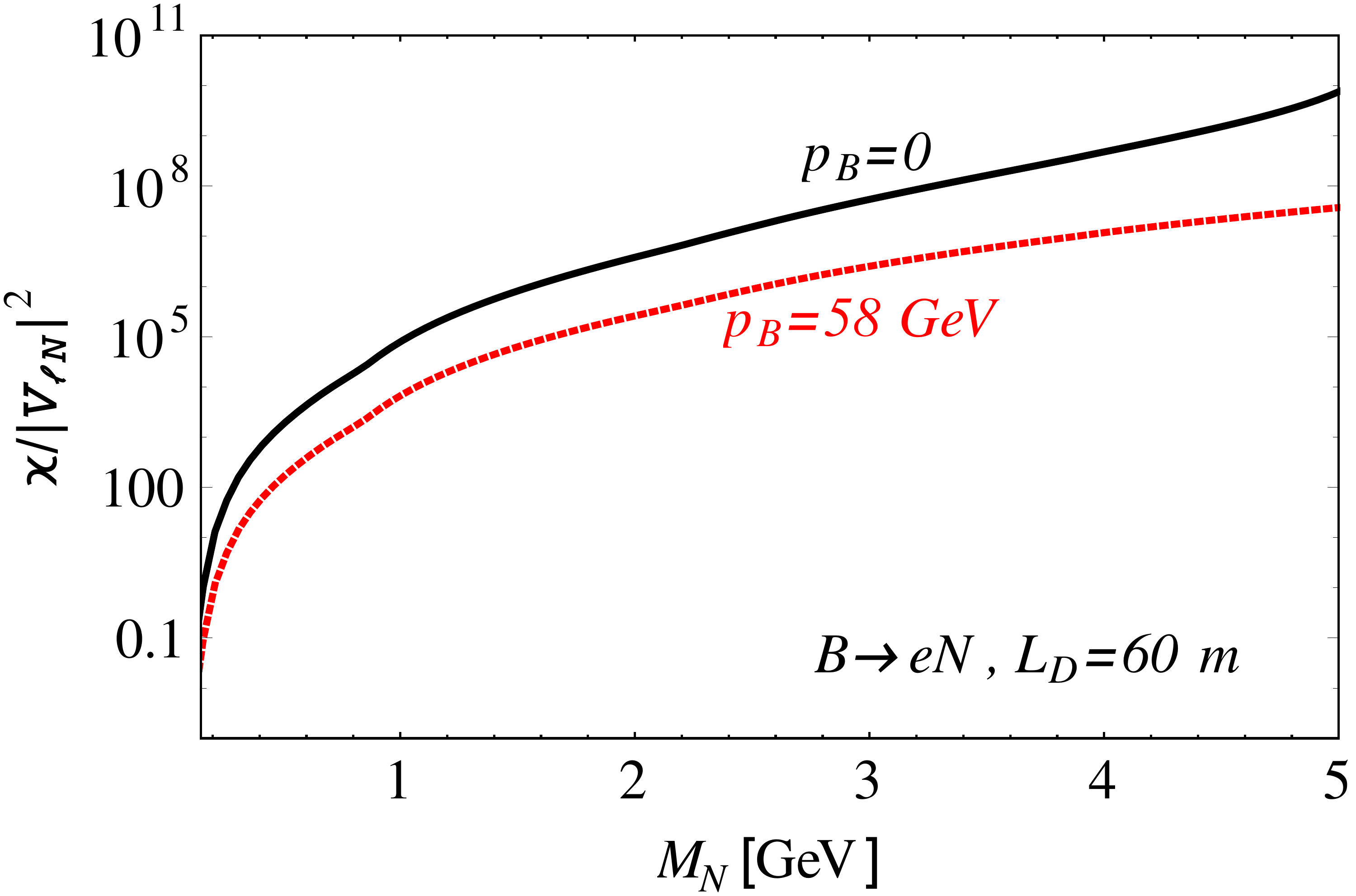}
		\includegraphics[width=0.45\textwidth]{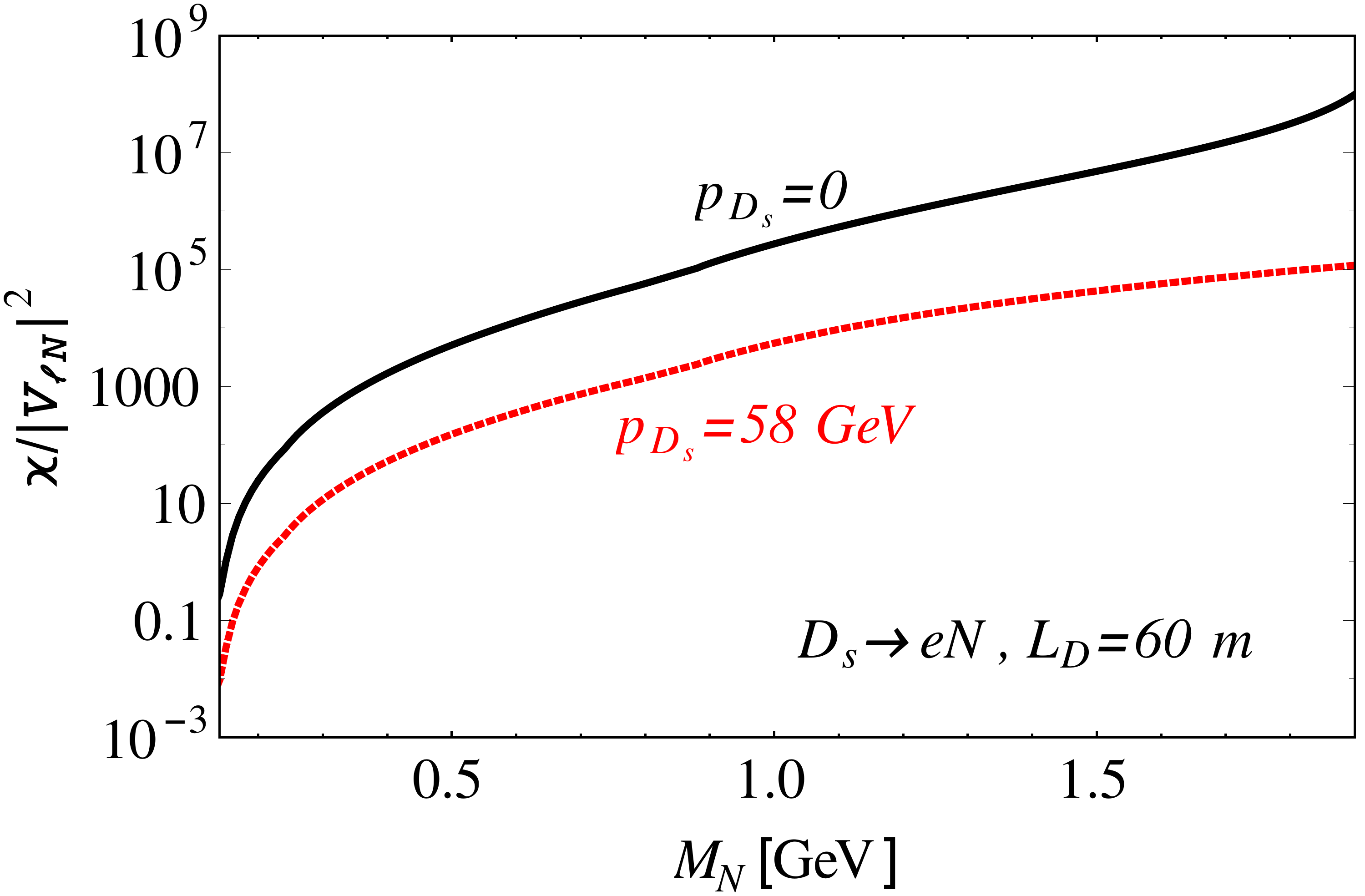}
	\includegraphics[width=0.45\textwidth]{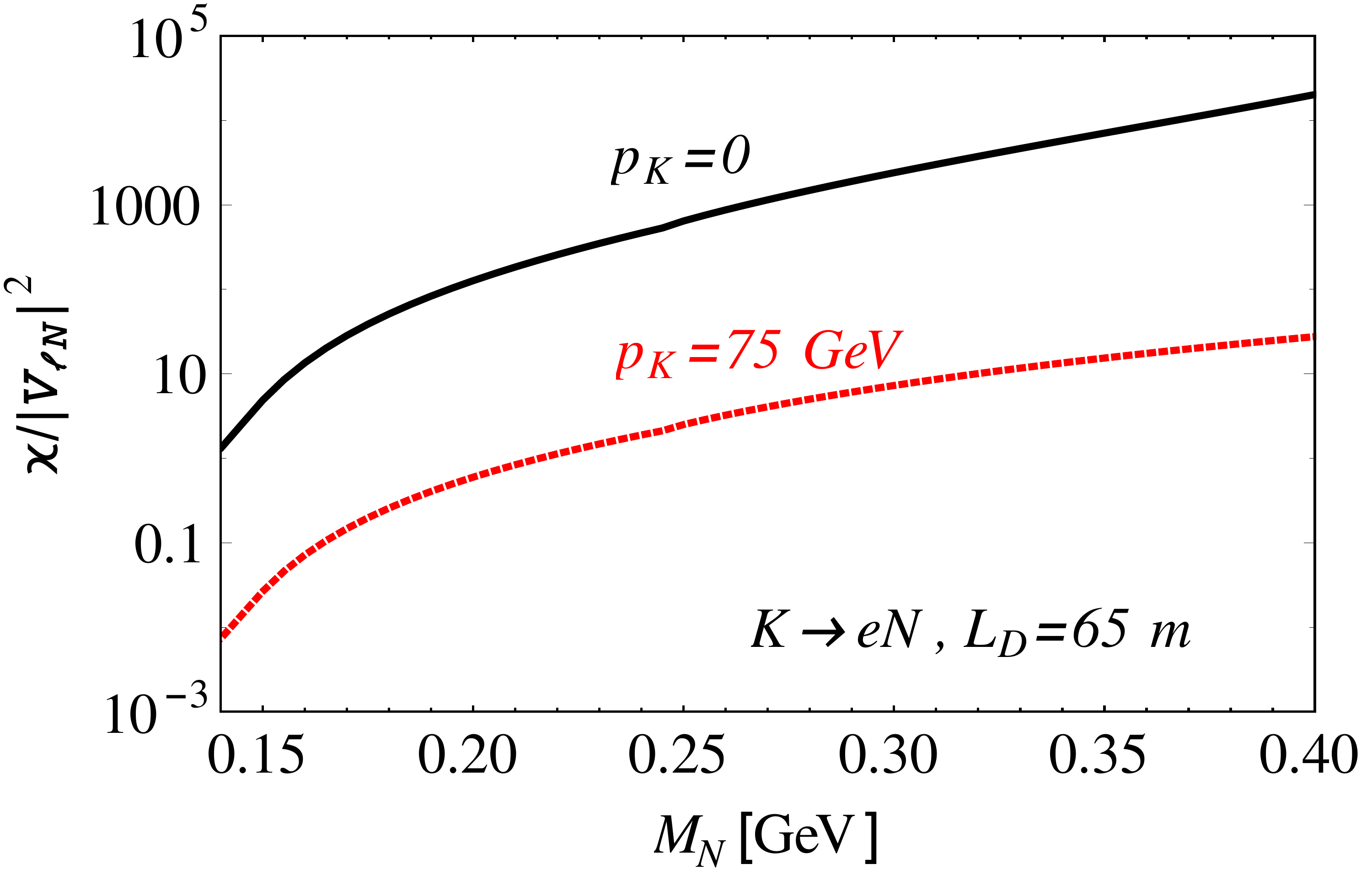}
	\caption{\small{Dependence of parameter $x=\frac{L_{D}}{L_{N}}$ on parent meson velocity as a function of RH neutrino mass $M_N$. The upper panel is for $B$ meson decay at Belle-II~(left) and SHiP~(right). The left figure of lower panel is for $D_s$ meson decay at SHiP and the right figure of lower panel is for $K$ meson decay at NA62.}}
	\label{LDByLN with velocity}
\end{figure}
In this section, we show how $x$, $L_{N}$ depends on parent meson velocity and compare to the case of parent meson decay at rest. For the case of meson decay at rest,
$p_{N}=p_{N}^{*}$ and for meson decay with non-zero momentum $p_{M_{1}}$, we take $p_{N}=\frac{p_{N}^{-}+p_{N}^{+}}{2}$ to compare.
With the assumption of $|V_{e N}|^{2}$=$|V_{\mu N}|^{2}$=$|V_{\tau N}|^{2}$, we can write the decay length $L_{N}$ and $x$ as
\begin{align}
 L_{N}&=\frac{p_{N}}{M_{N}\Gamma_{N}}=\frac{p_{N}}{M_{N}|V_{\ell N}|^{2}\left(a_e(M_{N})+a_{\mu}(M_{N})+a_{\tau}(M_{N})\right)}\\
 x&=\frac{L_{D}}{L_{N}}=\frac{L_{D}}{p_{N}}M_{N}|V_{\ell N}|^{2}\left(a_e(M_{N})+a_{\mu}(M_{N})+a_{\tau}(M_{N})\right)
 \label{Ln and x with meson velocity}
\end{align}
In Figs~\ref{LN with velocity} and \ref{LDByLN with velocity}, we have shown the variations of $L_{N}.|V_{\ell N}|^{2}$ and $\frac{x}{|V_{\ell N}|^{2}}$ as a function of RH neutrino
mass $M_{N}$ in $B$, $D_{s}$ and $K$ meson decays. To do the comparison with the meson decay at rest~($p_{M_1}=0$), we take $p_{B}=45$~GeV~(FCC-ee)~\cite{FCC}, 58 GeV~(SHiP)~\cite{Gorbunov:2007ak}; $p_{D_s}=58$ GeV~(SHiP)
and $p_{K}=75$ GeV~(NA62)~\cite{NA62page}. From these two figures it is clear that decay length increases~(hence $x$ decreases) for fixed mixing angle in the case of meson decays in flight compared to meson
decay at rest. Hence, the probability of RH neutrino $\mathcal{P}_N$ to decay inside the detector is smaller in the case of meson decay in flight comapare to meson decay at rest. As a result, compare to meson decay at rest, in the case of meson decay in flight we get a rather loose bound on
mixing angle from the expected signal events.
\section{Signal Events} 
\label{signal events}
The sensitivity reach for the LNV decay modes in a particular experiment depends on the number of the parent mesons $M_{1}$'s produced
($N_{M_1^-}$), their momentum ($\vec{p}_{M_{1}}$) 
and the branching ratio for these mesons to the LNV modes. Assuming the parent meson $M_1$ decays at rest, 
the expected number of signal events is~\cite{Asaka:2016rwd}:
\begin{align}\label{event signal in rest}
N_{\text{event}}&=2N_{M_{1}^{-}}\text{Br}\left(M_{1}^{-}\to\ell_{1}^{-}\ell_{2}^{-}M_{2}^{+}\right)\mathcal{P}_{N},\nonumber \\
&\approx 2N_{M_{1}^{-}}\text{Br}\left(M_{1}^{-}\to\ell_{1}^{-}N\right)
\frac{\Gamma(N\to\ell_{2}^{-}M_{2}^{+})}{\Gamma_{N}}\mathcal{P}_{N}, 
\end{align}
the factor 2 is due to inclusion of the charge conjugate process
$M^{+}_1 \to \ell_{1}^{+} N$ and $\mathcal{P}_{N}$ is the detector probability which is given by
\begin{align*}
\mathcal{P}_{N}=\left[1-exp\bigg(-\frac{M_{N}\Gamma_{N}L_{D}}{p^*_{N}}\bigg)\right].
\end{align*}
For the case of meson decay in flight the RH neutrino energy $E_{N}$ lies in range according to Eq.~\ref{range of EN}
and follows a flat distribution as: 
\begin{align*}
&f(E_N)=\frac{1}{E_N^+-E_N^-}=\frac{1}{2p_{N}^{*}\sqrt{\gamma^{2}-1}},
\end{align*}
Hence to calculate the total number of events for $M^{-}_1 \to \ell_{1}^{-} \ell_{2}^{-} M^{+}_2$ in the lab-frame we need to integrate within the range of $E_{N}$ as
\begin{align}\label{event signal in flight decay}
N_{\text{event}}\approx 2N_{M_{1}^{-}}\int_{E_{N}^{-}}^{E_{N}^{+}}dE_{N}\text{Br}\left(M_{1}^{-}\to\ell_{1}^{-}N\right)
\frac{m_{M_{1}}}{2p^{*}_{N}\left|\vec{p}_{M_{1}}\right|}
\frac{\Gamma(N\to\ell_{2}^{-}M_{2}^{+})}{\Gamma_{N}}\mathcal{P}^{\prime}_{N},
\end{align}
where $\mathcal{P}^{\prime}_{N}=\left[1-exp\bigg(-\frac{M_{N}\Gamma_{N}L_{D}}{\sqrt{E_{N}^{2}-M_{N}^{2}}}\bigg)\right]$ is the detector probability after taking into account the parent meson $M_1$ velocity. \\

Since the LNV meson decay rates will be extremely small, the expected number of signal events for
these processes can be assumed to follow a Poisson distribution. Following Ref.~\cite{Feldman:1997qc} and assuming
zero background events, we derive the average upper limit on the number of events at  
95$\%$ C.L., assuming number of signal events to be $N_{\text{event}}=3.09$.

Note that the number of events given in Eqs.~\ref{event signal in rest} or~\ref{event signal in flight decay} are functions of the 
mass parameters $M_{N}$ and mixing $V_{\ell N}$. Equating the numerical upper limit on the number of events to the
theoretical expressions, we get constraints
on mixing angle $V_{\ell N}$, corresponding to specific $M_{N}$ values for a particular experiments. We have assumed $|V_{e N}|^{2}=|V_{\mu N}|^{2}=|V_{\tau N}|^{2}$
in $\Gamma_{N}$ when deriving these bounds using Eq.~\ref{event signal in rest} and~\ref{event signal in flight decay}.
\section{Input for different experiments}
\label{input for different experiments}
\subsection{LHCb}
The LHCb detector is a forward spectrometer at the Large Hadron Collider~(LHC) at CERN.
A search for heavy Majorana neutrinos
in $B\to\mu\mu\pi$ decay mode had been performed by the LHCb collaboration using 7~TeV data~\cite{Aaij:2014aba} and bound on the mixing angle $|V_{\mu N}|^2$ is provided in the mass range $0.25\,\text{GeV}\leq M_N\leq 5\,\text{GeV}$\footnote{This bound has been recently revised in Ref.~\cite{Shuve:2016muy} by taking into account the correct life time calculation of $N$. }.
The cross-section for producing B, $D$ and $D_{s}$ mesons at $\sqrt{s}=13$~TeV within the LHCb
acceptance~($2<\eta<5$) are 86.6 $\mu$b, 834 $\mu$b and 353 $\mu$b, respectively~\cite{Aaij:2016avz, Aaij:2015bpa}. Hence, in LHCb upgrade with $300~\text{fb}^{-1}$ luminosity, expected number of B, $D$ and $D_{s}$ mesons are, $N_{B^{+}}=2.6\times 10^{13}$, $N_{D^{+}}=2.5\times 10^{14}$ and
$N_{D_{s}^{+}}=1.05\times 10^{14}$. LHCb will also produce a  large number of $B_c$ mesons.
A crude estimate~\cite{Belyaev} using the measured~\cite{Aaij:2014ija} ratio of production cross section times branching fractions between the $B_c\to J/\psi \pi^{+}$ and $B^{+}\to J/\psi K^{+}$ decays at $\sqrt{s}= 8~ \text{TeV}$, indicates $\mathcal{O}(10^{11})$ $B_c$ events with 300 $\text{fb}^{-1}$ luminosity at 14 TeV. 
Though the number of $B_c$ mesons at LHCb are smaller than the number of $B$ mesons, this mode being less suppressed with respect to $B^{+}\to\ell_{1}^{+}\ell_{2}^{+}\pi^{-}$,  gives tighter constraints on the mixing angles.
The produced mesons will decay in flight, carrying a momentum of order
of 100~GeV in forward direction~\cite{BmomentuminLHCb}. We take the detector length $L_{D}\approx 20$~m. 

\subsection{NA62}
NA62  is an ongoing experiment at CERN that will produce a large number of $K^{+}$ mesons~\cite{NA62page}. 
The primary SPS 400 GeV proton beam, aims on a target, produce a secondary high intensity hadron beam with an optimum content of $K^{+}(\approx 6\%)$.
The expected number of $K^{+}$ decays in the fiducial volume is $4.5\times10^{12}$ per year.
Assuming three years of running, $N_{K^{+}}=1.35\times 10^{13}$. The detector length $L_{D}\approx 65$~m and the  produced $K^{+}$ mesons will decay in flight,
carrying a momentum of $75$ GeV.

\subsection{SHiP}
The SHiP experiment is a newly proposed general purpose fixed target facility at the CERN SPS accelerator~\cite{ship}. A 400 GeV proton beam will be dumped on a heavy
target for the duration of five years. One of the primary goal of the experiment is to use decays of charmed mesons to search for
heavy sterile neutrinos using the decay mode $D^{+}_s/D^{+} \to \ell^{+} \ell^{+} \pi^{-}$. One can easily estimate the number of charmed meson pairs that are expected to be produced in this experiment as~\cite{Alekhin:2015byh},
\begin{align*}
 N_{meson}=X_{c\bar{c}}\times N_{POT}\times {\mathcal{R}},
\end{align*}
where $X_{c\bar{c}}$ is the $c\bar{c}$ production rate, $N_{POT}=2\times 10^{20}$ is the number of proton-target interaction. The relative abundances $\mathcal{R}$ of 
charmed mesons, such as, $D$ and $D_{s}$ are $30\%$ and $8\%$, respectively. Hence, the expected number of $D$ and $D_{s}$ mesons are $N_{D^{+}}=1.02\times 10^{17}$ and
$N_{D_{s}^{+}}=2.72\times10^{16}$, respectively. This very high intensity of the charmed mesons will permit 
to set tight constraints on mixing angle at SHiP. There will also be large number of $B$ and $B_c$ meson productions at SHiP. Following~\cite{SHiP:2018xqw}, we can estimate the number of $B$ and $B_c$ meson as $N_{B^+}=10^{13}$ and $N_{B_c}=10^{11}$, respectively.
The detector length is taken to be, $L_{D}=60$~m. For the 400~GeV CNGS proton beam on target, the expected momentum of the produced mesons is
$\sim$ 58~GeV~\cite{Gorbunov:2007ak}.\\

\subsection{MATHUSLA}
MATHUSLA~\cite{Curtin:2018mvb} is a newly proposed detector near ATLAS or CMS. Its main goal is to search for neutral long-lived particles~(LLP) produced in HL-LHC collisions by reconstructing displaced vertices. The detector is designed to have an area of $200\text{m}\times 200\text{m}$ and a height of 20m for the decay volume, which is displaced from ATLAS or CMS by 100m both horizontally and vertically. RH neutrino search is one of the primary goal of MATHUSLA and is most sensitive to the parameter space which yields a decay length~$\sim 200$m. RH neutrinos which are coming from the meson decays or $W$ and $Z$ boson decays has very large decay length and this has been already studied in~\cite{Curtin:2018mvb}. For the meson decay case, they have considered the decay modes $B\to D\ell N$, $B\to\ell N$ and $D\to K\ell N$ and after including the probability of RH neutrinos to decay visibly within the MATHUSLA detector, they derived the constraints on mixing angles. In this study, we have considered the meson decays $B\to\ell_1\ell_2\pi$ and $D\to\ell_1\ell_2\pi$ for MATHUSLA. For the number of $B$ and $D$ meson productions we followed the Ref.~\cite{Bondarenko:2019yob}. The result of their detailed simulation suggests that number of $B$ and $D$ meson production within the geometric acceptance of the MATHUSLA detector are $5.7\times 10^{14}$ and $5.4\times 10^{13}$, respectively. Their simulation also gives the average $\gamma$ factor of the $B$ and $D$ mesons as $\braket{\gamma_B}=2.3$ and $\braket{\gamma_D}=2.6$ from which we can derive the average momentum of the mesons. The detector length is taken to be 38m.
\subsection{Belle II} 
The asymmetric SuperKEKB facility is designed to collide electron and positron beams such that the centre of mass energy is in the region of the $\Upsilon$ resonances.
An upgrade of Belle, the newly completed Belle II detector is  expected to collect data samples corresponding to an integrated
luminosity of 50 ab$^{-1}$ by the end of 2024~\cite{Belle2}. The expected number of charged $B\bar{B}$ pairs to be produced is
{$5.5\times 10^{10}$}~\cite{Patrignani:2016xqp, Karim}. In addition, a large sample of charged $D, D_s$ mesons will also be accessible, with
$N_{D^+}=3.4\times 10^{10}$ and $N_{D_s^+}= 10^{10}$~\cite{Karim}.
A direct search for heavy Majorana neutrinos in B-meson decays was performed by Belle collaboration using a data sample
that contained $772\times 10^{6}$
$B\bar{B}$ pairs~(at 711 fb$^{-1}$)~\cite{Liventsev:2013zz}.
At KEKB as well as superKEKB, the energies of the $e^{+}$, $e^{-}$ beams are sufficiently low so that the momentum of the produced $B$ mesons as well as that for the charmed 
mesons will not be appreciable and the suppression from high momentum
of the decaying mesons in the number of events will be absent.

\subsection{FCC-ee}
The Future Circular Collider (FCC-ee)~\cite{FCC} will  collect multi-ab$^{-1}$ integrated luminosities for $e^{+}e^{-}$ collisions at c.m.energy $\sqrt{s}\approx 91$~GeV. 
The expected number of $Z$-bosons is $10^{12}-10^{13}$. The number of charged B mesons from $Z$ decays can be estimated as,
\begin{align*}
N_{B^{+}}=N_{Z}\times \text{Br}\left(Z\to b\bar{b}\right)\times f_{u},
\end{align*}
where $N_{Z} \sim 10^{13}$, Br$\left(Z\to b\bar{b}\right)=0.1512$~\cite{Agashe:2014kda}, $f_{u}=0.410$~\cite{Amhis:2014hma} is the fraction of $B^{+}$ from $\bar{b}$ quark
in $Z$ decay.
The B mesons produced at FCC-ee will have an energy distribution peaked at $E_{B^{+}}=\frac{M_{Z}}{2}$. Hence we can calculate the number
of signal events using Eq.~(\ref{event signal in flight decay}), where the detector length is taken to be $L_{D}=2$~m. 
\begin{figure}
	\centering
		\includegraphics[width=0.45\textwidth]{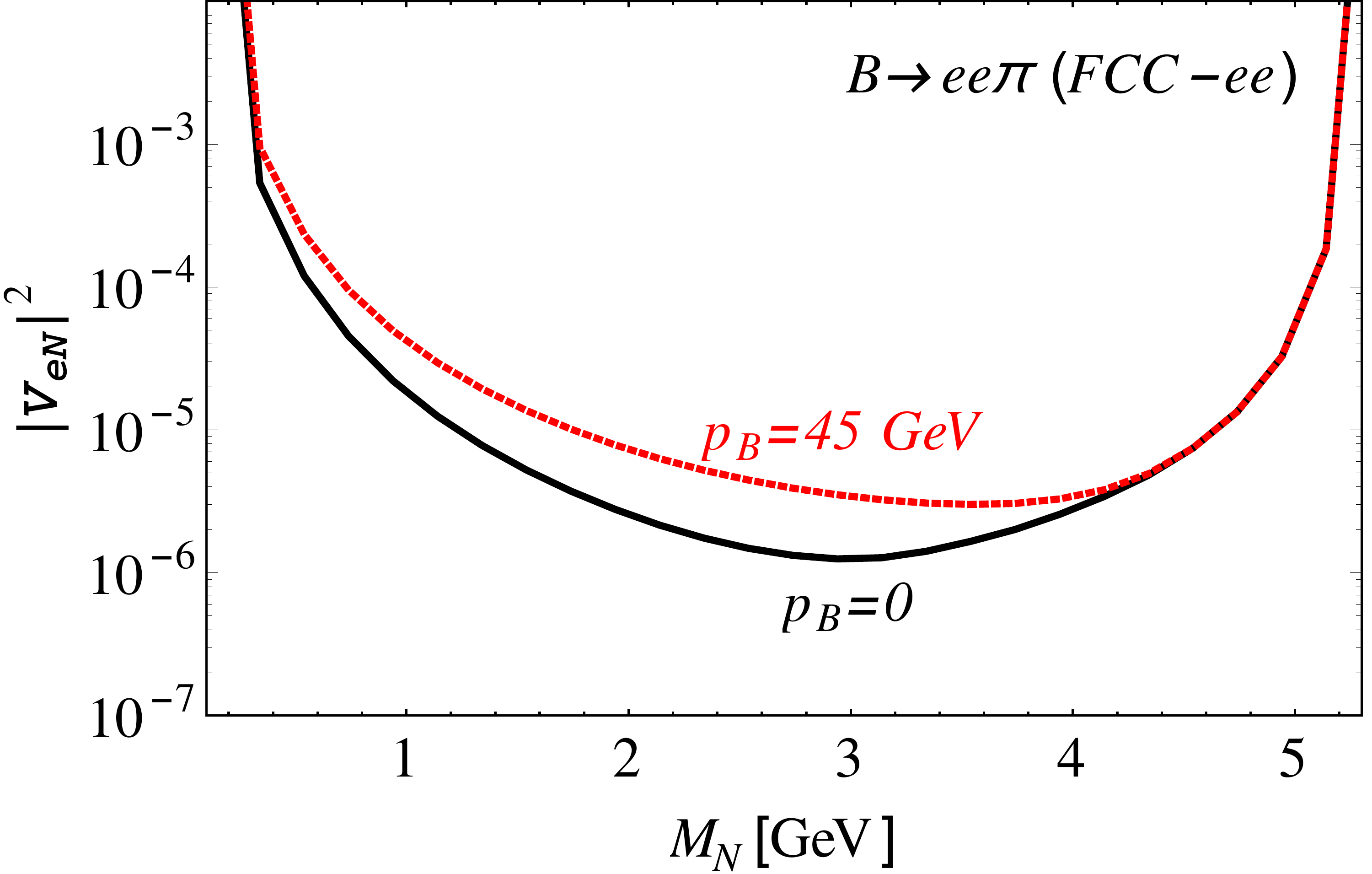}
	\includegraphics[width=0.45\textwidth]{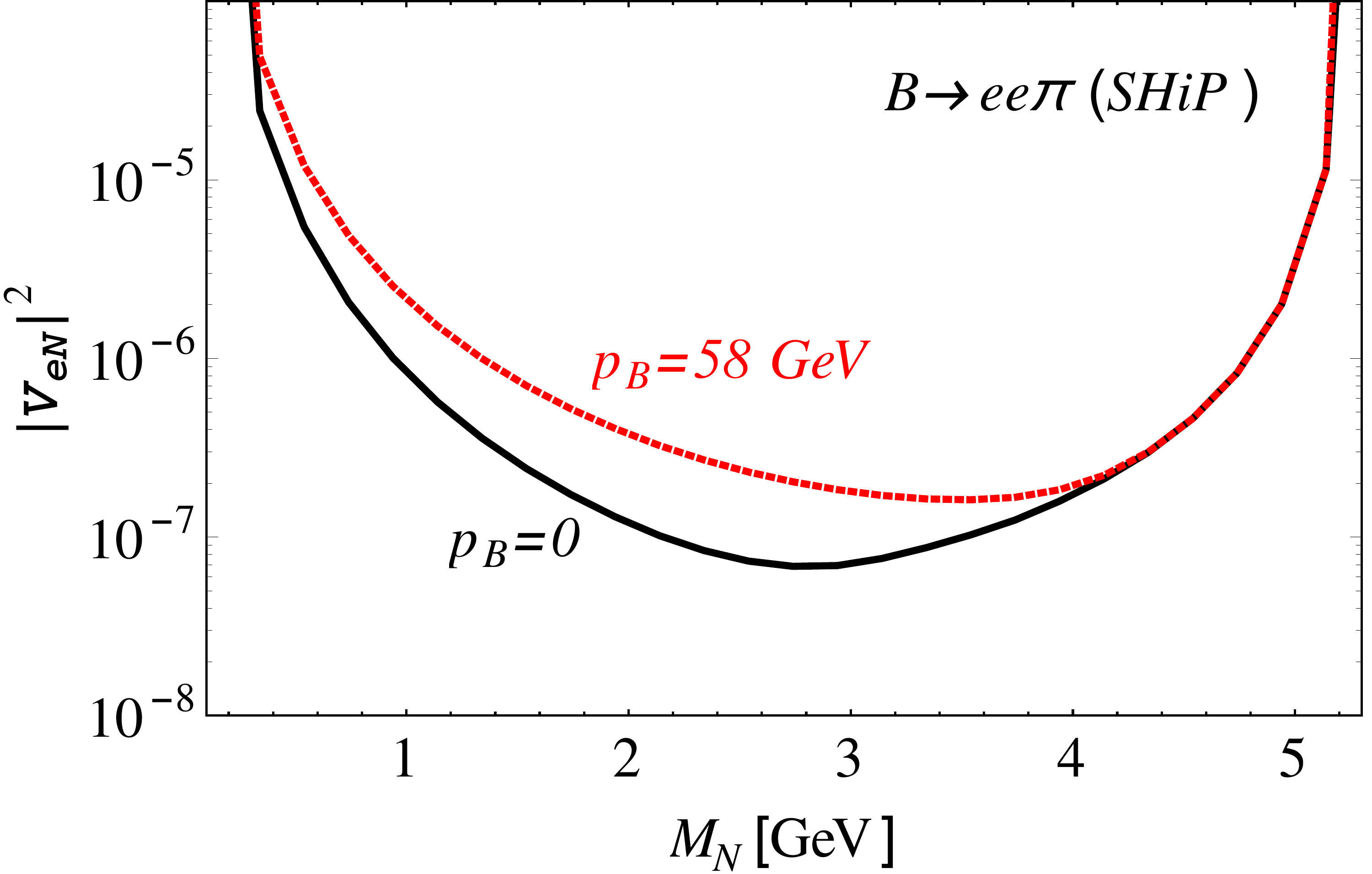}
		\includegraphics[width=0.45\textwidth]{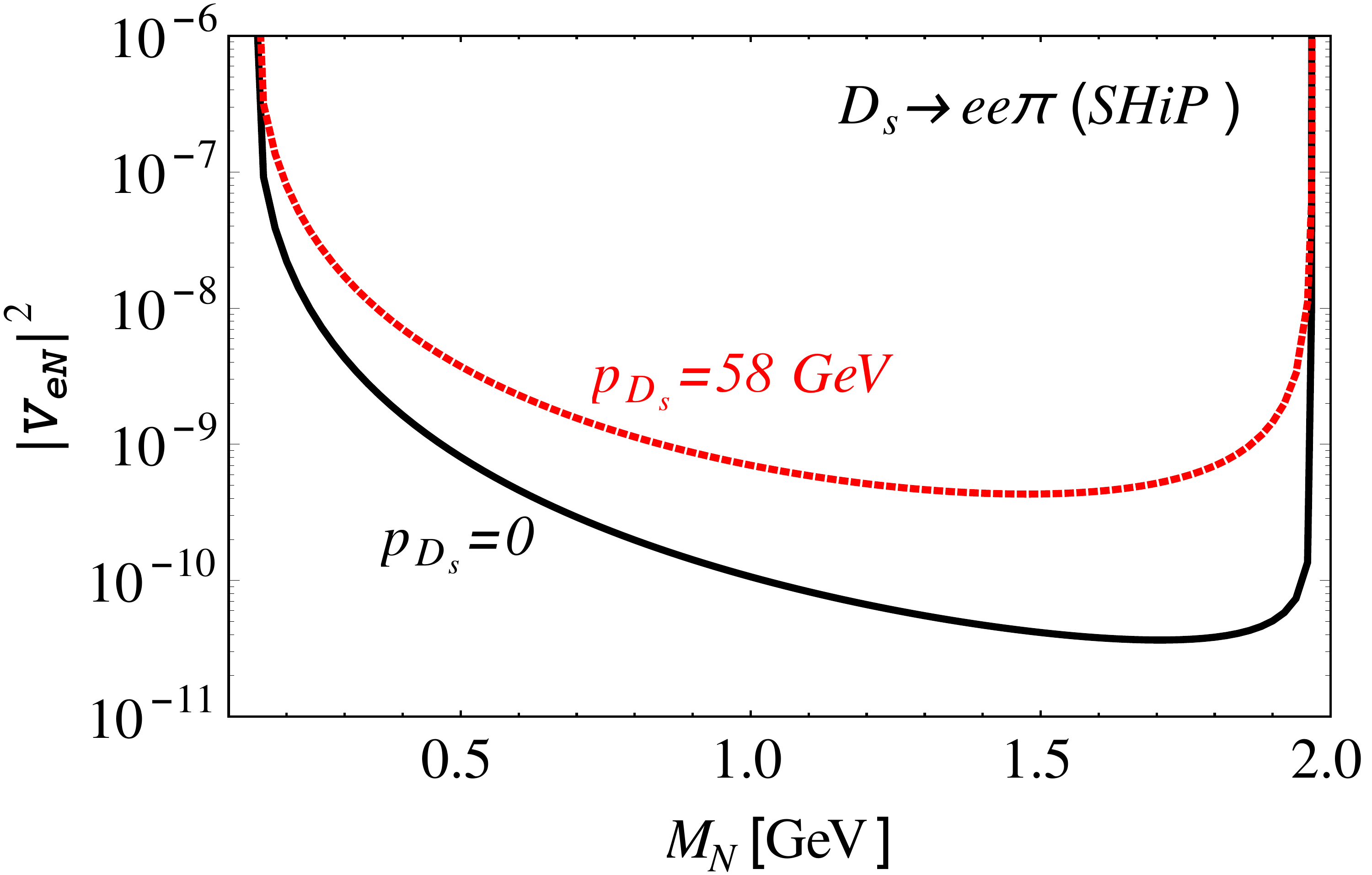}
	\includegraphics[width=0.45\textwidth]{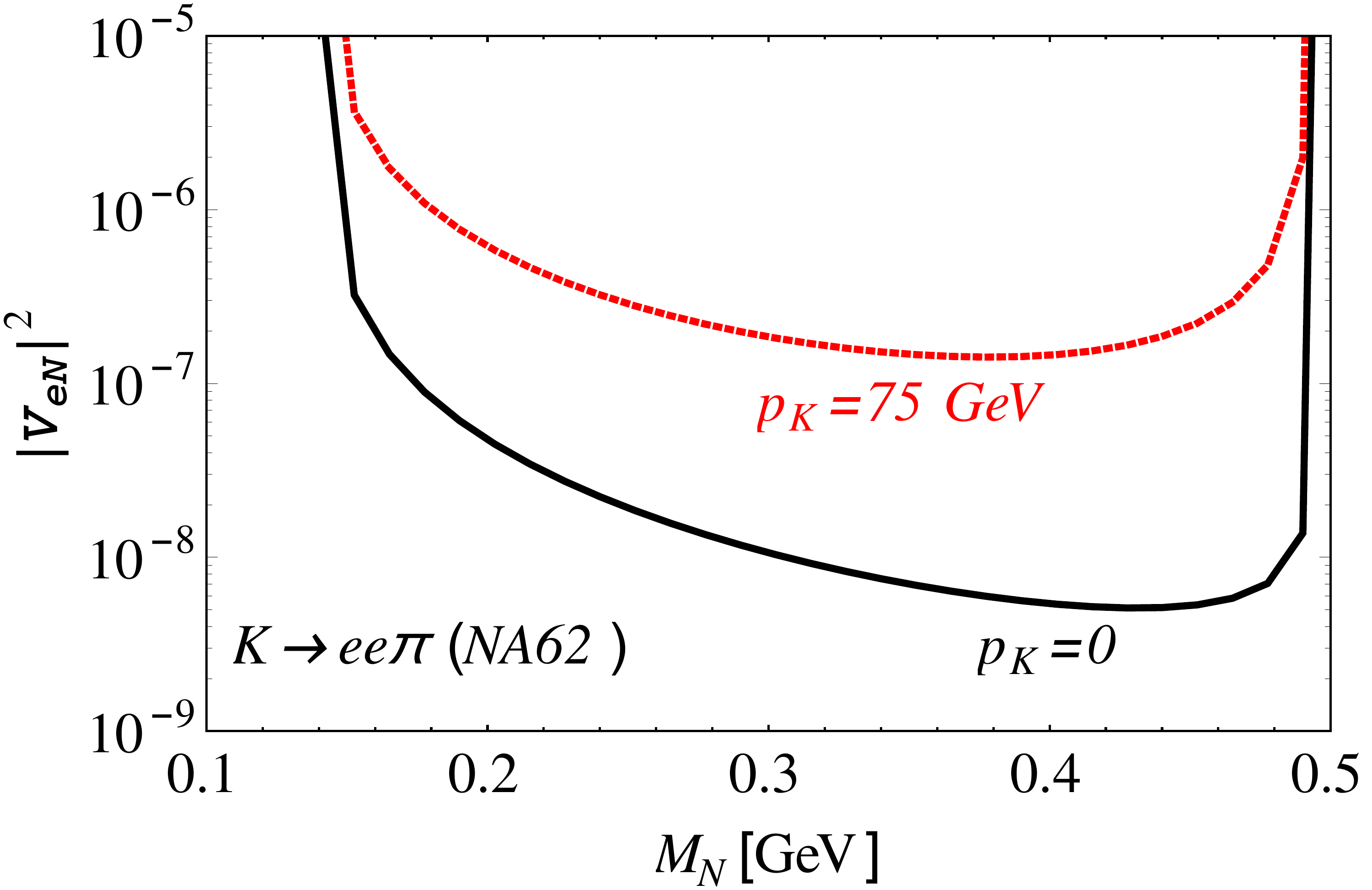}
	\includegraphics[width=0.45\textwidth]{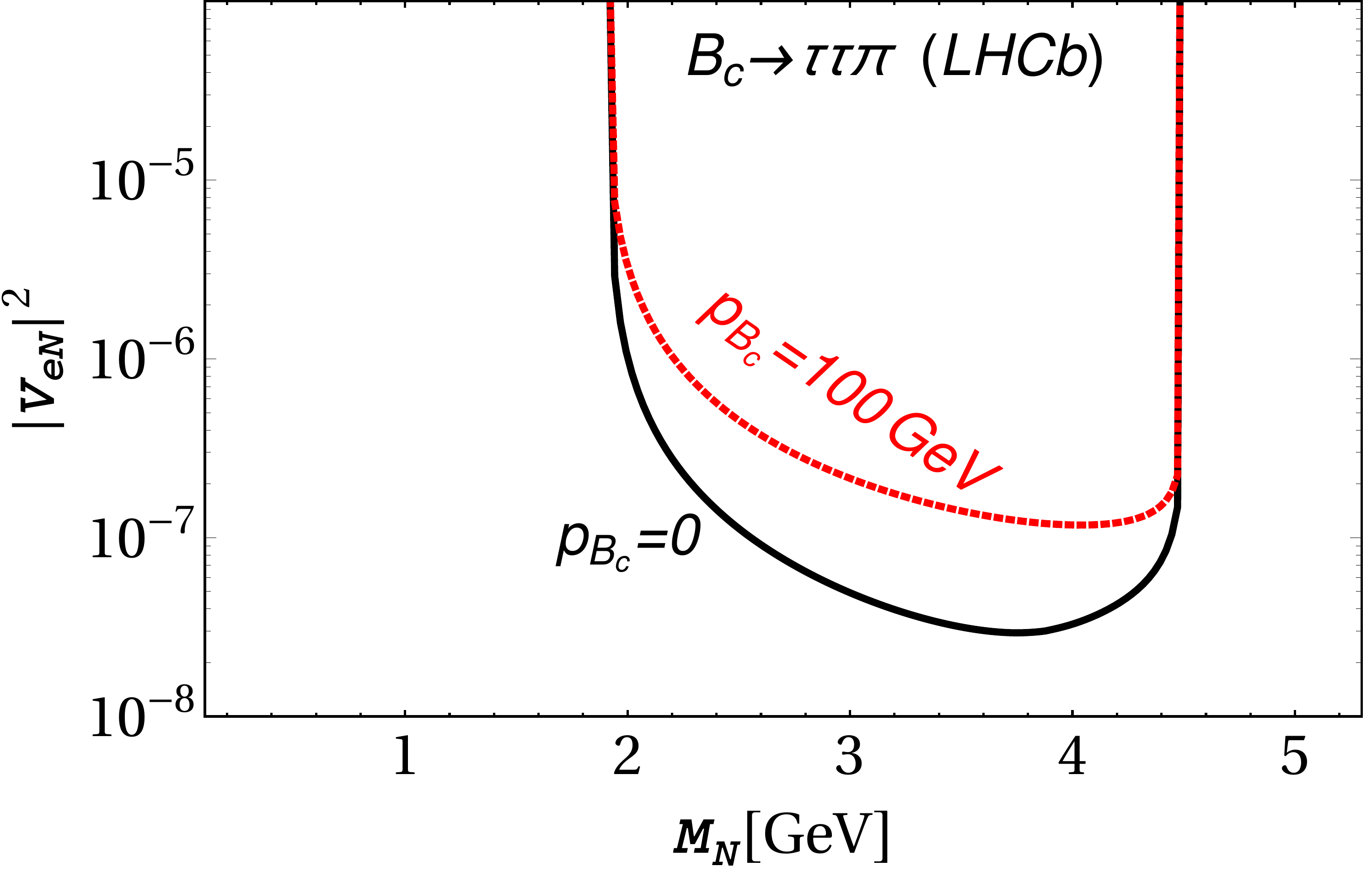}
	\includegraphics[width=0.45\textwidth]{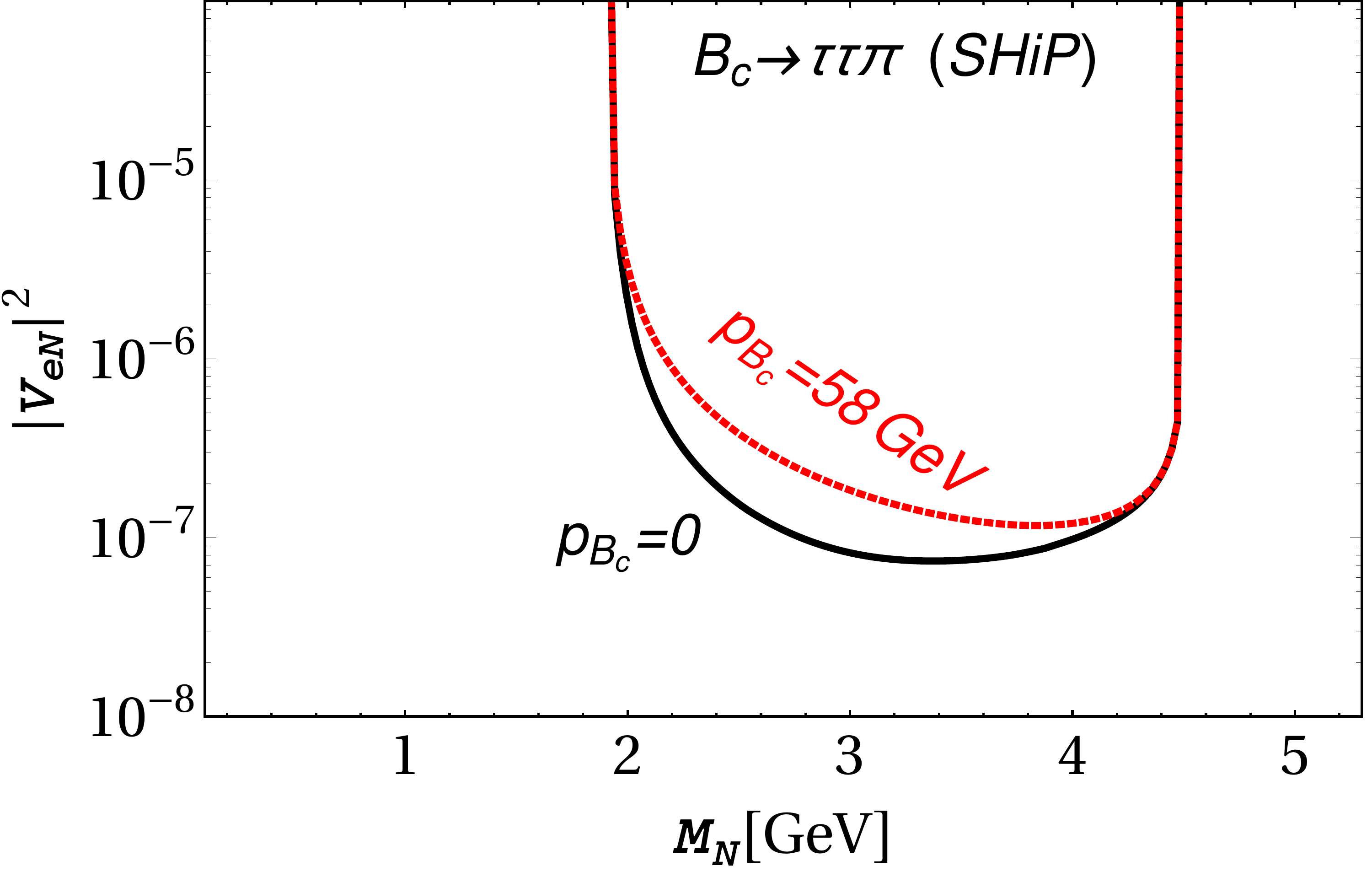}
	\caption{\small{Future sensitivity reach and present limits on the mixing angles as a function of RH neutrino mass $M_N$. The black and red lines stand for meson decay at rest and meson decay with finite momentum, respectively. The upper panel corresponds to the $B$ meson decay at FCC-ee(left) and SHiP~(right). The left figure of middle panel is for $D_s$ meson decay at SHiP and the right figure of middle panel is for $K$ meson decay at NA62. The lower panel represents $B_c$ meson decay at LHCb~(left) and SHiP~(right).}}
	\label{Comparison of Mixing with velocity}
\end{figure}
\section{Results}
\label{results}
In Fig.~\ref{Comparison of Mixing with velocity}, we show how the velocity of the parent mesons affect the sensitivity reach of
the mixing angles. We consider a number of ongoing and future experiments, such as, FCC-ee, SHiP to explore   $B\to ee\pi$, SHiP for   $D_s\to ee\pi$, NA62 for  $K\to ee\pi$,  and 
LHCb, SHiP for $B_c\to\tau\tau\pi$ meson decays. To derive the bounds/sensitivity  on the mixing angle as a function of RH neutrino mass
$M_{N}$, we use Eq.~\ref{event signal in rest}, and Eq.~\ref{event signal in flight decay}, for meson decay at rest and in flight, respectively.} For all of the above decays,  the obtained bounds/future sensitivity  on the mixing angles are 
rather loose in case of meson decays in
flight compared to meson decays at rest.
\begin{figure}[ht!]
	\centering
		\includegraphics[width=0.45\textwidth]{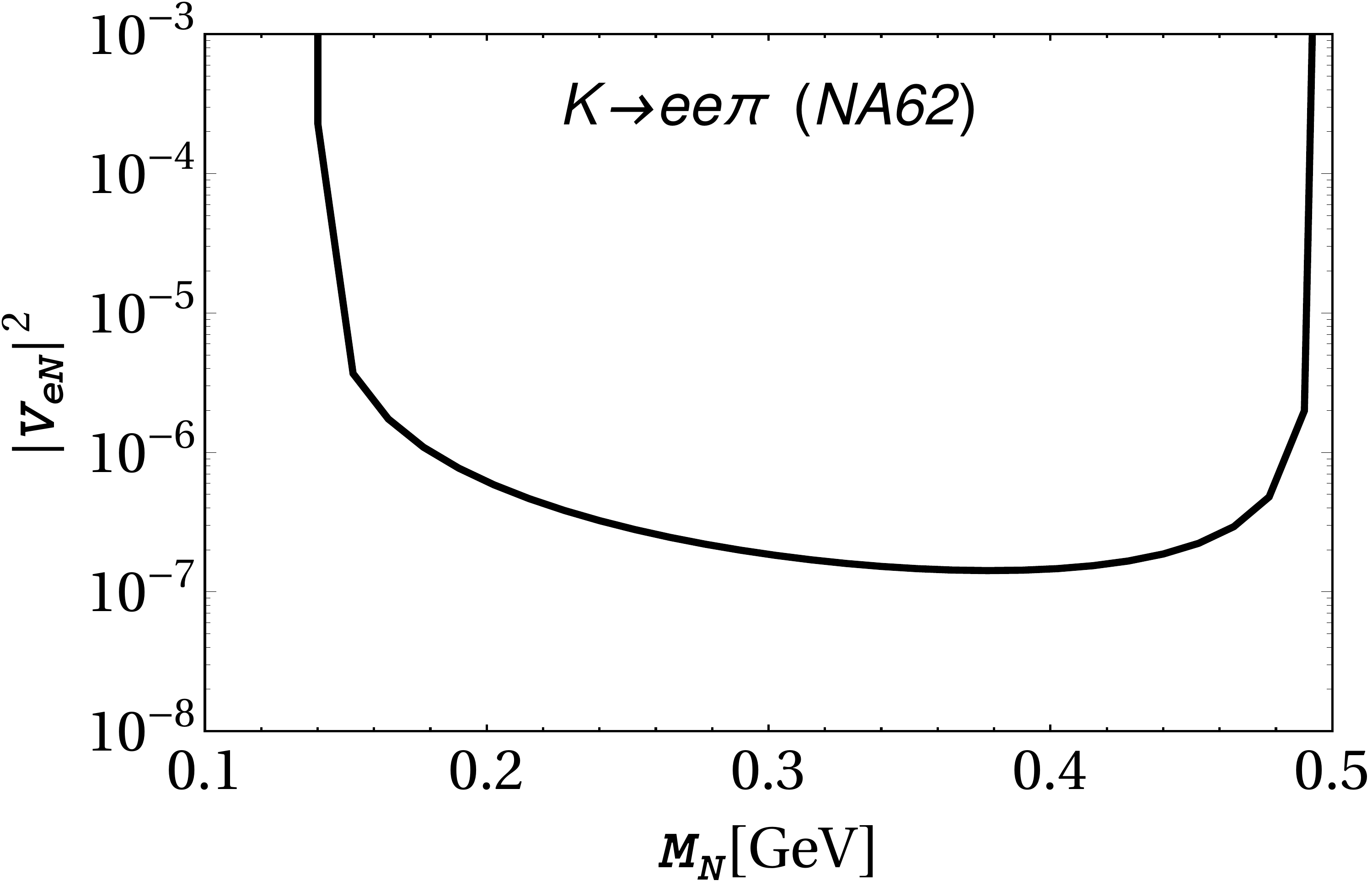}
	\includegraphics[width=0.45\textwidth]{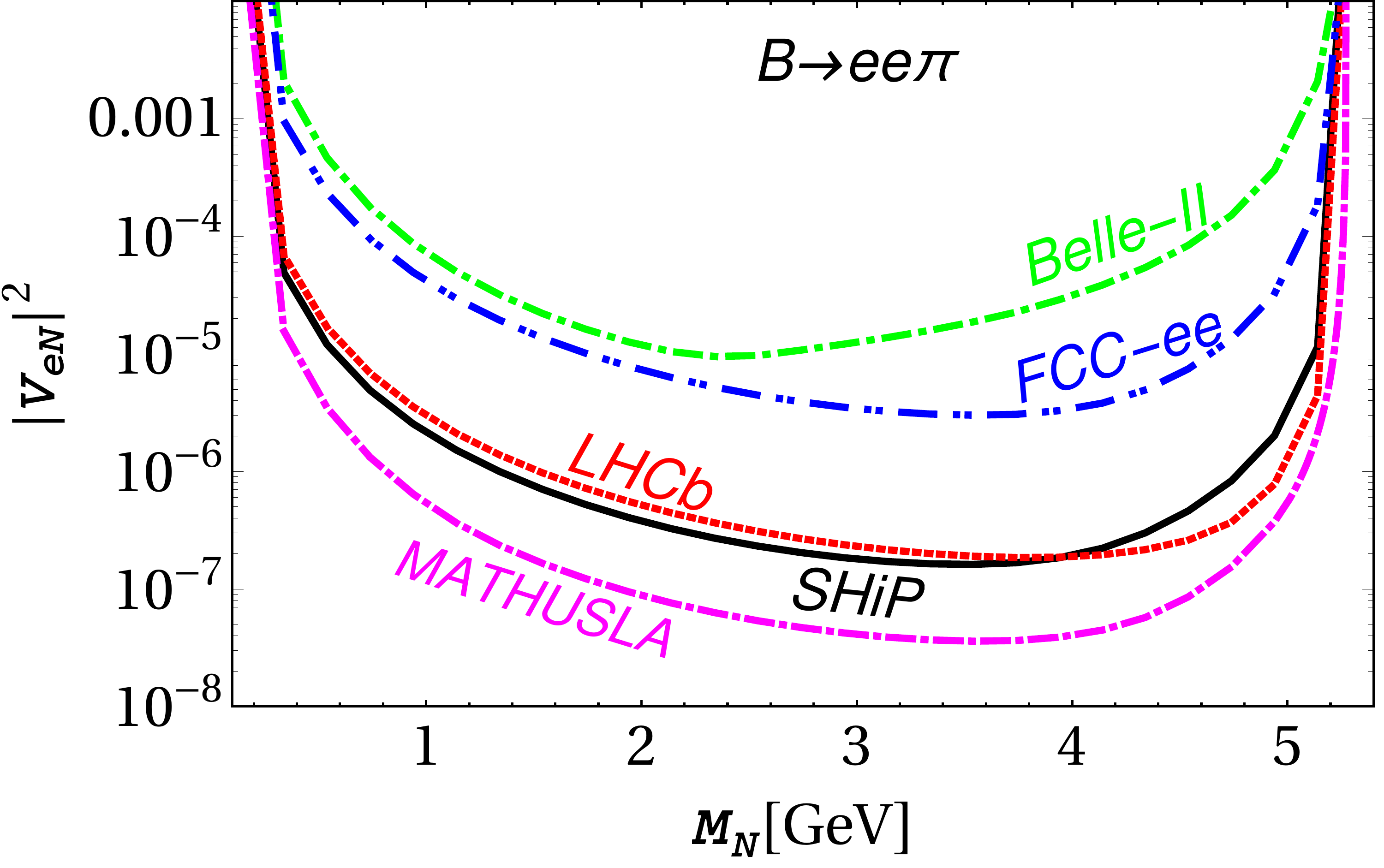}
	\includegraphics[width=0.45\textwidth]{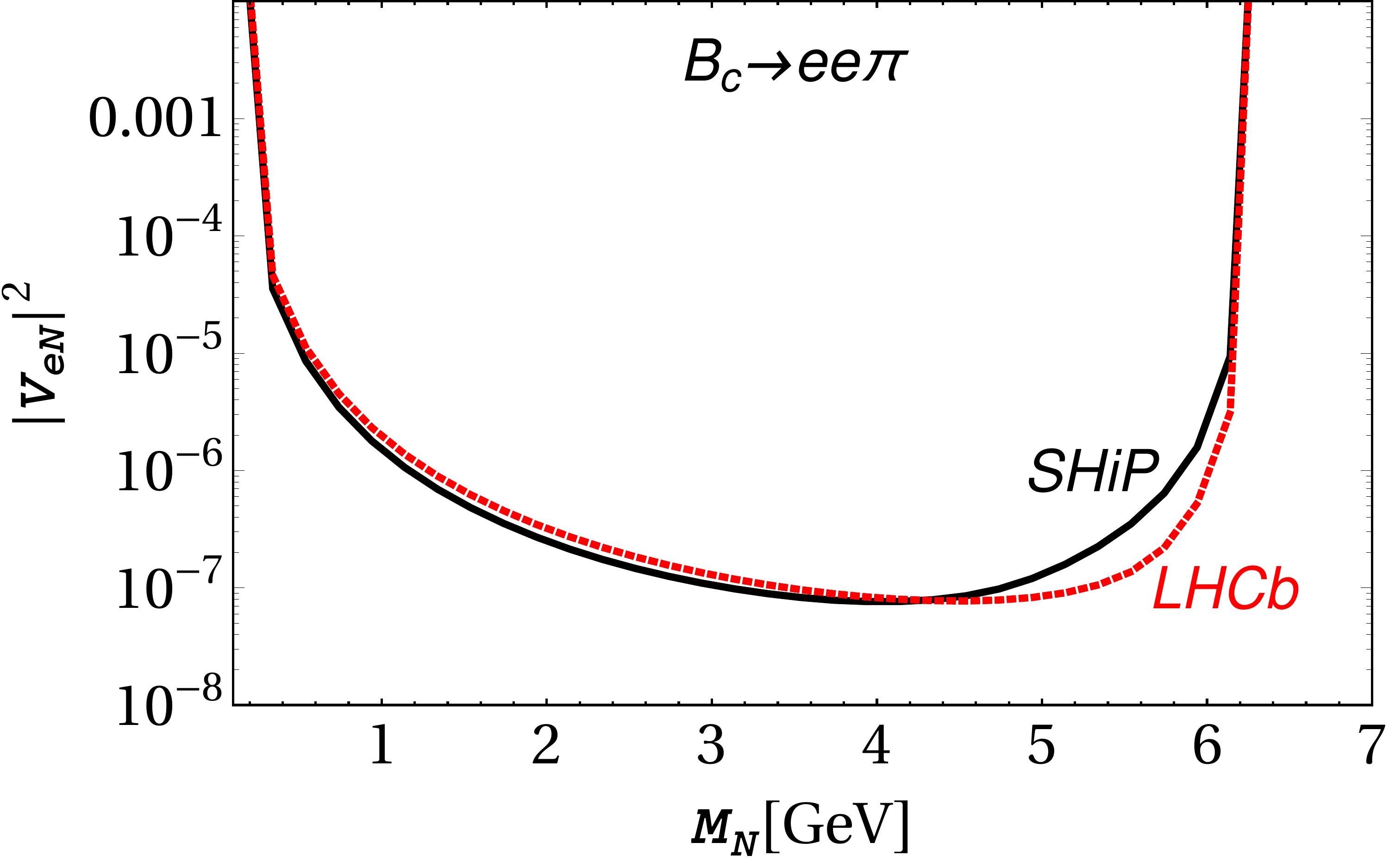}
		\includegraphics[width=0.45\textwidth]{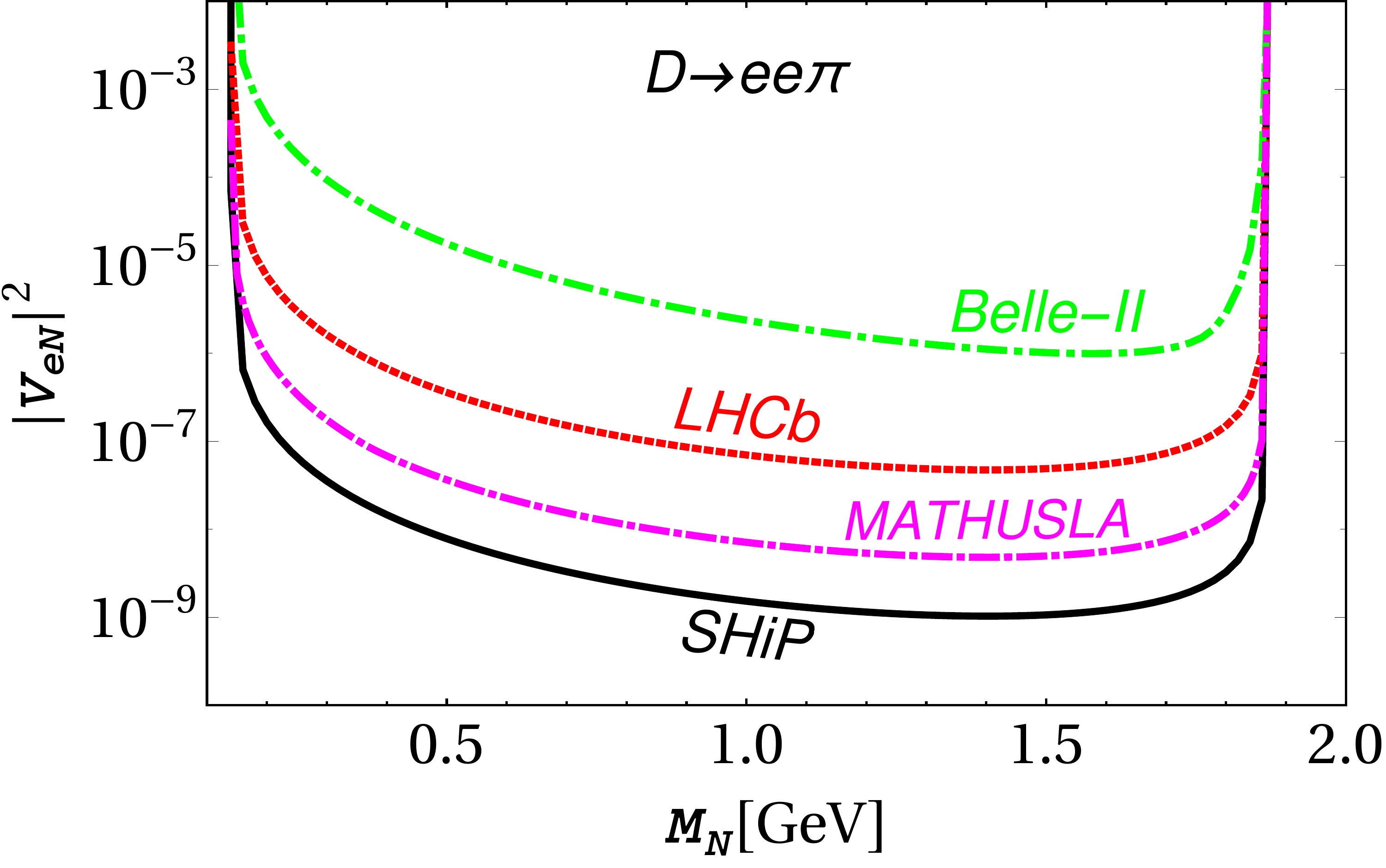}
	\includegraphics[width=0.45\textwidth]{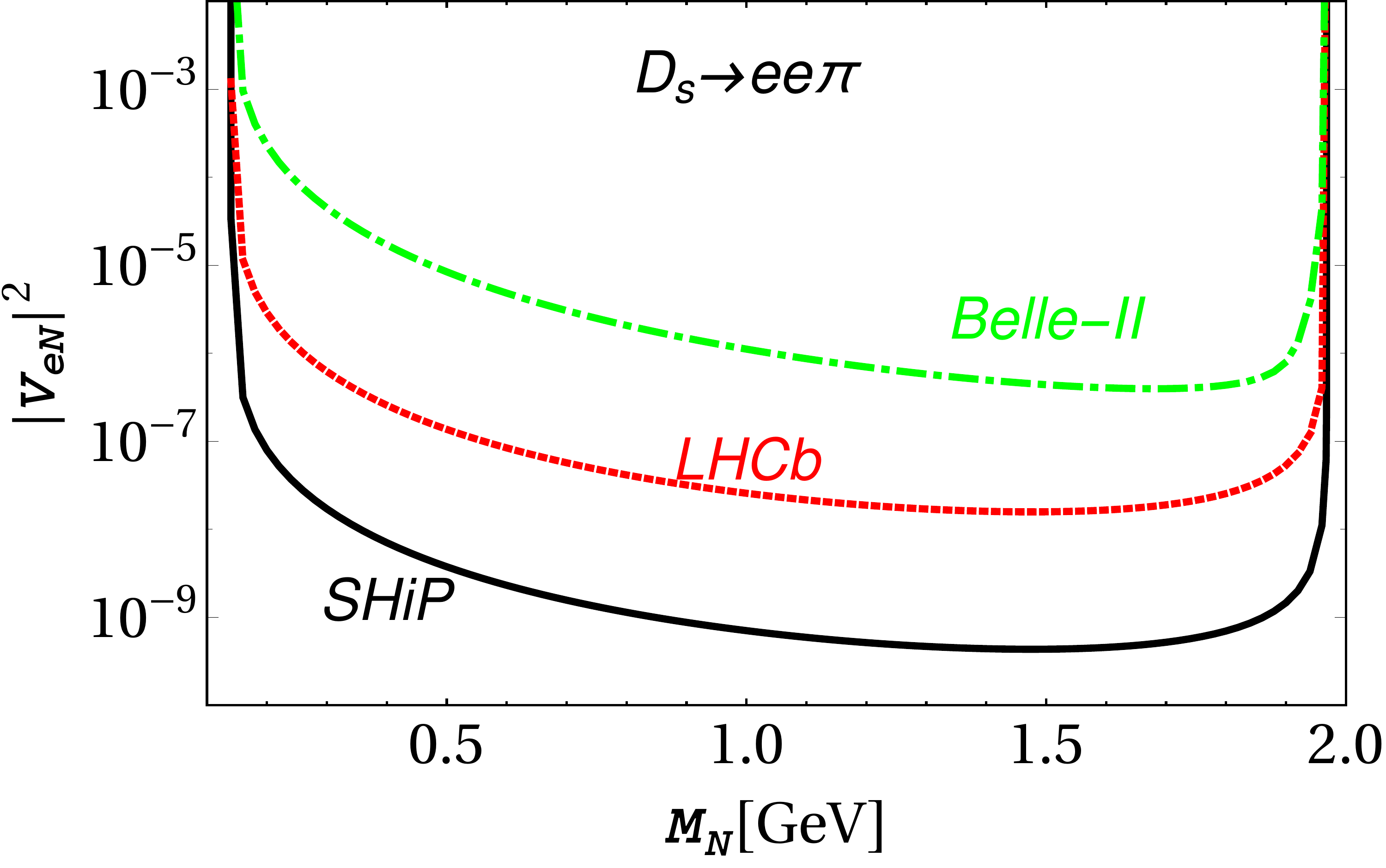}
	\caption{\small{Variation of the future sensitivity reach and present limits on the mixing angle $|V_{e N}|^{2}$ with respect to the mass $M_{N}$. We consider meson decay channel $M_{1}\to ee\pi$. The left figure in the upper panel is for $K$ meson decay at NA62. The right figure in the upper panel is for $B$ meson decay at SHiP~(black), MATHUSLA~(magenta), LHCb~(red), FCC-ee~(blue) and Belle-II~(green). The left figure in middle panel is for $B_c$ meson decay at SHiP~(black) and LHCb~(red). The right figure of the middle panel is for $D$ meson decay at SHiP~(black), MATHUSLA~(magenta), LHCb~(red) and Belle-II~(green). The last figure of $D_s$ meson decay at SHiP~(black), LHCb~(red) and Belle-II~(green). }}
	\label{Bounds on VeN}
\end{figure}
\begin{figure}
	\centering
		\includegraphics[width=0.45\textwidth]{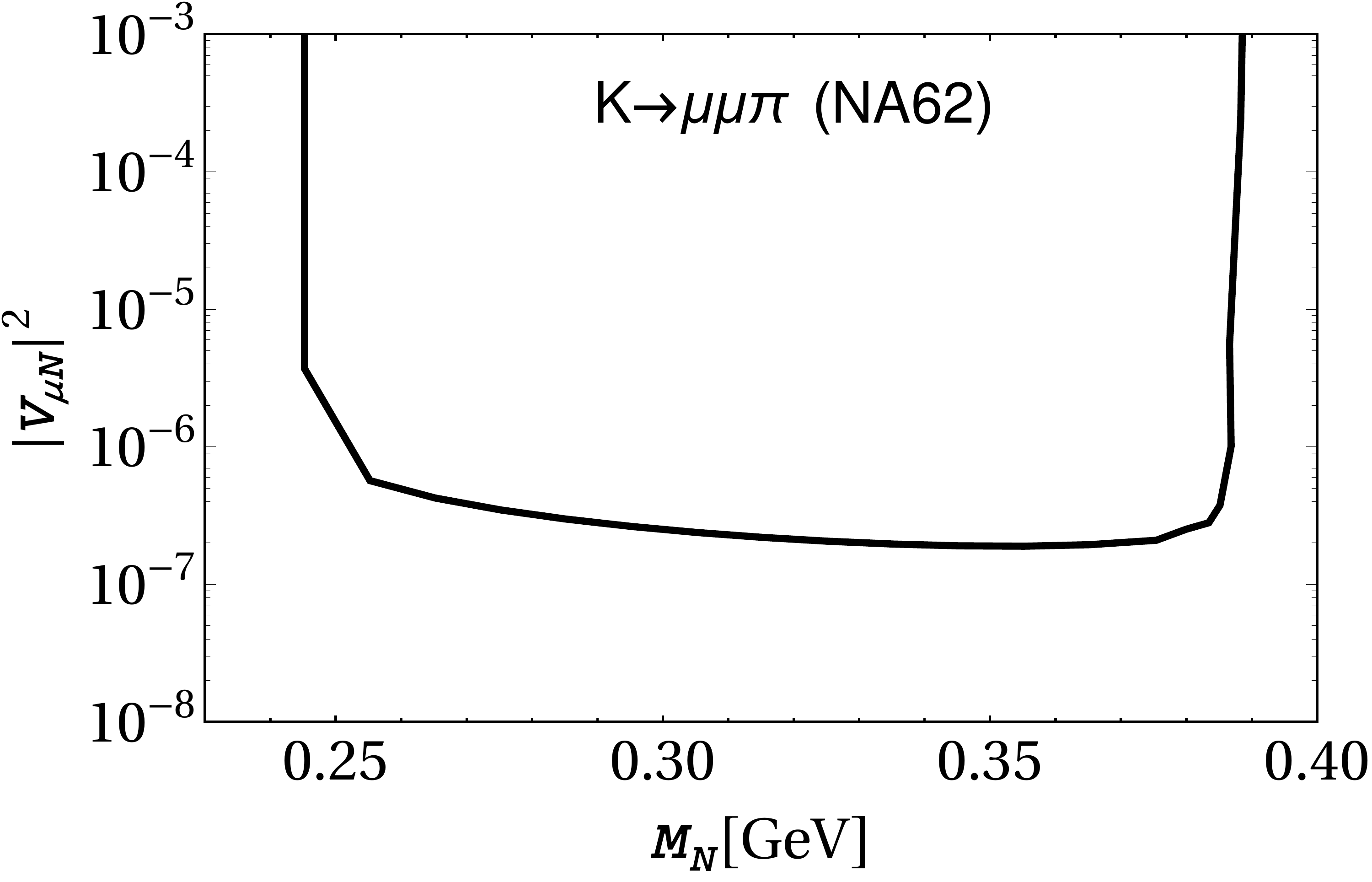}
	\includegraphics[width=0.45\textwidth]{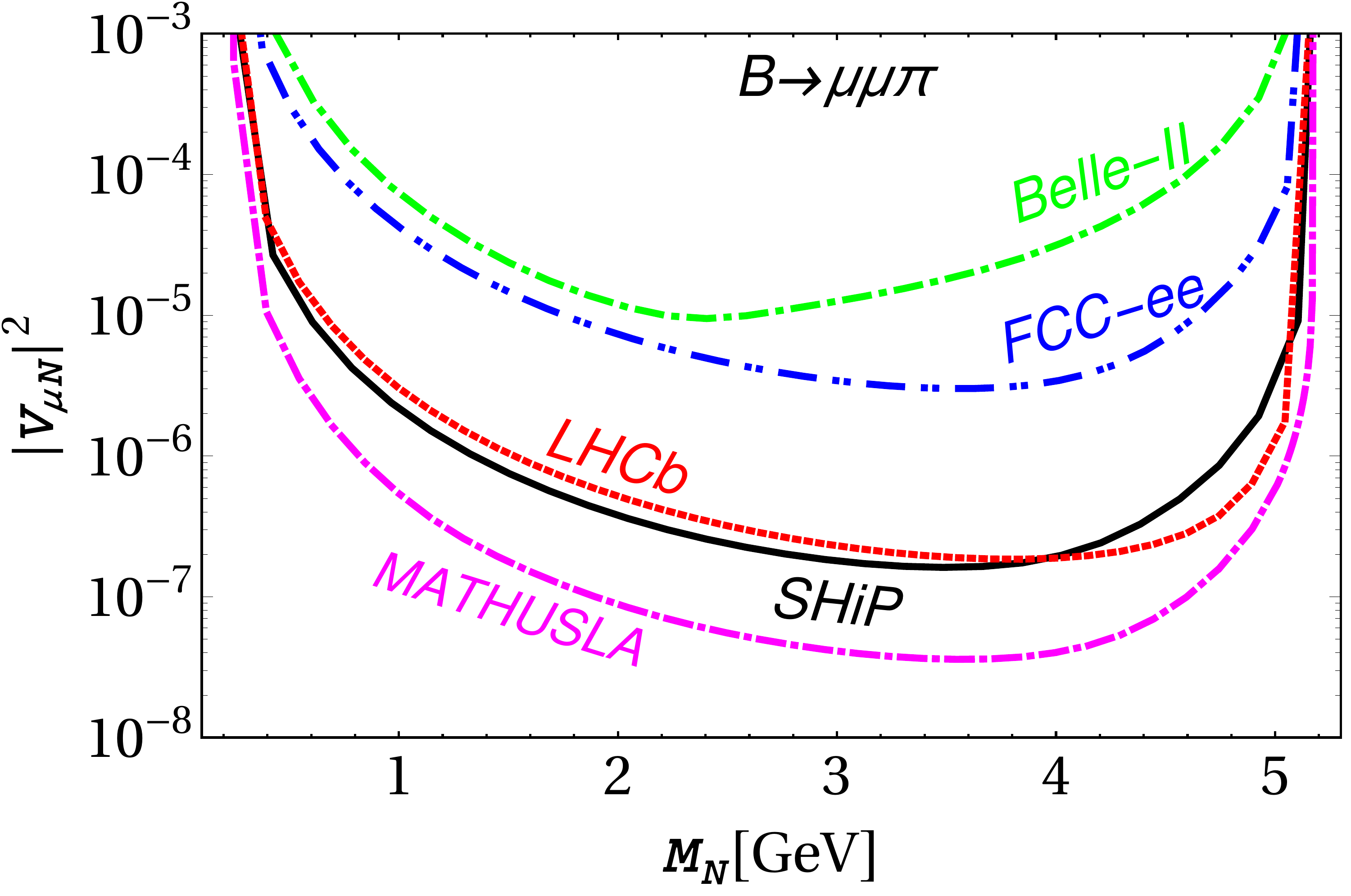}
		\includegraphics[width=0.45\textwidth]{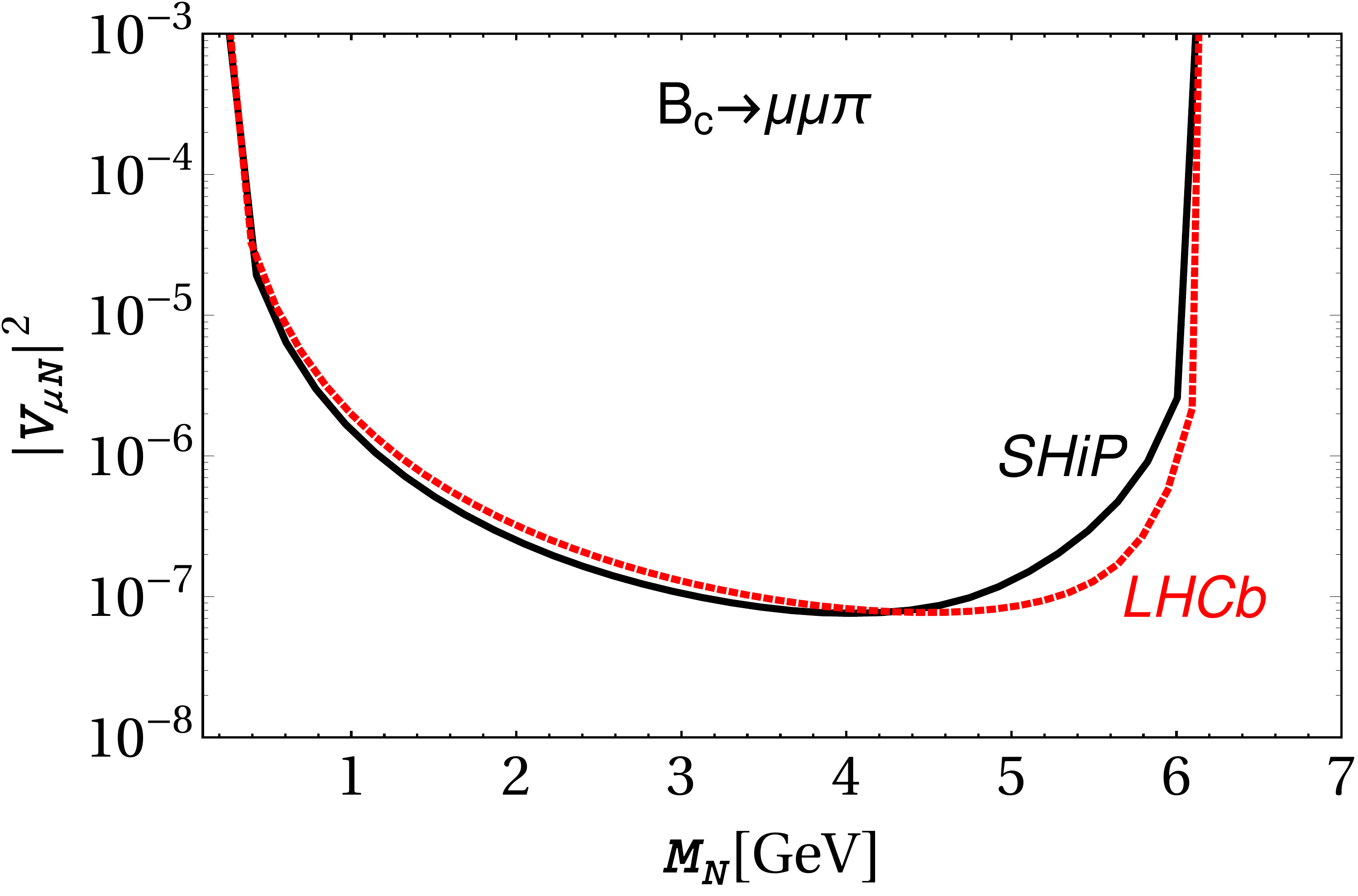}
		\includegraphics[width=0.45\textwidth]{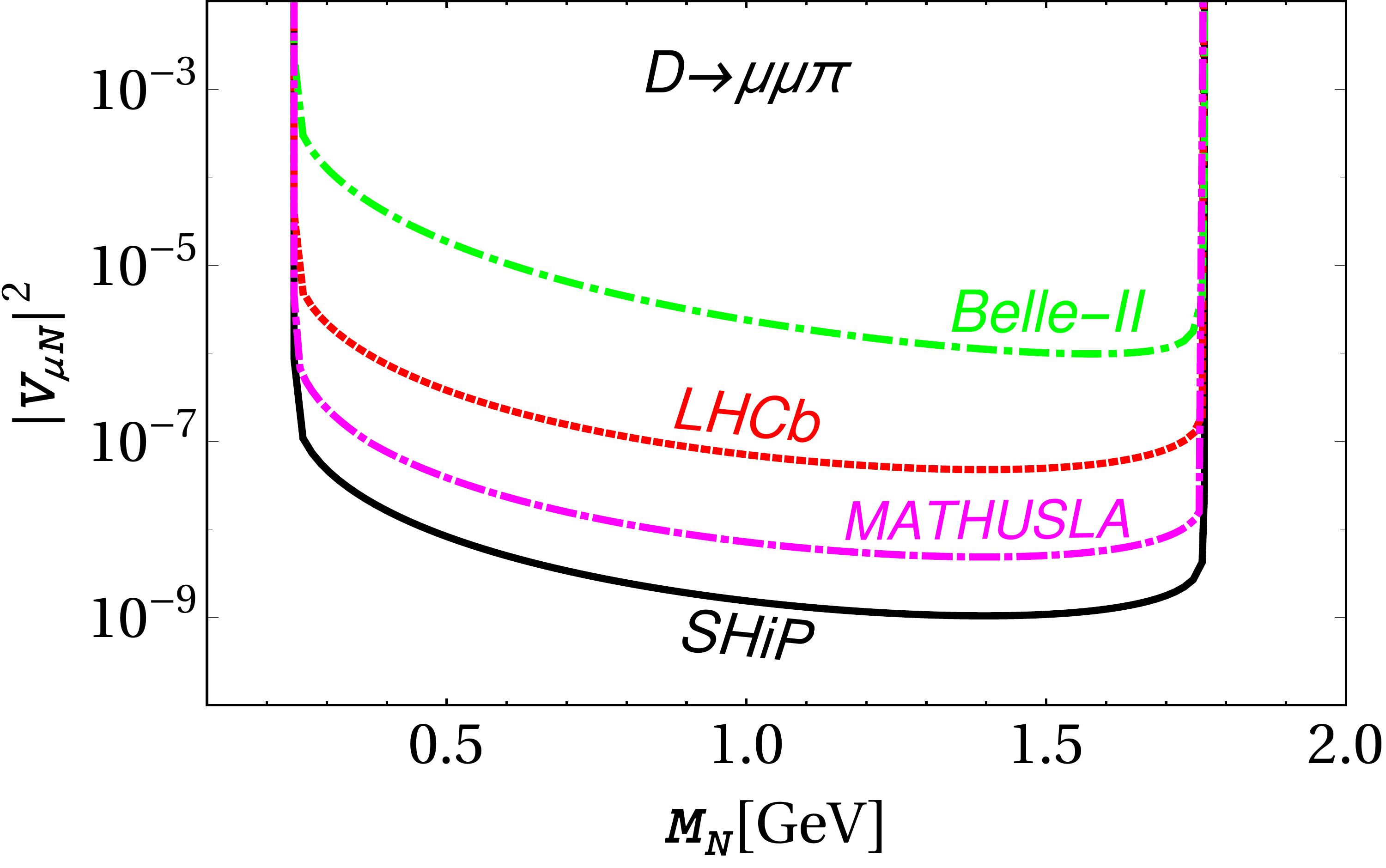}
	\includegraphics[width=0.45\textwidth]{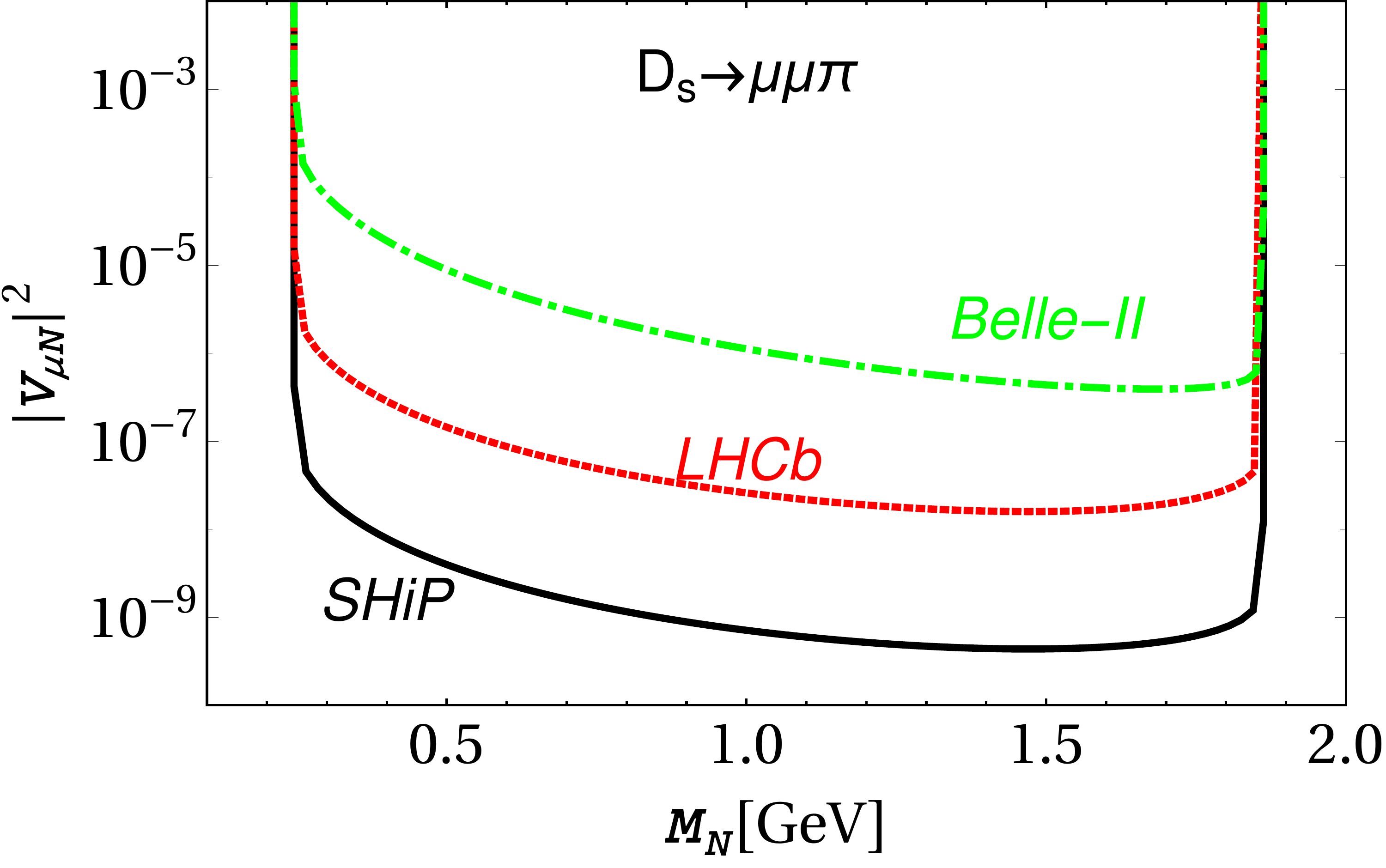}
	\caption{\small{Same as Fig.~\ref{Bounds on VeN} except now the limits are on the mixing angle $|V_{\mu N}|^2$ 
	from the meson decays $M_{1}\to \mu\mu\pi$.}}
	\label{Bounds on VmuN}
\end{figure}
\begin{figure}
	\centering
		\includegraphics[width=0.45\textwidth]{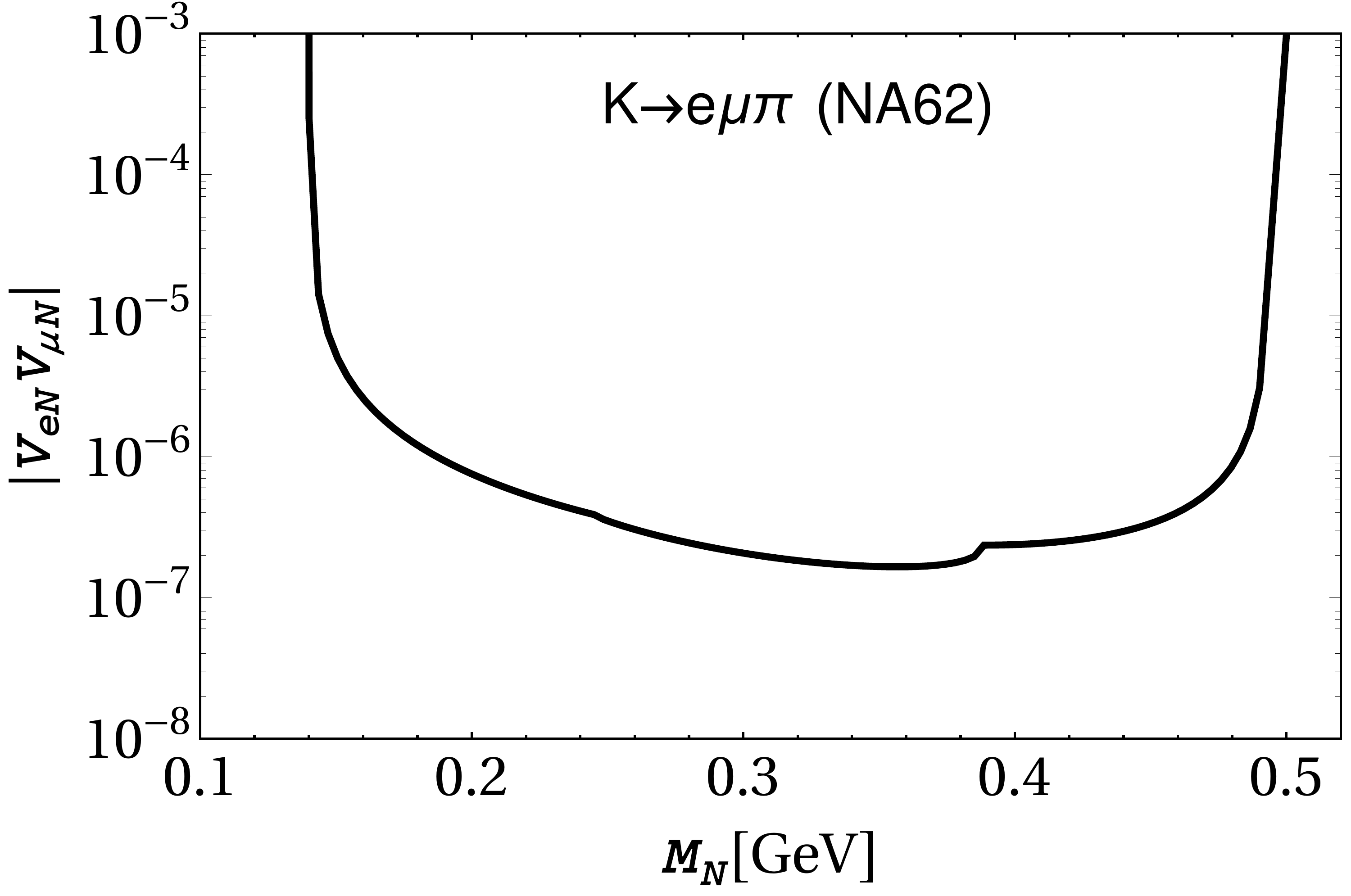}
	\includegraphics[width=0.45\textwidth]{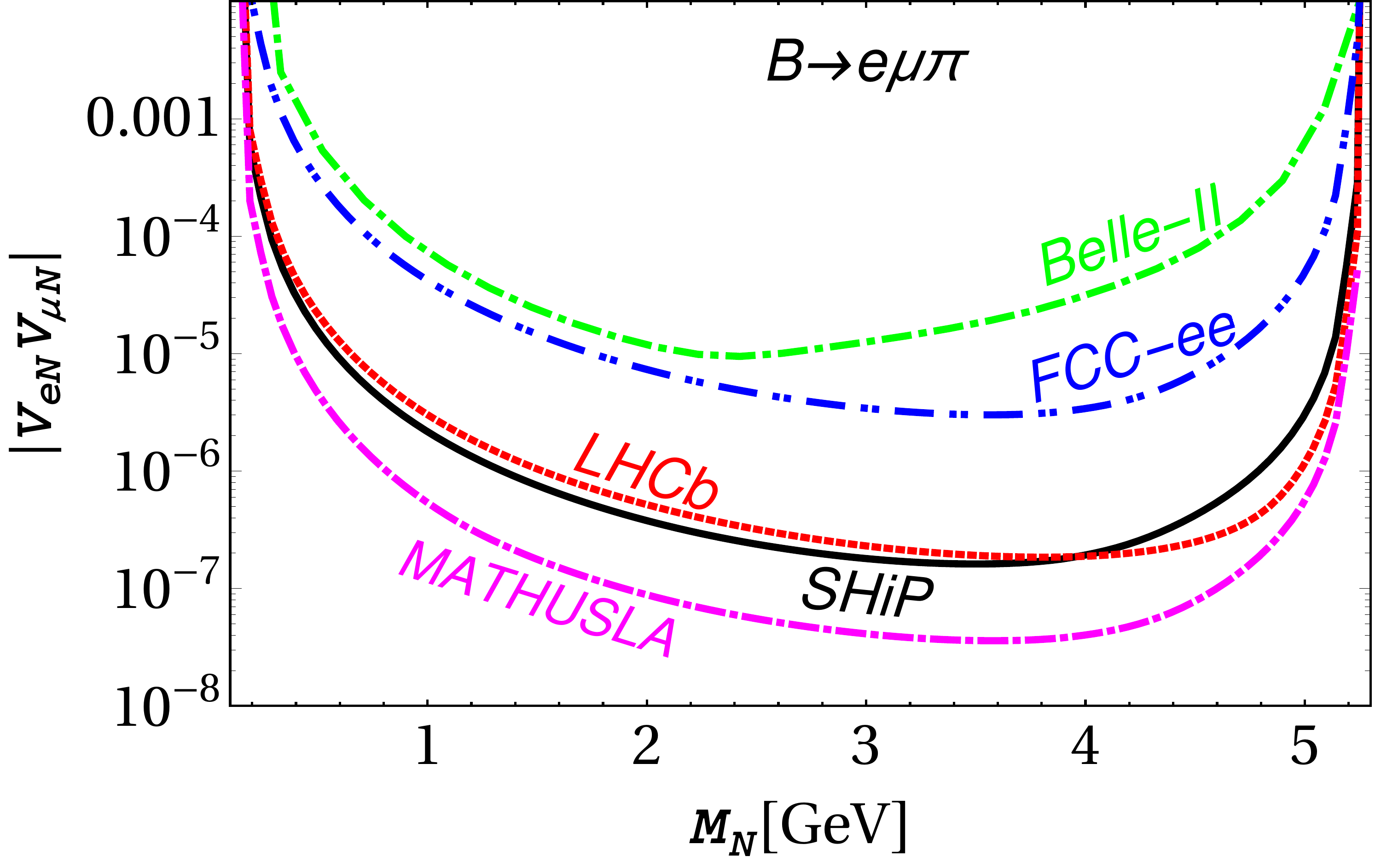}
		\includegraphics[width=0.45\textwidth]{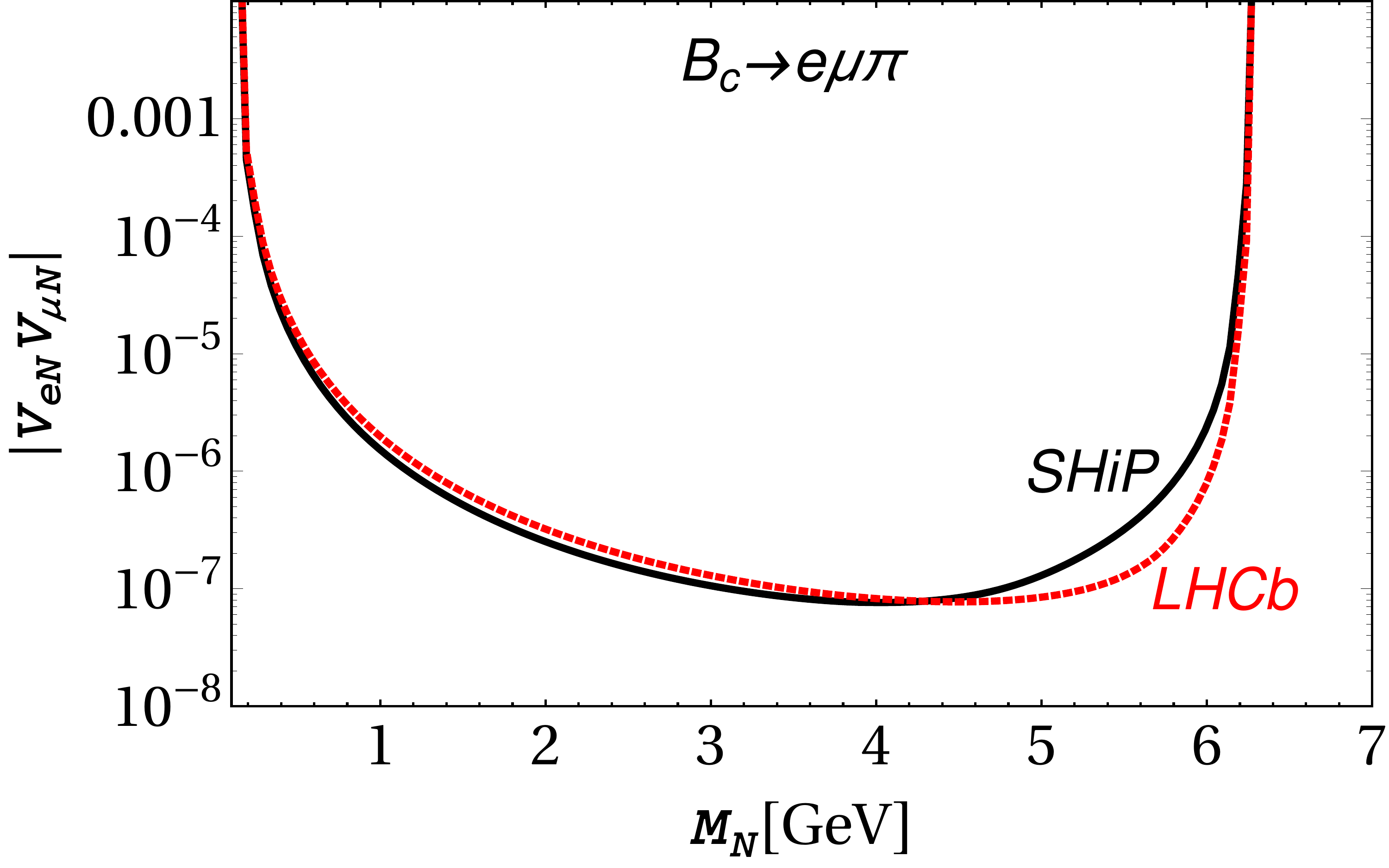}
		\includegraphics[width=0.45\textwidth]{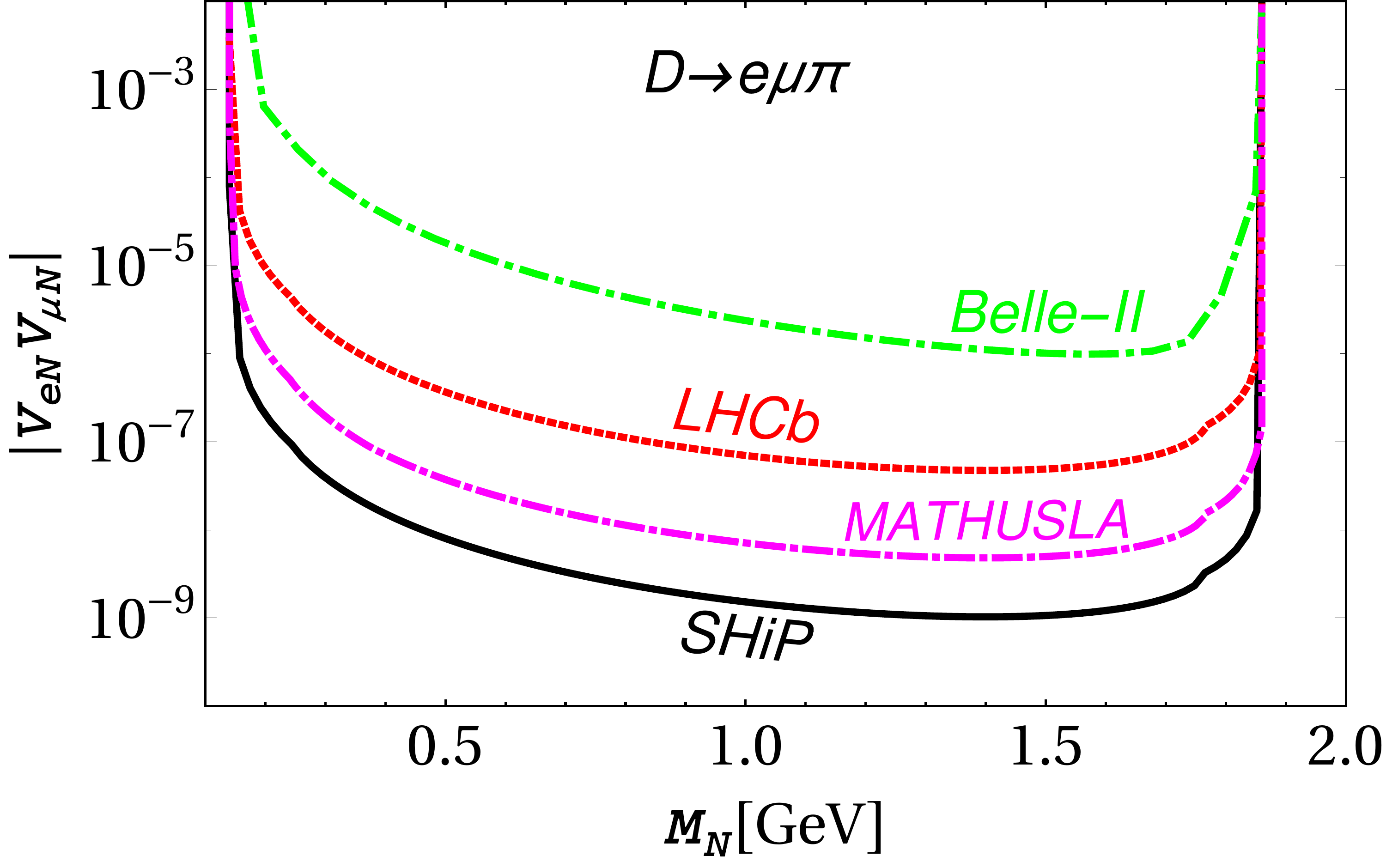}
	\includegraphics[width=0.45\textwidth]{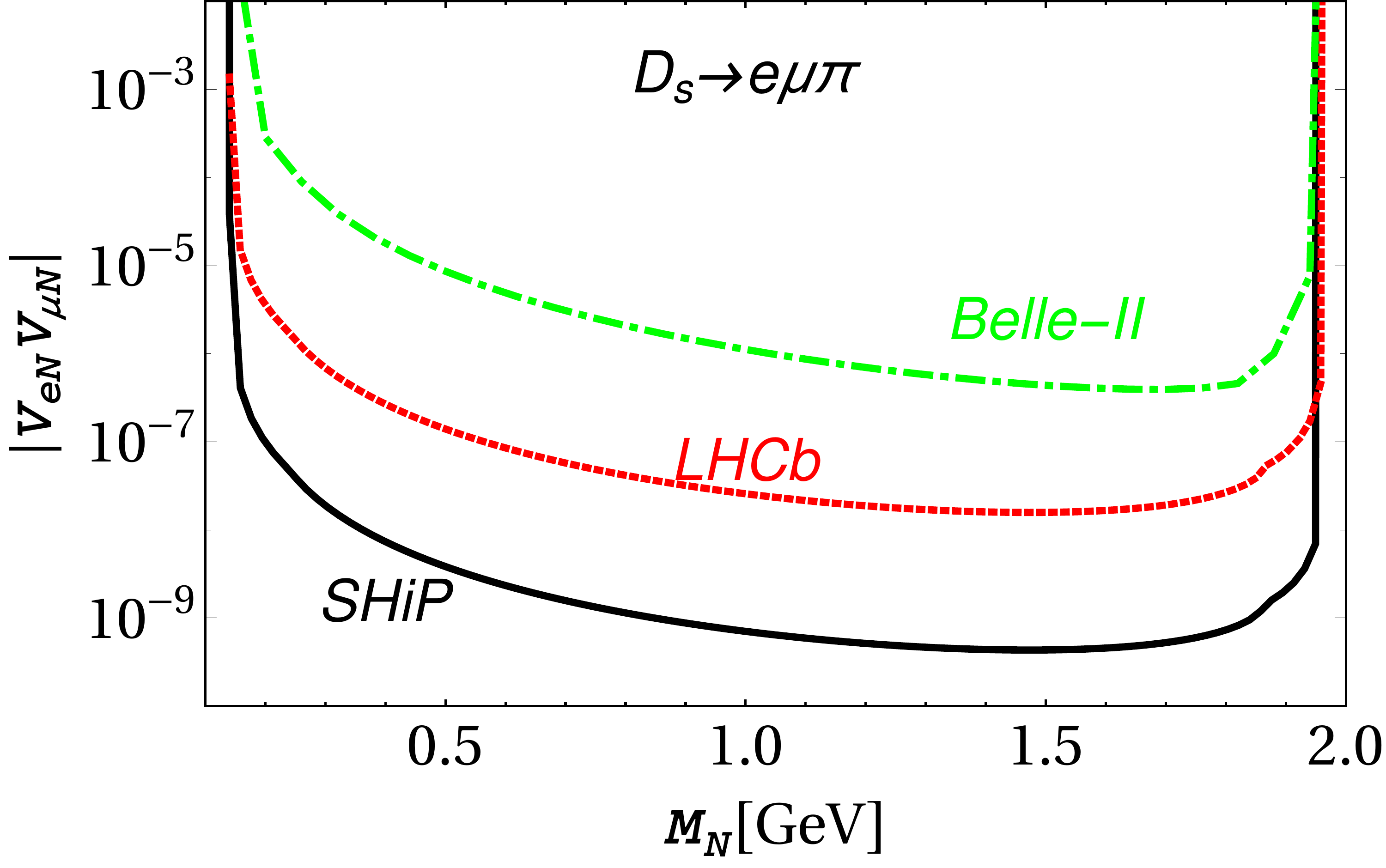}
	\caption{\small{Same as Fig.~\ref{Bounds on VeN}. The plots in different panels show the sensitivity reach of the mixing angle $|V_{eN}V_{\mu N}|$ from the meson decays $M_{1}\to e\mu\pi$.}}
	\label{Bounds on VeNVmuN}
\end{figure}
As an example, for the case of $B$~(FCC-ee, SHiP) and $D_s$~(SHiP) meson decays,  there is  approximately one  order of magnitude difference between the two results. The result for  $K$ meson decay at NA62 differs by two order of magnitude. Hence, the inclusion of parent meson velocity 
is indeed very important when calculating the bounds on the mixing angles. 
\begin{figure}[ht!]
	\centering
		\includegraphics[width=0.45\textwidth]{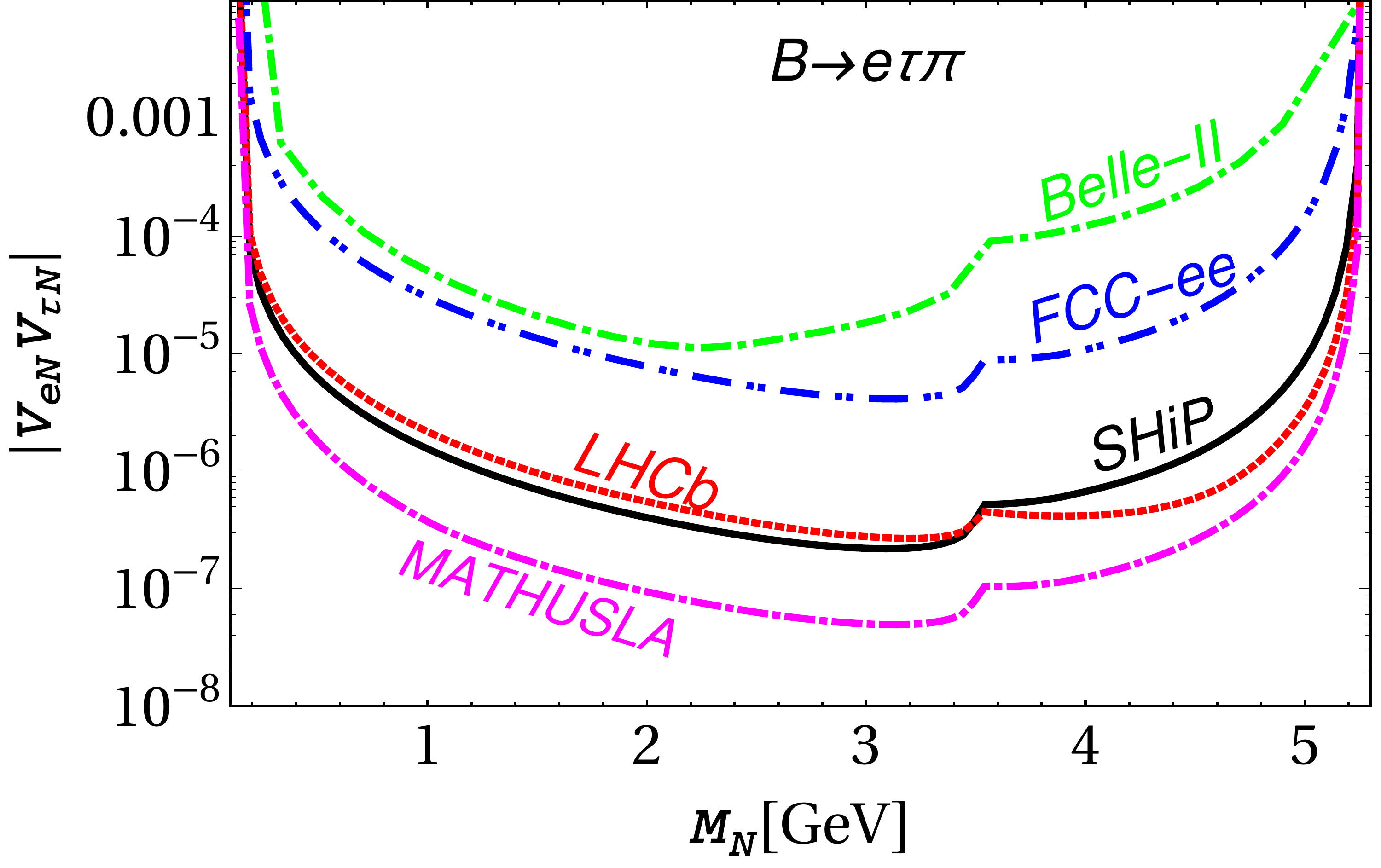}
	\includegraphics[width=0.45\textwidth]{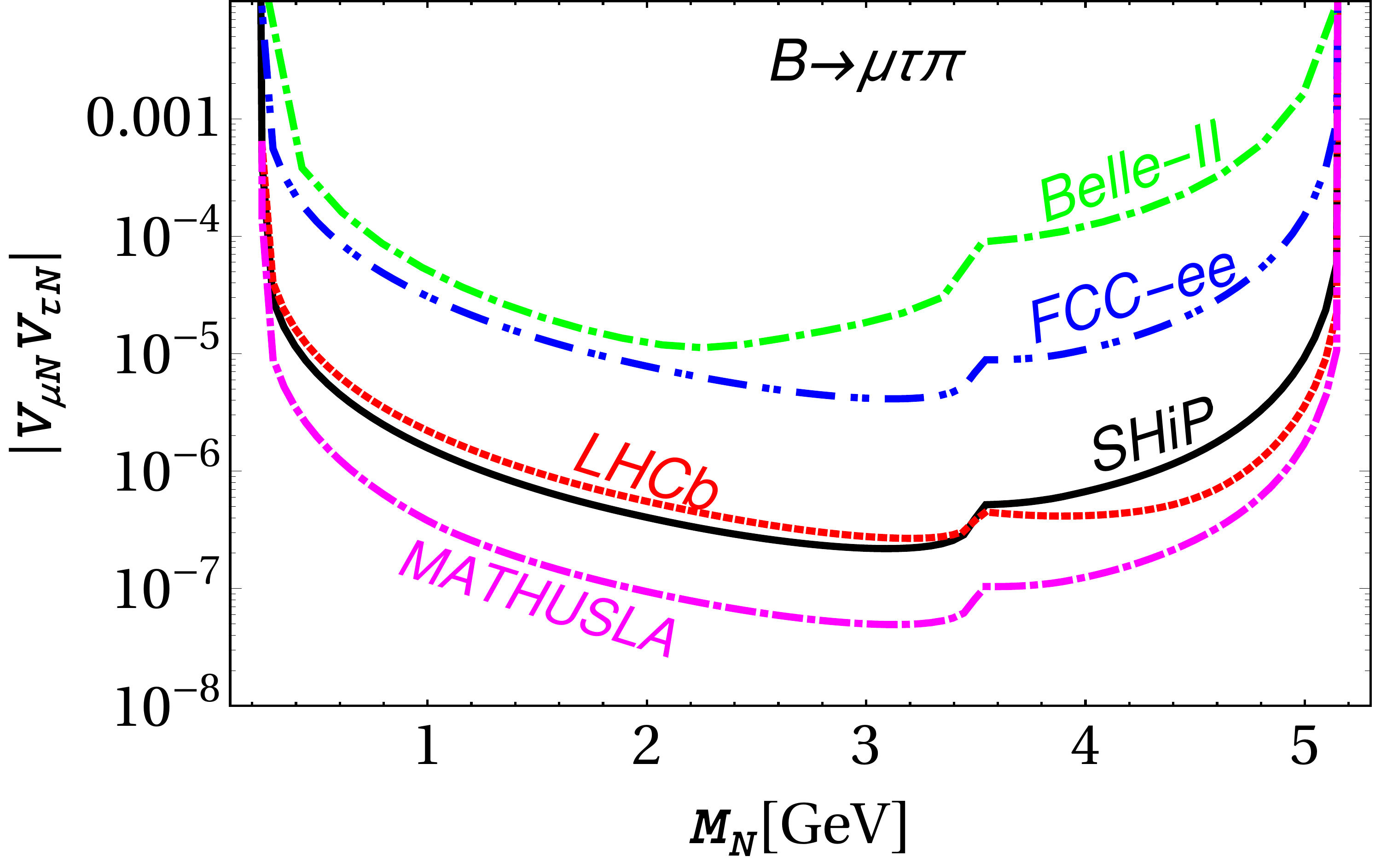}
		\includegraphics[width=0.45\textwidth]{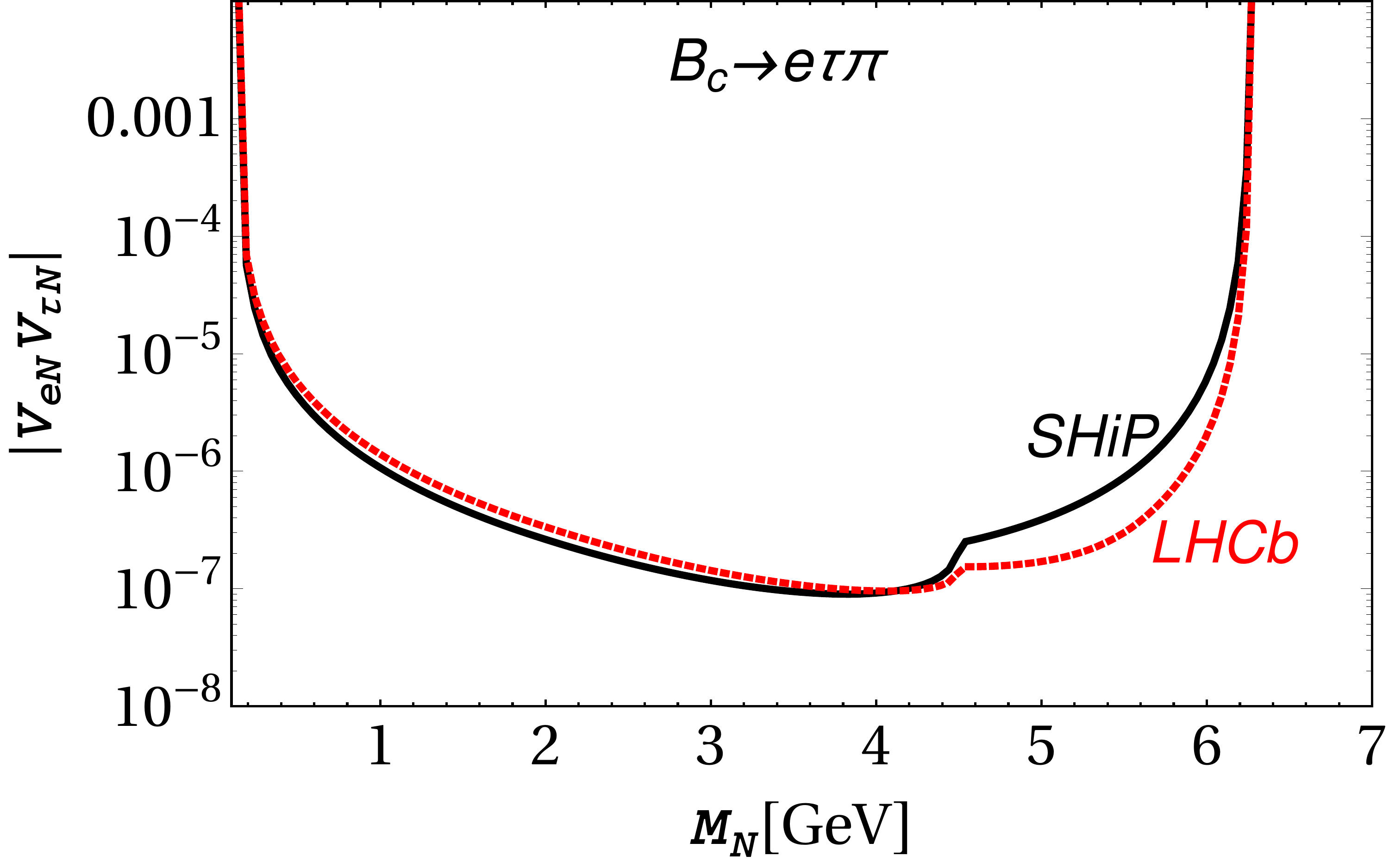}
		\includegraphics[width=0.45\textwidth]{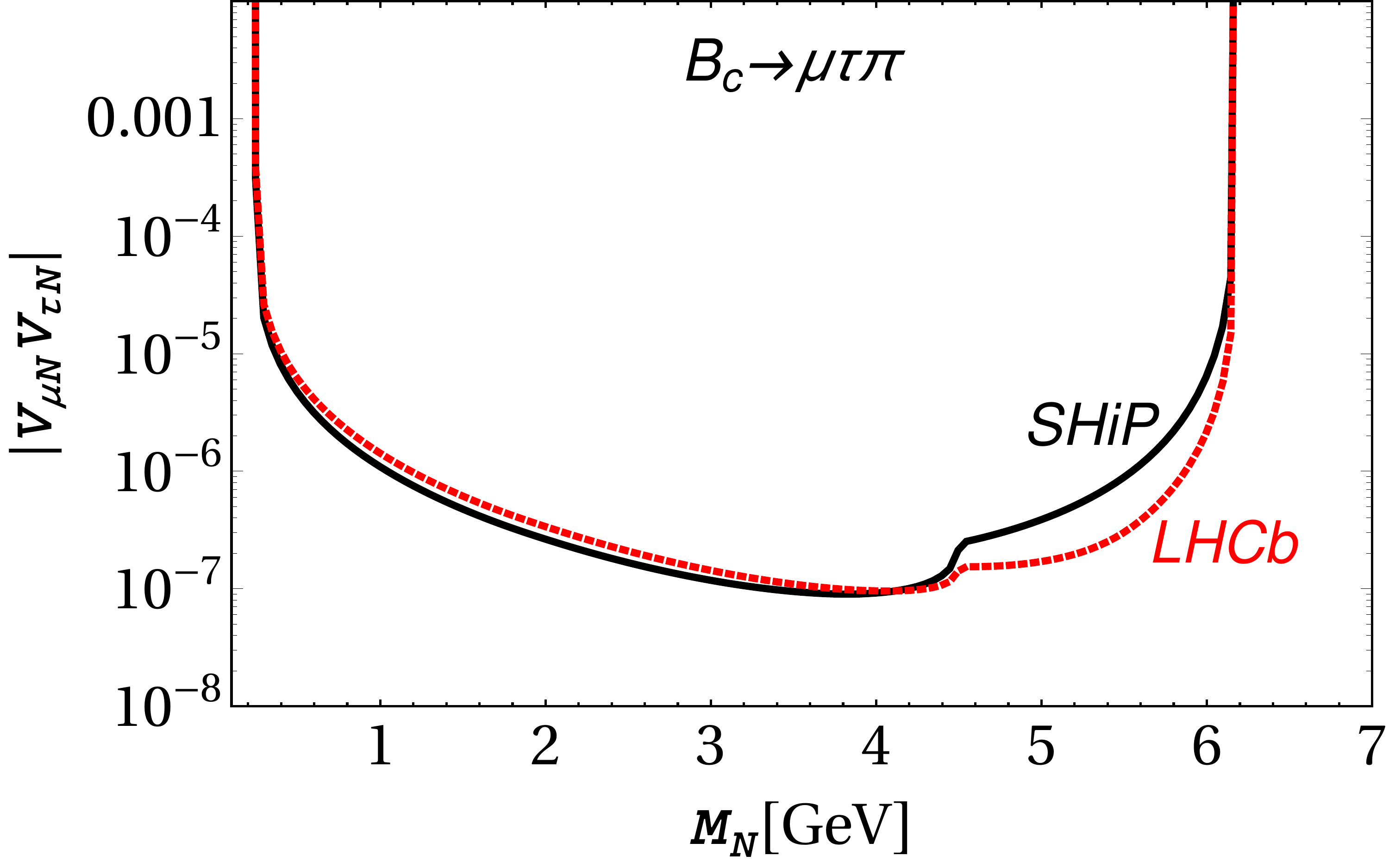}
	\includegraphics[width=0.45\textwidth]{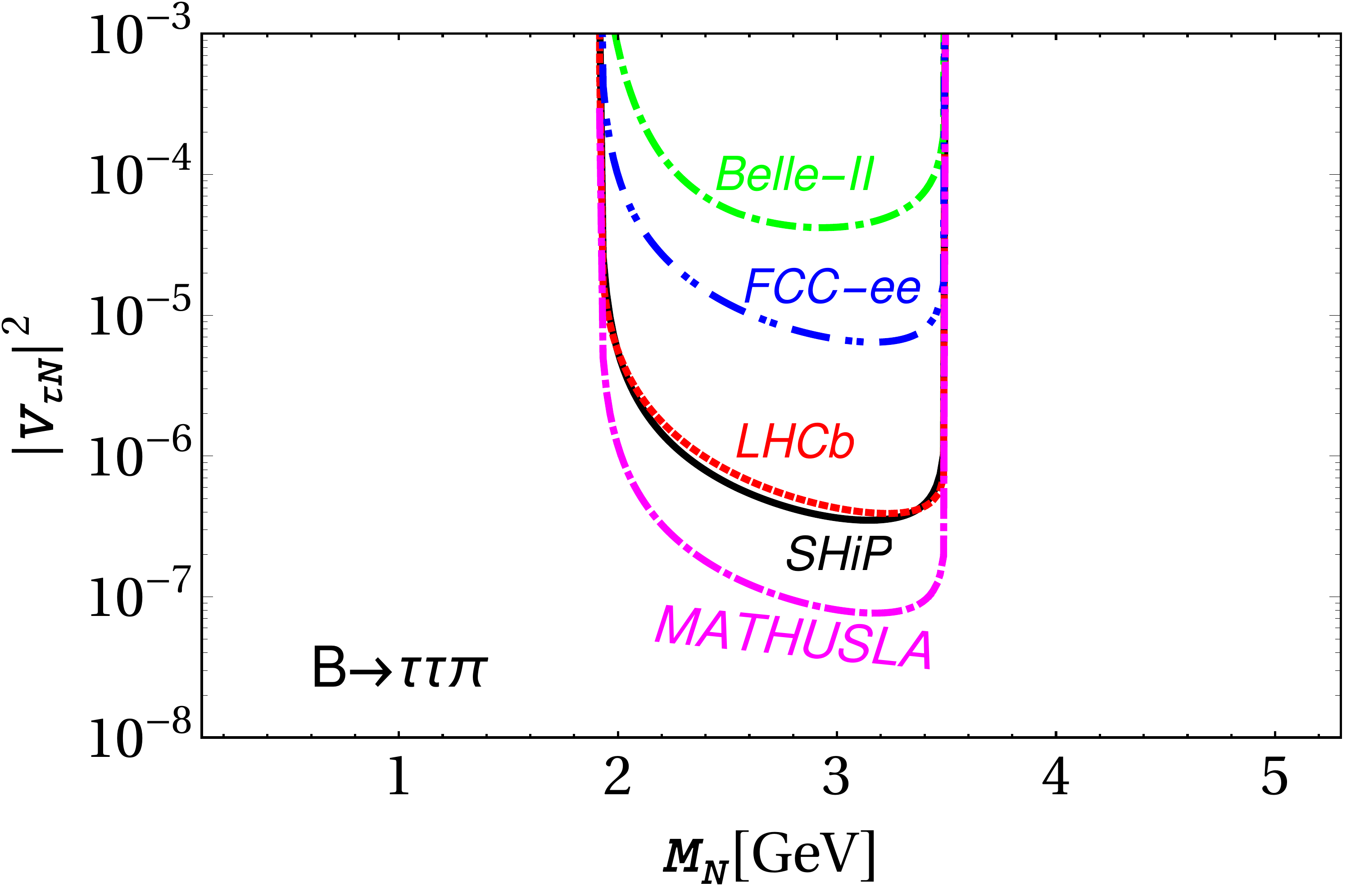}
		\includegraphics[width=0.45\textwidth]{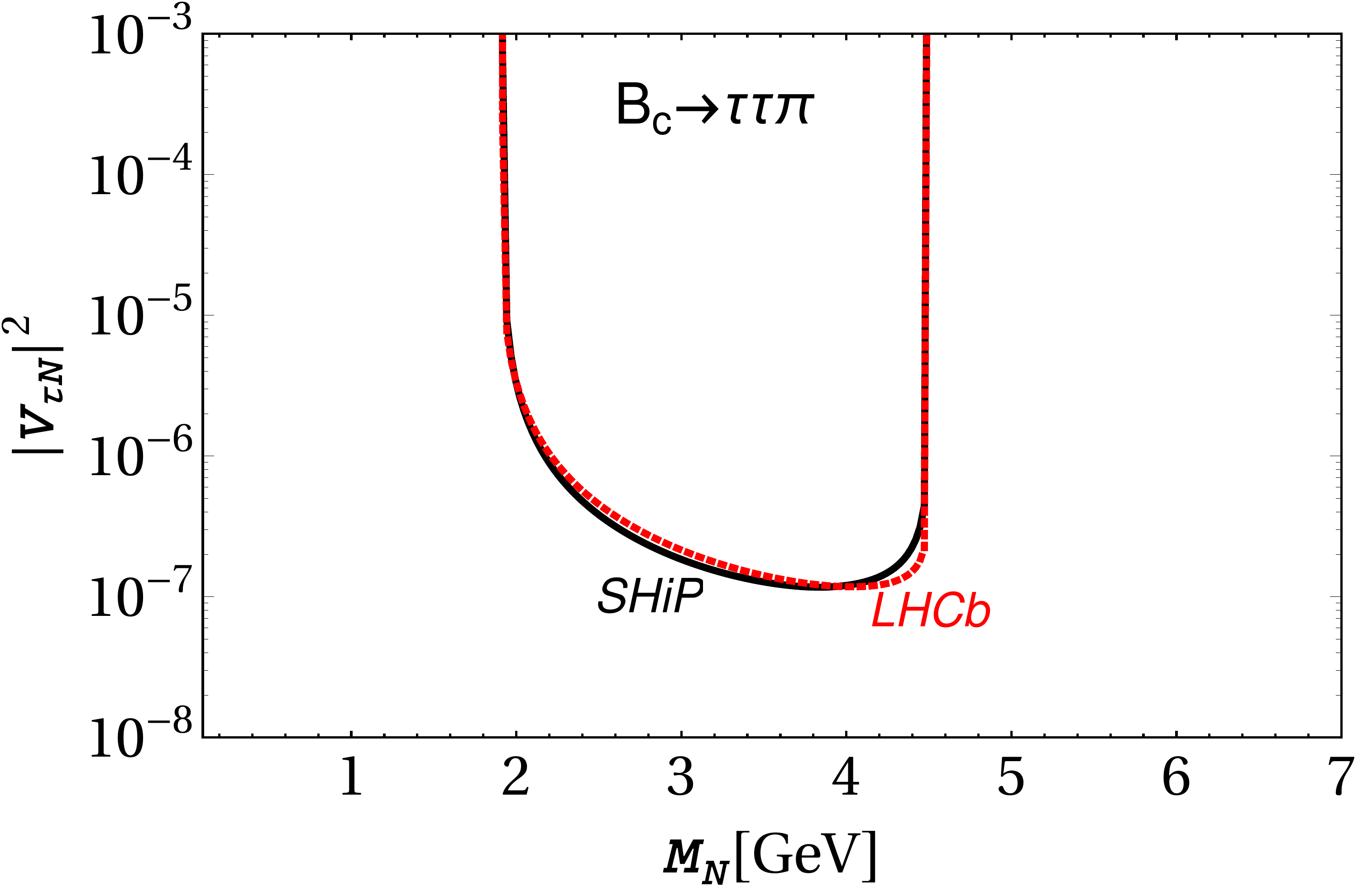}
	\caption{\small{Future sensitivity reach and present limits on the mixing angles $|V_{eN}V_{\tau N}|$, $|V_{\mu N}V_{\tau N}|$ and $|V_{\tau N}|^{2}$ with respect to the mass $M_{N}$ from meson decays 
	$B,B_c\to e\tau\pi$, $B,B_c\to\mu\tau\pi$ and $B,B_c\to\tau\tau\pi$, respectively at various experiments. The upper panel is for $B$ meson decay at SHiP~(black), MATHUSLA~(magenta), LHCb~(red), FCC-ee~(blue) and Belle-II~(green). The middle panel is for $B_c$ meson decay at SHiP~(black) and LHCb~(red). The left figure of the lower panel is for the bound on $|V_{\tau N}|^2$ from $B\to\tau\tau\pi$ decay at SHiP~(black), MATHUSLA~(magenta), LHCb~(red), FCC-ee~(blue) and Belle-II~(green). The right figure of the lower panel is for the bound on $|V_{\tau N}|^2$ from $B_c\to\tau\tau\pi$ decay at SHiP~(black) and LHCb~(red).}}
	\label{Bounds on VeNVtauN}
\end{figure}
\\
In Figs.~\ref{Bounds on VeN}-\ref{Bounds on VeNVtauN}, we show the final  
bounds and future sensitivity on various mixing angles such as $|V_{\ell N}|^{2}$, $\ell=e,\mu,\tau$ and $|V_{\ell_1 N}V_{\ell_2 N}|$, $\ell_1,
\ell_2=e,\mu,\tau$, $\ell_1\ne\ell_2$ as a function of RH neutrino mass $M_N$. When calculating the limits on these mixing angles, we are using Eq.~\ref{event signal in rest} for Belle-II and Eq.~\ref{event signal in flight decay} for the other experiments. The latter properly takes into account parent meson velocity effect in the RH neutrino decay probability inside the detector. Due to huge number of charmed meson productions, the future experiment SHiP will be able to probe $|V_{eN}|^2$, $|V_{\mu N}|^2\sim\mathcal{O}(10^{-9})$ and $|V_{eN}V_{\mu N}|\sim\mathcal{O}(10^{-9})$ in the mass range $0.14\,\text{GeV}<M_N<1.9\,\text{GeV}$. Fig.~\ref{Comparison of Mixing with velocity} shows that without considering the $K$ meson velocity, tightest bounds on mixing angles $|V_{eN}|^2$~($0.14\,\text{GeV}<M_N<0.49\,\text{GeV}$),
$|V_{\mu N}|^2$~($0.24\,\text{GeV}<M_N<0.38\,\text{GeV}$) and 
$|V_{eN}V_{\mu N}|$~($0.14\,\text{GeV}<M_N<0.49\,\text{GeV}$) are obtained from the meson decays $K\to ee\pi$, $K\to\mu\mu\pi$ and $K\to e\mu\pi$, respectively. Taking into account parent meson velocity, the tightest bound in the above mass ranges can instead be obtained from the $D_s$ meson decays at SHiP. For relatively higher mass range $2\,\text{GeV}<M_N<5\,\text{GeV}$, the tightest bound on mixing angles $|V_{eN}|^2$, $|V_{\mu N}|^2\sim\mathcal{O}(10^{-7})$ and $|V_{e N}V_{\mu N}|\sim\mathcal{O}(10^{-7})$ can be obtained from $B$ meson decays at MATHUSLA. Finally for the mass range $5\,\text{GeV}<M_N<6\,\text{GeV}$, the tightest limit on mixing angles $|V_{eN}|^2$, $|V_{\mu N}|^2\sim\mathcal{O}(10^{-7})$ and $|V_{e N}V_{\mu N}|\sim\mathcal{O}(10^{-7})$ will be provided by the $B_c$ meson decay at LHCb. 

Furthermore, the large mass gap between $B(B_c)$ and $\pi$ meson 
allows one or both final leptons to be tau. Hence, we have included in our study additional final states like $B$, $B_c\to e\tau\pi$, $\mu\tau\pi$
and $\tau\tau\pi$. 
The highest sensitivity reach on $|V_{eN}V_{\tau N}|\sim\mathcal{O}(10^{-7})$ and $|V_{\mu N}V_{\tau N}|\sim\mathcal{O}(10^{-7})$ can be provided from the $B\to e(\mu)\tau\pi$~($0.2\,\text{GeV}<M_N<5\,\text{GeV}$) at MATHUSLA and $B_c\to e(\mu)\tau\pi$~($5\,\text{GeV}<M_N<6\,\text{GeV}$) at LHCb, respectively. Additionally, the tightest bound on $|V_{\tau N}|^2\sim\mathcal{O}(2\times 10^{-7})$ can also be provided by $B\to \tau\tau\pi$ decay mode~($2\,\text{GeV}<M_N<3.4\,\text{GeV}$) at MATHUSLA and $B_c\to \tau\tau\pi$ decay mode~($3.4\,\text{GeV}<M_N<4.5\,\text{GeV}$) at LHCb, respectively. Note that, $B$, $B_c\to\tau\tau\pi$ meson decays constraint the mixing angle $|V_{\tau N}|^2$ in the mass range, where
it has so far been unconstrained by any of the $\tau$ or other meson decays. In spite of larger number of $D$ production, compared to $D_s$ meson at SHiP,
the suppression from the weak annihilation vertex in the case of $D_s$ meson $V_{cs}$ is less compared to $D$ meson $V_{cd}$. 
As a result of this, tightest limits on the mixing angles will be provided by the $D_s$ meson decays in the relatively lower mass range. 

Note that, if both the like sign di-leptons are not of the same flavor~($\ell_1\ne\ell_2$), 
then the process is not only lepton number violating, but also lepton flavor violating. Further, 
if the distance between $N$ production and decay points is large enough, then the two processes, 
$M_1\to\ell_1 N$ followed by $N\to\ell_2\pi$ will be separated. Assuming this separation, the two processes 
$M_1\to\ell_1\ell_2\pi$ and $M_1\to\ell_2\ell_1\pi$ can be distinguished. While deriving the bounds on the mixing angles
for the case of $\ell_1\ne\ell_2$,
we are assuming this separation in our study. This is justified as the decay width $\Gamma_N$ is very small~(hence the lifetime is very large) in the mass range of interest.  The allowed mass range of $N$ for the decay modes 
$M_1\to\ell_1\ell_2\pi$ and $M_1\to\ell_2\ell_1\pi$ are $m_{\ell_2}+m_\pi < M_N < m_{M_1}-m_{\ell_1}$ and
$m_{\ell_1}+m_\pi < M_N < m_{M_1}-m_{\ell_2}$, respectively. We consider both of the channels $M_1\to\ell_1\ell_2\pi$ and $M_1\to\ell_2\ell_1\pi$ to derive the bound on the mixing angle $|V_{\ell_1 N}V_{\ell_2 N}|$.\\
One important point to note is that we have considered idealized 
detector with $100\%$ detection, reconstruction efficiencies etc to derive the constraints on the mixing angles. The realistic constraints are expected to be weaker
and will only be feasible through searches by the experimental collaborations, incorporating the detection, reconstruction
efficiencies in actual experiment.
\begin{figure}
	\centering
		\includegraphics[width=0.8\textwidth]{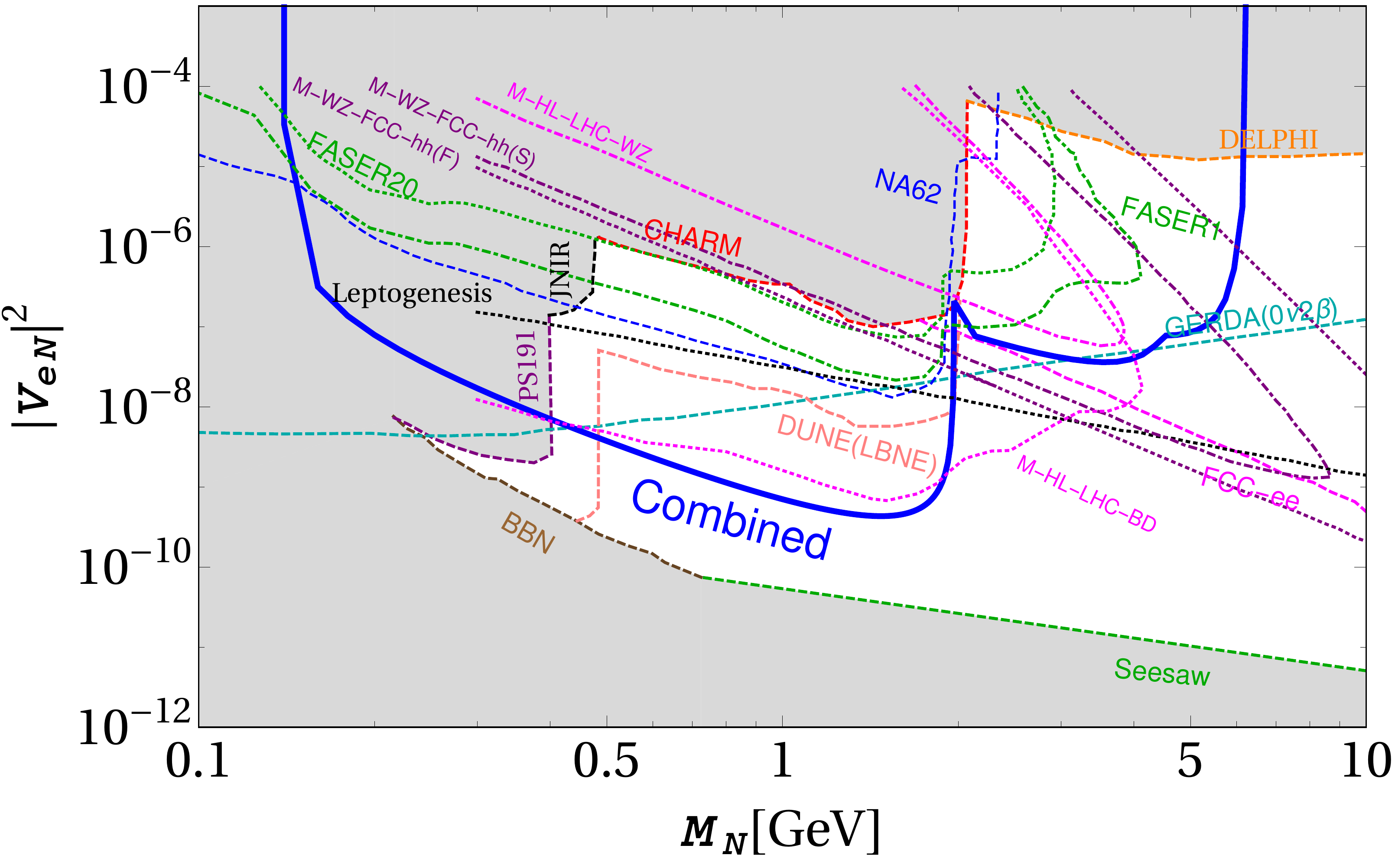}
	\caption{\small{Combined bounds (thick-blue solid) on mixing angle $|V_{e N}|^{2}$ as a function of mass $M_{N}$ from all the meson decays. Strongest lower limits from the Seesaw (green-dashed) and BBN (brown-dahsed) on $|V_{eN}|^2$ are shown in this plot. 
	Strongest upper bounds on $|V_{eN}|^2$ are obtained from PS191 (magenta-dashed), JNIR (black-dashed), CHARM (red-dashed), DELPHI (Orange-dashed) and Leptogenesis (black-dotted) are shown in this plot. The shaded region is ruled out by these results.
	The projected upper limits from the NA62 (blue dashed), GERDA (dark-cyan dashed), FCC-ee (magenta-dashed) and DUNE (pink-dashed) on $|V_{eN}|^2$ are shown. Prospective bounds are shown from the FASER with detector radius $R=20$ cm is shown by FASER20 (green-dotted)
	whereas the limits from the $R=1$ m is shown by FASER1 (green-dot-dashed). Prospective upper limits from the MATHUSLA at the FCC-hh for the $W/Z$ boson decays at the FCC-hh for MATHUSLA standard benchmark surface version are represented by 
	M-WZ-FCC-hh(S) (magenta-dot-dashed) and forward version M-WZ-FCC-hh(F) (magenta-dotted). Prospective limits for the heavy neutrinos produced from the $W/Z$ decays are represented by M-HL-LHC-WZ (magenta-dot-dashed) and $B/D$ meson decays are represented by 
	M-HL-LHC-BD (magenta-dotted) at the HL-LHC.}}
	\label{combinedbound1}
\end{figure}
\begin{figure}
	\centering
	\includegraphics[width=0.8\textwidth]{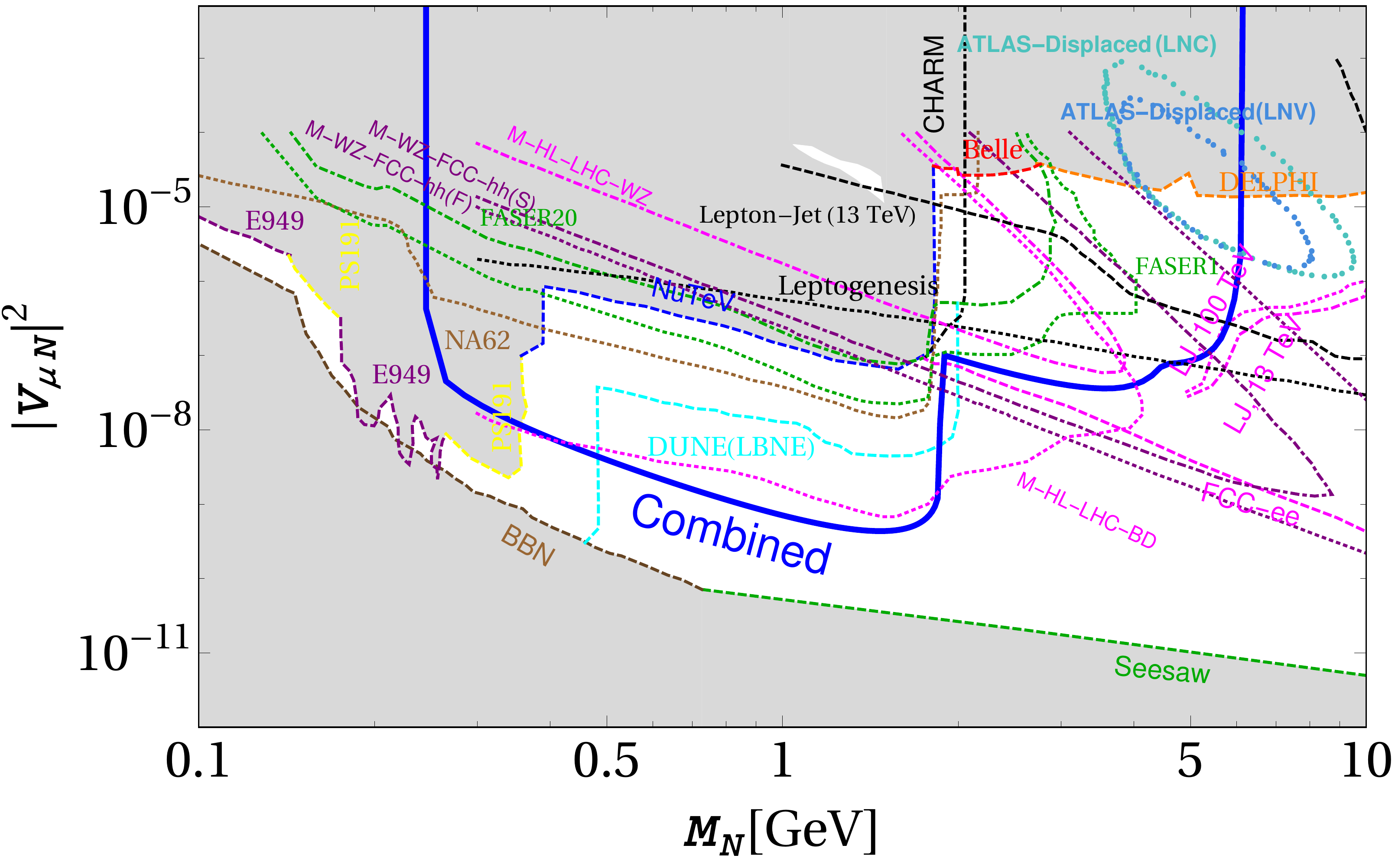}
	\caption{\small{Combined bounds (thick-blue solid) on mixing angle $|V_{\mu N}|^{2}$ as a function of mass $M_{N}$ from all the meson decays. Strongest lower limits from the Seesaw (green-dashed) and BBN (brown-dahsed) and  
	strongest upper limits from PS191 (yellow-dashed), E949 (magenta-dashed), NuTeV (blue-dashed), CHARM (black-dashed), Belle (red-dashed) and DELPHI (Orange-dashed) are shown in this plot. 
	The shaded region is ruled out by these results.
	The projected upper limits from the NA62 (brown-dotted), Leptogenesis (black-dotted), FCC-ee (magenta-dashed) and DUNE (cyan-dashed) and  FASER with detector radius $R=20$ cm is shown by FASER20 (green-dot-dahsed)
	whereas the limits from the $R=1$ m is shown by FASER1 (green-dotted). Prospective upper limits from the MATHUSLA at the FCC-hh for the $W/Z$ boson decays at the FCC-hh for MATHUSLA standard benchmark surface version are represented by 
	M-WZ-FCC-hh(S) (magenta-dot-dashed) and forward version M-WZ-FCC-hh(F) (magenta-dotted). Prospective limits for the heavy neutrinos produced from the $W/Z$ decays are represented by M-HL-LHC-WZ (magenta-dot-dashed) and $B/D$ meson decays are represented by 
	M-HL-LHC-BD (magenta-dotted) at the HL-LHC. Experimental bounds from the ATLAS displaced vertex searches of the Majorana heavy neutrino for the Lepton Number Violating (LNV) channel are represented by ATLAS-Displaced (LNV) (light-blue-dotted) and the limits obtained from the 
	Lepton Number Conserving (LNC) channel are represented by ATLAS-Displaced (LNC) (sea-blue-dotted). Theoretical limits from lepton-jet search for $1$ GeV $\leq M_N \leq 10$ GeV has been represented by Lepton-Jet ($13$ TeV)(Black, dashed). Corresponding limits for $M_N \geq 5$ GeV for the $13$ TeV LHC and $100$ TeV are shown by LJ, $13$ TeV (Magenta, dot-dashed) and LJ, $100$ TeV (Magenta, dotted) respectively.}}
	\label{combinedbound2}
\end{figure}
\begin{figure}
	\centering
		\includegraphics[width=0.8\textwidth]{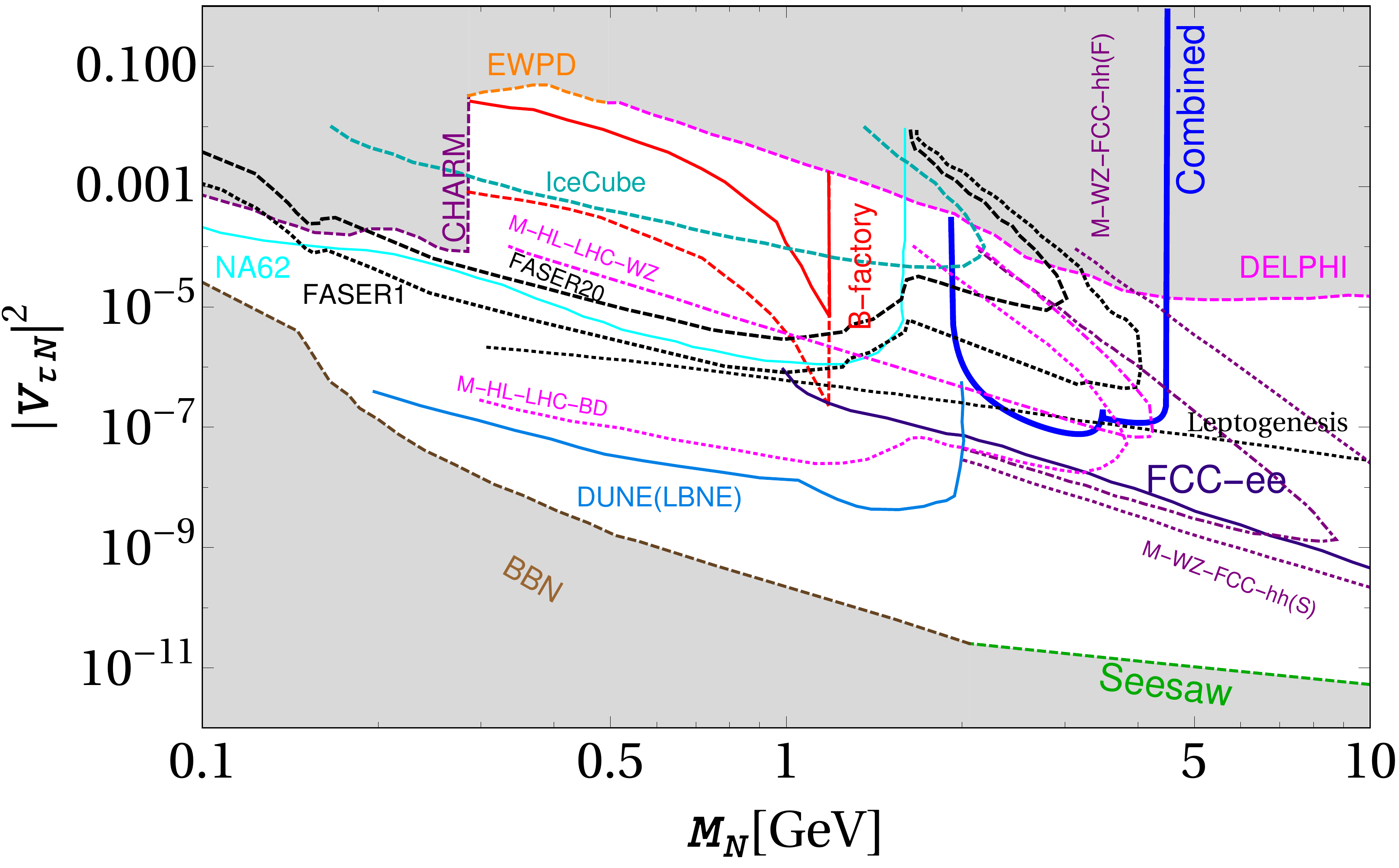}
	\caption{\small{Combined bounds (thick-blue solid) on mixing angle $|V_{\tau N}|^{2}$ as a function of mass $M_{N}$ from all the meson decays. Strongest lower limits from the Seesaw (green-dashed) and BBN (brown-dahsed) on $|V_{\tau N}|^2$ are shown in this plot. 
	Strongest upper bounds on $|V_{\tau N}|^2$ are obtained from  CHARM (purple-dashed), EWPD (orange-dashed), Leptogenesis (black-dotted), DELPHI (magenta-dashed) are shown in this plot. The shaded region is ruled out by these results.
	The projected upper limits from the NA62 (cyan-solid), FCC-ee (darker-blue-solid) and DUNE (light-blue-solid) on $|V_{\tau N}|^2$ are shown. Prospective bounds are shown from the FASER with detector radius $R=20$ cm is shown by FASER20 (black-dashed)
	whereas the limits from the $R=1$ m is shown by FASER1 (black-dotted). Prospective upper limits from the MATHUSLA at the FCC-hh for the $W/Z$ boson decays at the FCC-hh for MATHUSLA standard benchmark surface version are represented by 
	M-WZ-FCC-hh(S) (magenta-dot-dashed) and forward version M-WZ-FCC-hh(F) (magenta-dotted). Prospective limits for the heavy neutrinos produced from the $W/Z$ decays are represented by M-HL-LHC-WZ (magenta-dot-dashed) and $B/D$ meson decays are represented by 
	M-HL-LHC-BD (magenta-dotted) at the HL-LHC. The prospective limits from the B-factory (red-dashed and red-solid) and IceCube (darker-cyan-dashed) are also shown in this figure.}}
	\label{combinedbound3}
\end{figure}
\section{Combined  Sensitivity Reach from meson decays and Comparison with Existing Constraints}
\label{combined limit}
In this section, we discuss  the  future sensitivity reach from LNV three body meson decays.  The combined limits represent the strongest limits obtained in  different mass ranges of $N$.   
In Fig.~\ref{combinedbound1}, we show  the   combined sensitivity reach  for $|V_{eN}|^2$ by  dark blue solid line.  
This  corresponds to the tightest constraints  that can be obtained from $D_s \to e e \pi$ mode (by  SHiP) in the  lower mass range $0.14\,\text{GeV}<M_N<2\,\text{GeV}$, and from  $B \to e e \pi$~(by MATHUSLA), $B_c\to ee\pi$~(by LHCb) in the  higher 
mass range $2\,\text{GeV}<M_N<6\,\text{GeV}$. Note that, for the lower mass range,  SHiP can probe $|V_{eN}|^2 \sim  10^{-9}$, while for higher mass range,  MATHUSLA and LHCb can probe $|V_{eN}|^2 \sim 10^{-7}$.  In particular, the  very near future  accumulation of data ($\mathcal{L}=300 \, \rm{fb}^{-1}$ ) in  LHCb   can probe $|V_{eN}|^2 \sim 10^{-7}$, around $M_N \sim 5$ GeV.
The sensitivity reach of $|V_{\mu N}|^2$, as shown in  Fig.~\ref{combinedbound2}  is very similar.  The combined limit  represents  the constraint  from $D_s \to \mu \mu \pi$, $B\to\mu\mu\pi$ and $B_c \to \mu \mu \pi$ decay modes, that can again be  probed in  SHiP, MATHUSLA and LHCb. For $|V_{\tau N}|^2$,  the best sensitivity reach $|V_{\tau N}|^2 \sim 10^{-7}$ can be provided by MATHUSLA in  $B \to \tau \tau \pi$ mode, while SHiP and LHCb can give similar 
 sensitivity reach with the mode $B_c \to \tau \tau \pi$.  The combined sensitivity reach has been shown in Fig.~\ref{combinedbound3}.
 
\begin{figure}
	\centering
		\includegraphics[width=0.75\textwidth]{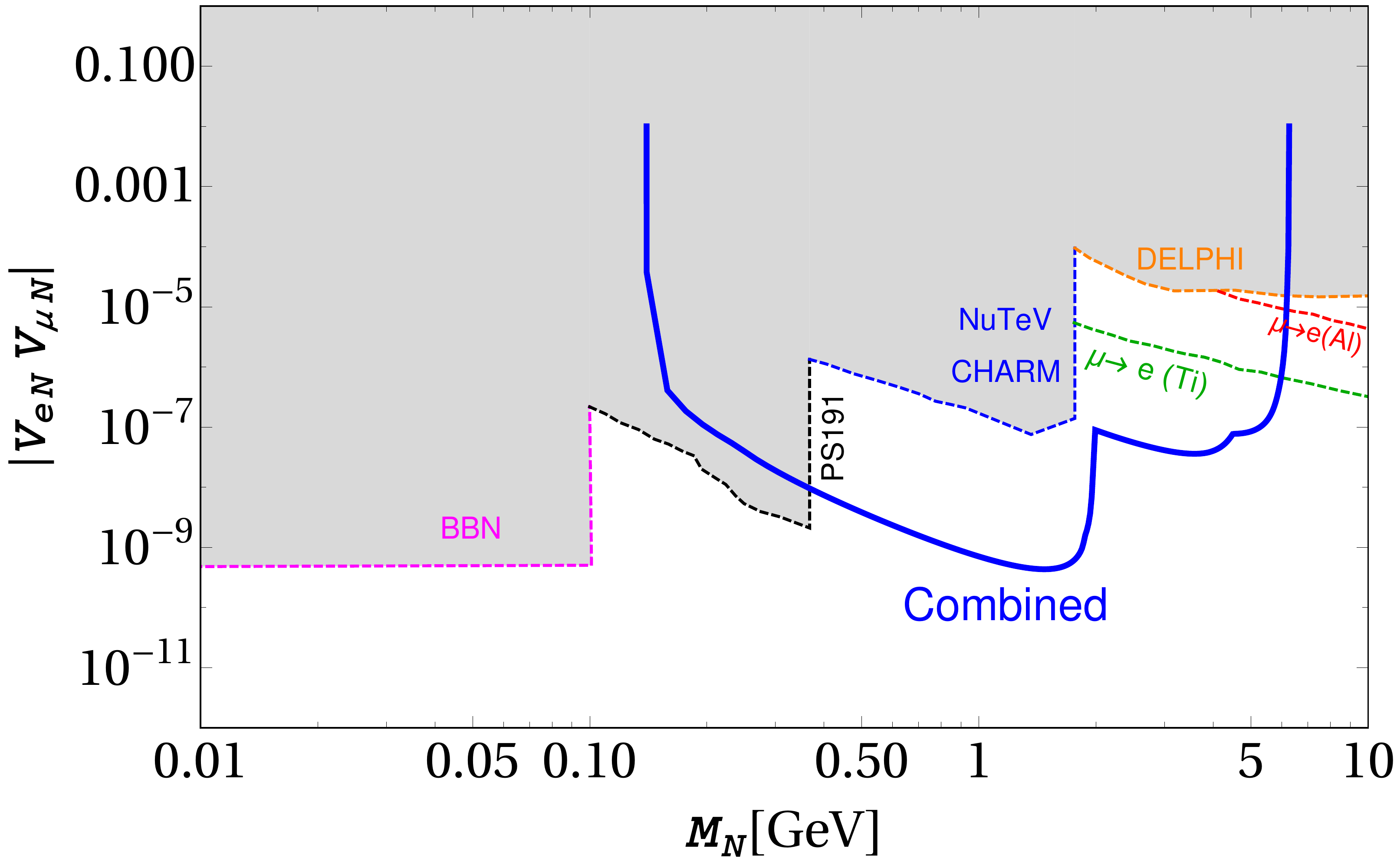}
	\caption{\small{Combined bounds on mixing angle $|V_{e N}V_{\mu N}|$ as a function of mass $M_{N}$ from all the meson decays. 
	The other strongest bounds on $|V_{eN} V_{\mu N}|$ from BBN (magenta-dashed)}, PS191 (black-dahsed), NuTeV, CHARM (blue-dashed, DELPHI (ornage-dahsed)) are also shown in this figure.
	Prospective bounds on $|V_{e N}V_{\mu N}|$ from the $\mu \to e$(Ti) (green-dashed) and $\mu \to e $(Al) (red-dashed) are also shown for $M_N \leq 10$ GeV. }
	\label{combinedbound4}
\end{figure}

\begin{figure}
	\centering
	\includegraphics[width=0.75\textwidth]{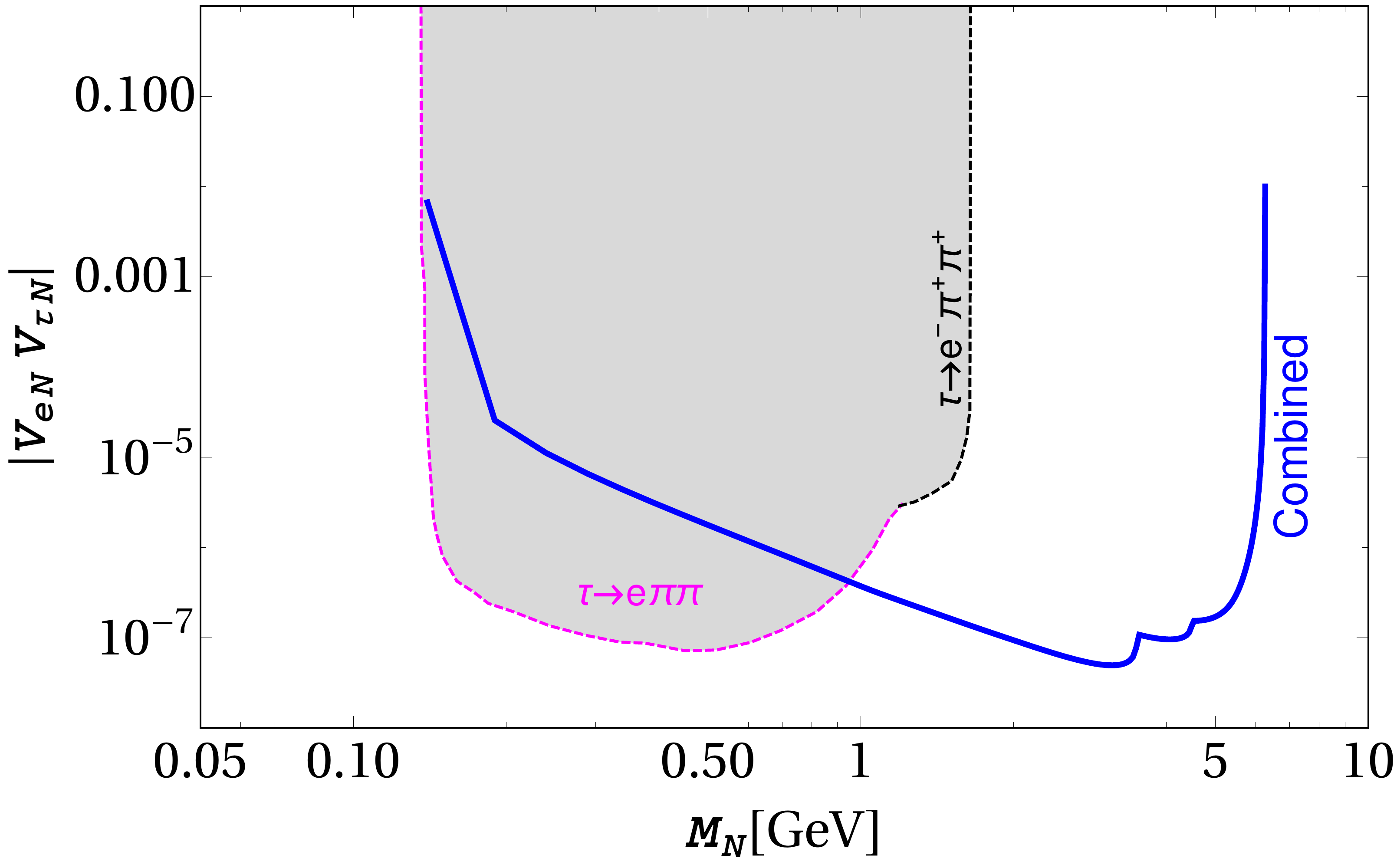}
	\caption{\small{Combined bounds (thick-blue solid) on mixing angle $|V_{e N}V_{\tau N}|$ as a function of mass $M_{N}$ from all the meson decays. 
	Upper limits from $\tau\to e\pi\pi$ (magenta-dashed) and $\tau \to e^-\pi^+\pi^+$ (black-dashed) are also shown for $M_N \leq 10$ GeV.}}
	\label{combinedbound5}
\end{figure}

\begin{figure}
	\centering
		\includegraphics[width=0.75\textwidth]{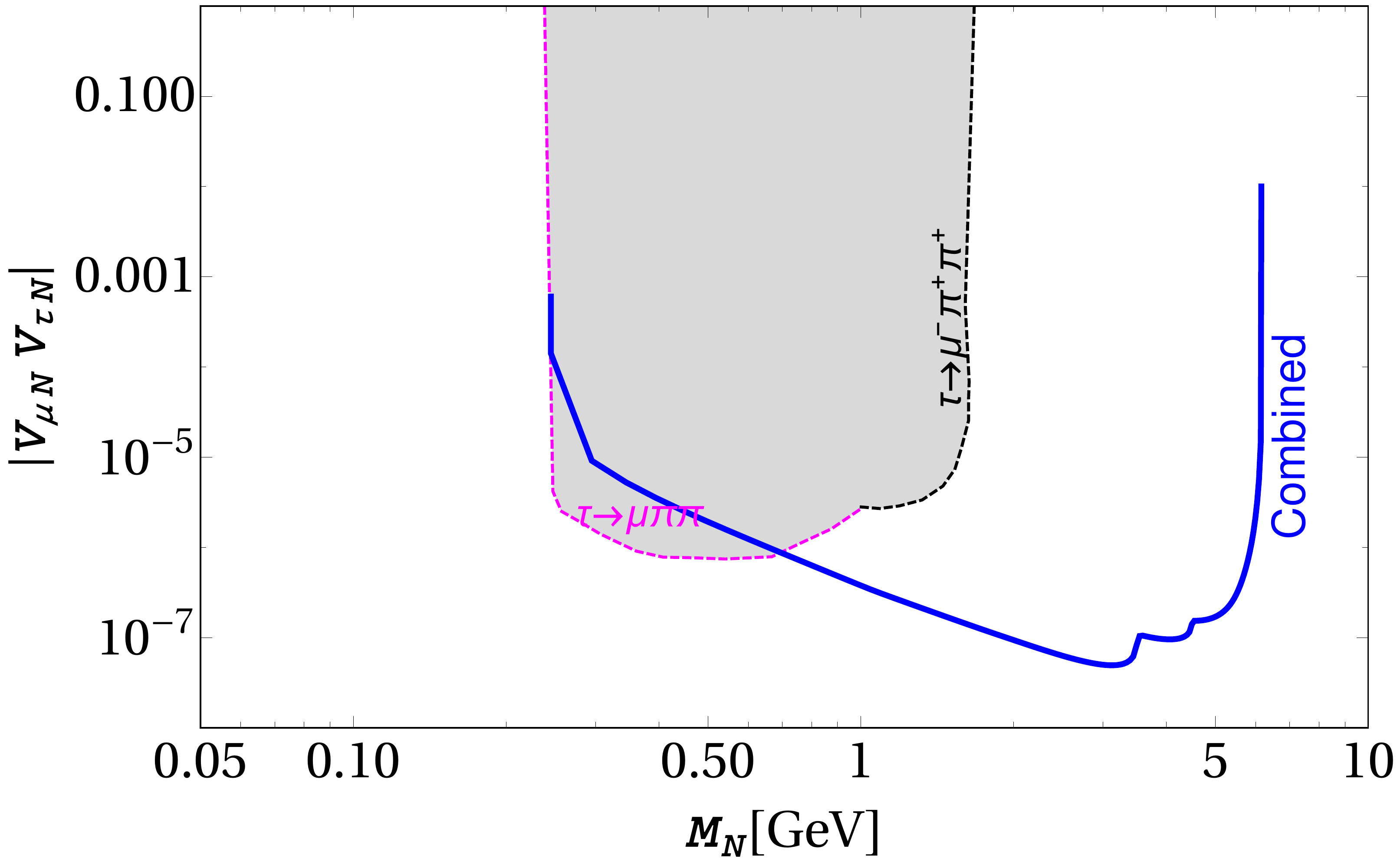}
	\caption{\small{Combined bounds (thick-blue solid) on mixing angle $|V_{\mu N}V_{\tau N}|$ as a function of mass $M_{N}$ from all the meson decays.
	Upper limits from $\tau\to \mu \pi\pi$ (magenta-dashed) and $\tau \to \mu^-\pi^+\pi^+$ (black-dashed) are also shown for $M_N \leq 10$ GeV. }}
	\label{combinedbound6}
\end{figure}

The future sensitivity of $|V_{eN} V_{\mu N}|$, $|V_{\mu N} V_{\tau N}|$ and $|V_{eN} V_{\tau N}|$ are shown in Fig.~\ref{combinedbound4}, Fig.~\ref{combinedbound5} and Fig.~\ref{combinedbound6}, respectively.  For $|V_{eN} V_{\mu N}|$ mode, 
lower mass range up to $M_N \sim $2 GeV  can be probed by the channel $D_s \to e \mu \pi$ ( by SHiP) where  sensitivity  down to $|V_{eN} V_{\mu N}| \sim  10^{-9}$ can be obtained.  
RH neutrino of higher mass  $M_N \sim 5$ GeV and $M_N\sim 6$ GeV can be probed by $B \to e\mu \pi$ mode (by MATHUSLA) and $B_c\to e\mu\pi$ mode (by LHCb), with sensitivity reach  $|V_{eN} V_{\mu N}| \sim  10^{-7}$. For $|V_{eN} V_{\tau N}|$ and $|V_{\mu N} V_{\tau N}|$ mixings, 
the sensitivity for the active-sterile mixing angles are  similar, as depicted in  the Fig.~\ref{combinedbound5} and Fig.~\ref{combinedbound6}. These can be probed in  LHCb,  MATHUSLA and SHiP.
We note that, the future limits from LNV meson decays will be most sensitive in between 0.5 GeV $< M_N < $ 2 GeV for $|V_{eN}|^2$, $|V_{\mu N}|^2$, $|V_{eN} V_{\mu N}|$. For  other mixing angles that 
involves $\tau$ in the final state,  best limit can be obtained in  relatively higher mass range $M_N \sim 5$ GeV.  We stress that, LHCb, and future experiments -  SHiP and MATHUSLA can probe  mixing angle of  the 
$\tau$ sector in a region, that is very loosely constrained. 

A number of other  constraints on the active-sterile mixing have been obtained from peak searches, pion decays, collider searches etc. A variety of choices of the heavy neutrino mass have been made in different articles \cite{Drewes:2013gca,Deppisch:2015qwa, Cai:2017mow,Das:2018hph,Caputo:2016ojx,Cvetic:2018elt,Kang:2019xuq,Cvetic:2019rms,Zamora-Saa:2019naq,Bondarenko:2018ptm,Antusch:2016ejd,Das:2017kkm,Rasmussen:2016njh,Das:2014jxa,Dev:2013wba,Das:2017zjc,Das:2017rsu,Banerjee:2015gca,BhupalDev:2012zg,Deppisch:2015qwa}, 
which discussed the limits on the heavy neutrino mass and mixing. We summarize the existing limits from these articles. We only show the constraints for  the heavy neutrinos lighter than $10$ GeV. See Fig.~\ref{combinedbound1} 
  for the two electron,  Fig.~\ref{combinedbound2} for two muon,  and Fig.~\ref{combinedbound3} for  two tau final states, that constrain $|V_{eN}|^2, |V_{\mu N}|^2 $ and $|V_{\tau N}|^2$.  These  bounds in the mass vs mixing plane of Figs.~\ref{combinedbound1} to \ref{combinedbound3} represent the theory  constraint  from  the seesaw (Seesaw) \cite{deGouvea:2009fp,deGouvea:2005er,Cirelli:2004cz}, big bang nucleosynthesis (BBN) \cite{Gorbunov:2007ak,Boyarsky:2009ix,Ruchayskiy:2012si},  experimental constraint from CHARM \cite{Bergsma:1985is,Vilain:1994vg,Orloff:2002de}, DELPHI \cite{Abreu:1996pa}. We also show the future sensitivity reach of  FCC-ee \cite{Blondel:2014bra, Abada:2014cca}, and DUNE (LBNE ) \cite{Adams:2013qkq}.  The PS191 \cite{Bernardi:1987ek} limits for the electron and muon flavors are shown in Figs.~\ref{combinedbound1} and \ref{combinedbound2}, respectively. The JNIR \cite{Baranov:1992vq} limit is represented by the black dashed line for the electron flavor in Fig.~\ref{combinedbound1}. The regions excluded by the present constraints  are shaded in gray. The limits from GERDA \cite{Agostini:2013mzu} on the mass mixing plane in search of Majorana neutrinos from the neutrinoless double beta decay is represented by the dark cyan line in Fig.~\ref{combinedbound1}. Majorana heavy neutrino searches from the Meson decay in E949 \cite{Artamonov:2014urb} and NuTeV \cite{Vaitaitis:1999wq} can produce strong bounds on the heavy neutrino mass-mixing plane. Lepton-jet theoretical searches \cite{Dube:2017jgo} for the Majorana neutrinos with muon flavor can also put strong bounds. These bounds are shown in Fig.~\ref{combinedbound2} for the muon flavors. For the tau lepton, the bound in the corresponding mass region from EWPD~\cite{delAguila:2008pw,Akhmedov:2013hec,deBlas:2013gla} has been shown in Fig.~\ref{combinedbound3}. The decay of tau lepton into heavy Majorana neutrino and meson can also put bounds on the mass-mixing plane and can have prospective limits marked as 
B-factory \cite{Deppisch:2015qwa,Cvetic:2018elt}\footnote{Recently, Ref~\cite{Dib:2019tuj} also put bound on $|V_{\tau N}|^2$ using large samples of $e^{+}e^{-}\to\tau^{+}\tau^{-}$ collected at B-factory experiments.}. The NA62 \cite{NA62:2017rwk,CortinaGil:2017mqf,Lanfranchi:2017wzl} projection lines from the electron, muon and tau are shown in Figs.~\ref{combinedbound1}, \ref{combinedbound2} and \ref{combinedbound3}, respectively. Such experiment can be performed in the kaon mode and beam dump mode \cite{Drewes:2018gkc}. This search is sensitive to the heavy neutrinos that are produced in weak decays \cite{Shrock:1980ct,Shrock:1981wq} of mesons or tau leptons \cite{Gorbunov:2007ak}. The upper limit on the mass mixing plane from the Leptogenesis \cite{Drewes:2016jae} for the minimal scenario with two right handed neutrinos are shown in Figs.~\ref{combinedbound1} and \ref{combinedbound2}, respectively for the electrons and muons assuming the normal hierarchy of the light neutrino masses. The upper limit on the mixing angle $|V_{\mu N}|^2$ from Belle \cite{Liventsev:2013zz} are shown in Fig.~\ref{combinedbound2}.
Projected sensitivities (for the 4 events) of the MATHUSLA \cite{Curtin:2018mvb} detector in the mass-mixing plane for the heavy neutrinos produced from the $W/Z$ decays at the FCC-hh for MATHUSLA standard benchmark surface version (M-WZ-FCC-hh(S)) and forward version ((M-WZ-FCC-hh(F))) for the electron, muon and tau are shown in the Figs.~\ref{combinedbound1}, \ref{combinedbound2} and \ref{combinedbound3}, respectively for $M_N > 2$ GeV. We have shown the projected sensitivities in the mass-mixing plane for the heavy neutrinos produced from the $W/Z$ decays (M-HL-LHC-WZ) and $B/D$ meson decays (M-HL-LHC-BD) for electron, muon and tau lepton at the HL-LHC in Figs.~\ref{combinedbound1}, \ref{combinedbound2} and \ref{combinedbound3}, respectively. The bounds on the mass-mixing plane from the FASER \cite{Kling:2018wct}  with detector radius $R=20$ cm has been represented by FASER20 and $R=1$ m has been represented by FASER1, respectively for the electron, muon and tau leptons in Figs.~\ref{combinedbound1}, \ref{combinedbound2} and \ref{combinedbound3}, respectively. ATLAS displaced vertex bounds on the $|V_{\mu N}|^2$ for the Lepton Number Violating (LNV) and Lepton Number Conserving (LNC) searches are given in \ref{combinedbound2} \cite{Aad:2019kiz}. The prospective upper bounds in the mass-mixing plane for the tau lepton from the Ice-Cube \cite{Aartsen:2015dlt,Aartsen:2016nxy,Aartsen:2017nmd,Coloma:2017ppo} for $M_N < 10$ GeV are shown in Fig.~\ref{combinedbound3}. The projected sensitivity (theoretical) on $|V_{\mu N}|^2$ for the lepton-jet search for the for $1$ GeV $\leq M_N \leq 10$ GeV at the $13$ TeV are shown in Fig.~\ref{combinedbound2},  by the black dashed line, and the line labelled by  Lepton-Jet ($13$ TeV).
The corresponding limits (theoretical) on $|V_{\mu N}|^2$ from another lepton-jet search \cite{Dube:2017jgo} for $M_N \geq 5$ GeV at $13$ TeV (LJ, $13$ TeV) and $100$ TeV (LJ, $100$ TeV) are also shown in Fig.~\ref{combinedbound2}.

We briefly summarize the current strongest experimental bounds on the mixing angles such as $|V_{eN} V_{\mu N}|$, $|V_{eN} V_{\tau N}|$, and $|V_{\mu N} V_{ \tau N}|$ for the Majorana heavy neutrinos in 
Fig.~\ref{combinedbound4}, \ref{combinedbound5} and \ref{combinedbound6}, respectively for $M_N < 10$ GeV. Strongest bounds from the BBN \cite{Gorbunov:2007ak,Boyarsky:2009ix,Ruchayskiy:2012si}, 
PS191 \cite{Bernardi:1987ek}, NuTeV \cite{Vaitaitis:1999wq}, CHARM \cite{Bergsma:1985is,Vilain:1994vg,Orloff:2002de}, DELPHI \cite{Abreu:1996pa} are obtained from the Majorana heavy neutrino search for $M_N \leq 10$ GeV.
The shaded region is excluded by the results obtained from these experiments. The prospective bounds from the $\mu \to e$ (Ti) and $\mu \to e$ (Al) are shown in Fig.~\ref{combinedbound4} from \cite{Alonso:2012ji}.
The limits on the mixings from the $\tau$ decay into hadrons in association with electron and muon are shown in Figs.~\ref{combinedbound5} and \ref{combinedbound6}, respectively from BABAR \cite{Aubert:2005tp}. 
The limits from the $\tau \to e \pi \pi$ and $\tau \to e^-\pi^+\pi^+$ are shown in Fig.~\ref{combinedbound5} and those obtained from $\tau \to \mu \pi \pi$ and $\tau \to \mu^-\pi^+\pi^+$ are shown in Fig.~\ref{combinedbound6}, respectively \cite{Zamora-Saa:2016ito}.

We stress that, in the relatively lower mass range,  among the experimental constraints, the   tightest constraint on the   mixing angles $|V_{eN}|^2$ and $|V_{\mu N}|^2$ can be obtained  from LNV meson decays. These are 
however still one order of magnitude weaker than the theory constraints   from  BBN and Seesaw. For relatively higher mass range, our combined bounds on the mixing angles $|V_{eN}V_{\mu N}|$, $|V_{eN}V_{\tau N}|$ and $|V_{\mu N}V_{\tau N}|$ are the tightest bounds. As we have discussed before, the LNV meson decays can probe  the product of the mixings  $|V_{\mu N}V_{\tau N}|$, $|V_{eN}V_{\tau N}|$ in higher mass ranges $M_N \sim 5 $ GeV, that are so far unconstrained.
\section{Conclusion}
\label{conclusion}
We analyse discovery prospect of a heavy Majorana neutrino via    lepton number violating meson decays $M_{1}^{-}\to\ell_{1}^{-}\ell_{2}^{-}\pi^{+}$  at various ongoing and future experiments, such as,  NA62, LHCb, FCC-ee, Belle-II, SHiP and MATHUSLA. The large number of decaying mesons in these experiments may possibly result in an observation of the different  rare lepton number violating decays. Even their non-observation can be used to set constraints on the mixing coefficients between the standard flavour neutrinos and the heavy mass eigenstates. We explore in detail the effect of   parent meson's velocity  on the  sensitivity reach of the active-sterile mixing angles in the ongoing and future experiments. We compare the  resulting constraints on the mixing angles for the case of meson decay at rest with that of meson decaying in flight, with the  former being much tighter.  We stress that, 
significant difference in the mixing angles can occur  for  experiments NA62, LHCb, and the future experiment SHiP. Due to non-zero velocity of the parent meson, the probability of the generated RH neutrino  to decay inside the detector changes. This 
alters the sensitivity reach by more than an order of magnitude for the  above mentioned experiments.

We explore a number of channels, $B/D/D_s \to \mu \mu \pi, B/D/D_s \to e e \pi, B_c/D_s \to e  \mu \pi, B \to \tau \tau \pi, B/B_c \to e \tau \pi, B/B_c \to \mu \tau \pi $, and few others.
We find that,  for the mass range $M_{N}\sim 1$~GeV, future experiment SHiP can probe $|V_{eN}|^{2}\sim 10^{-9}$, while for mass range $M_N\sim 5$~GeV, future experiment MATHUSLA, and LHCb with 300 $\rm{fb}^{-1}$ integrated luminosity  can probe $|V_{eN}|^2\sim 10^{-7}$. The sensitivity reach of $|V_{\mu N}|^2$ is very similar to $|V_{eN}|^2$. For $|V_{\tau N}|^2$, the best sensitivity reach $|V_{\tau N}|^2\sim 10^{-7}$ can be provided by MATHUSLA in the mass range $2\,\text{GeV}<M_N<3.4\,\text{GeV}$, while in the mass range $3.4\,\text{GeV}<M_N<4.4\,\text{GeV}$, SHiP and LHCb
gives similar sensitivity reach. For $|V_{eN}V_{\mu N}|$, mass range up to $M_N\sim 2$ GeV can be probed at SHiP with the sensitivity reach $|V_{eN}V_{\mu N}|\sim 10^{-9}$, while higher mass $M_{N}\sim 5$~GeV and $M_N\sim 6$~GeV can be probed at MATHUSLA and LHCb with sensitivity reach $|V_{eN}V_{\mu N}|\sim 10^{-7}$. The highest sensitivity reach on $|V_{eN}V_{\tau N}|$, $|V_{\mu N}V_{\tau N}|\sim 10^{-7}$ can be provided by MATHUSLA and LHCb. The combined sensitivity of the mixing angles $|V_{eN}V_{\mu N}|$, $|V_{eN}V_{\tau N}|$ and $|V_{\mu N}V_{\tau N}|$ from meson decays are tighter than the other constraints available in a large range of heavy neutrino mass.
\section*{Acknowledgements}
The work of A. D. is supported by the Japan Society for the Promotion of Science (JSPS) Post- doctoral Fellowship for Research in Japan. MM acknowledges the support of DST INSPIRE Research Grant IFA14-PH-99.
\section*{Appendix} 
The different partial decay widths of the RH neutrinos $N_{i}$ are,
\begin{align}
\Gamma (N_{j}\rightarrow\ell^{-}P^{+})&=\frac{G_{F}^{2}M_{N_{j}}^{3}}{16\pi}f_{p}^{2}\left|V^{CKM}_{q\bar{q}^{'}}\right|^{2}\left|V_{\ell_1 N_{j}}\right|^{2}F_{P}\left(x_{\ell},x_{P}\right),
\end{align}
\begin{align}
\Gamma\left(N_{j}\rightarrow\ell^{-}V^{+}\right)&=\frac{G_{F}^{2}M_{N_{j}}^{3}}{16\pi}f_{V}^{2}\left|V^{CKM}_{q\bar{q}^{'}}\right|^{2}\left|V_{\ell_1 N_{j}}\right|^{2}F_{V}\left(x_{\ell},x_{V}\right),
\end{align}
\begin{align}
 \Gamma\left(N_{j}\rightarrow\nu_{\ell}P^{0}\right)&=\frac{G_{F}^{2}M_{N_{j}}^{3}}{4\pi}f_{P}^{2}\sum_{i}\left|U_{\ell i}\right|^{2}\left|V_{\ell_1N_{j}}\right|^{2}K_{P}^{2}F_{P}\left(x_{\nu_{\ell}},x_{P}\right),
\end{align}
\begin{align}
\Gamma\left(N_{j}\rightarrow\nu_{\ell}V^{0}\right)&=\frac{G_{F}^{2}M_{N_{j}}^{3}}{4\pi}f_{V}^{2}\sum_{i}\left|U_{\ell i}\right|^{2}\left|V_{\ell_1N_{j}}\right|^{2}K_{V}^{2}F_{V}\left(x_{\nu_{\ell}},x_{P}\right),
\end{align}
\begin{align}
 \Gamma\left(N_{j}\rightarrow\ell_{1}^{-}\ell_{2}^{+}\nu_{\ell_{2}}\right)&=\frac{G_{F}^{2}M_{N_{j}}^{5}}{16\pi^{3}}\left|V_{\ell_{1}N_{j}}\right|^{2}\sum_{i}\left|U_{\ell_{2}i}\right|^{2}
 I_{1}\left(x_{\ell_{1}},x_{\nu_{\ell_{2}}},x_{\ell_{2}}\right),
\end{align}
\begin{align}
\Gamma\left(N_{j}\rightarrow\nu_{\ell_{2}}\ell_{2}^{-}\ell_{2}^{+}\right)&=\frac{G_{F}^{2}M_{N_{j}}^{5}}{16\pi^{3}}\left|V_{\ell_{2}N_{j}}\right|^{2}\sum_{i}\left|U_{\ell_{2}i}\right|^{2}\bigg[I_{1}
\left(x_{\nu_{\ell_{2}}},x_{\ell_{2}},x_{\ell_{2}}\right)+2\left((g_{V}^{\ell})^{2}+(g_{A}^{\ell})^{2}\right)\nonumber\\
&I_{1}
\left(x_{\nu_{\ell_{2}}},x_{\ell_{2}},x_{\ell_{2}}\right)
+2\left((g_{V}^{\ell})^{2}-(g_{A}^{\ell})^{2}\right)I_{2}
\left(x_{\nu_{\ell_{2}}},x_{\ell_{2}},x_{\ell_{2}}\right)\bigg]
\end{align}
\begin{align}
\Gamma\left(N_{j}\rightarrow\nu_{\ell_{1}}\ell_{2}^{-}\ell_{2}^{+}\right)&=\frac{G_{F}^{2}M_{N_{j}}^{5}}{8\pi^{3}}\left|V_{\ell_{1}N_{j}}\right|^{2}\sum_{i}\left|U_{\ell_{1}i}\right|^{2}\bigg[\big((g_{V}^{\ell})^{2}
+(g_{A}^{\ell})^{2}\big)I_{1} \left(x_{\nu_{\ell_{1}}},x_{\ell_{2}},x_{\ell_{2}}\right)\nonumber\\
&+\big((g_{V}^{\ell})^{2}-(g_{A}^{\ell})^{2}\big)I_{2} \left(x_{\nu_{\ell_{1}}},x_{\ell_{2}},x_{\ell_{2}}\right)\bigg]\\
\end{align}
In the above decay mode $\ell_{1}\ne\ell_{2}$.
\begin{align*}
\Gamma\left(N_{j}\rightarrow\nu_{\ell}\nu\overline{\nu}\right)&=\frac{G_{F}^{2}M_{N_{j}}^{5}}{192\pi^{3}}\left|V_{\ell N_{j}}\right|^{2}\sum_{i}\left|U_{\ell i}\right|^{2},
\end{align*}
where $x_{i}=\frac{m_{i}}{M_{N}}$ with $m_{i}=m_{\ell}, m_{P^{0}},m_{V^{0}},m_{P^{+}},m_{V}^{+}$.
The kinematical function are given by,
\begin{align*}
I_{1}(x,y,z)&=\int_{(x+y)^{2}}^{(1-z)^{2}}\frac{ds}{s}(s-x^{2}-y^{2})(1+z^{2}-s)\lambda^{\frac{1}{2}}(s,x^{2},y^{2})\lambda^{\frac{1}{2}}(1,s,z^{2});\\
I_{2}(x,y,z)&=yz\int_{(y+z)^{2}}^{(1-x)^{2}}\frac{ds}{s}(1+x^{2}-s)\lambda^{\frac{1}{2}}(s,y^{2},z^{2})\lambda^{\frac{1}{2}}(1,s,x^{2});\\
F_{P}(x,y)&=\big((1+x^{2})(1+x^{2}-y^{2})-4x^{2}\big)\lambda^{\frac{1}{2}}(1,x^{2},y^{2});\\
F_{V}(x,y)&=\big((1-x^{2})^{2}+(1+x^{2})y^{2}-2y^{4}\big)\lambda^{\frac{1}{2}}(1,x^{2},y^{2}).
\end{align*}
Neutral current couplings of leptons are given by,
\begin{eqnarray}
&g_{V}^{\ell}=-\frac{1}{4}+\text{sin}^{2}\theta_{w},\,\,\,g_{A}^{\ell}=\frac{1}{4},
\end{eqnarray}
Neutral current coupling of pseudoscalar and vector mesons are given by,
\begin{align*}
&K_{\pi^{0}}=-\frac{1}{2\sqrt{2}},\,\,K_{\eta}=-\frac{1}{2\sqrt{6}},\,\,K_{\eta^{\prime}}=\frac{1}{4\sqrt{3}},\,\,K_{\eta_{c}}=-\frac{1}{4},\,\,K_{\rho^{0}}=\frac{1}{\sqrt{2}}(\frac{1}{2}-\text{sin}^{2}\theta_{w}),\nonumber\\
&K_{\omega}=-\frac{1}{3\sqrt{2}}\text{sin}^{2}\theta_{w},\,\,K_{\phi}=(-\frac{1}{4}+\frac{1}{3}\text{sin}^{2}\theta_{w}),\,\,K_{J/\psi}=(\frac{1}{4}-\frac{2}{3}\text{sin}^{2}\theta_{w})\\
\end{align*}
\bibliography{bibitem}
\bibliographystyle{utphys}
\end{document}